\documentclass[12pt,a4paper]{article}

\usepackage[T1]{fontenc}
\usepackage{lmodern}
\usepackage[a4paper,margin=28mm]{geometry}
\usepackage{microtype}
\usepackage{setspace}
\usepackage{amsmath,amssymb,amsfonts,mathtools}
\usepackage{booktabs}
\usepackage{array}
\usepackage{makecell}
\usepackage{rotating}
\usepackage[inline]{enumitem}
\usepackage{algorithm}
\usepackage{algpseudocode}
\usepackage[titletoc,title]{appendix}
\usepackage{graphicx}
\usepackage[section]{placeins}
\usepackage{chngcntr}
\usepackage{xcolor}
\usepackage[authoryear]{natbib}
\usepackage[hidelinks]{hyperref}
\usepackage[capitalise,nameinlink,noabbrev]{cleveref}

\allowdisplaybreaks

\newcommand{\R}{\mathbb{R}}
\newcommand{\E}{\mathbb{E}}
\newcommand{\I}{\mathbf{1}}
\newcommand{\D}{\mathcal{D}}
\newcommand{\Hset}{\mathcal{H}}
\newcommand{\given}{\,|\,}
\newcommand{\dd}{\,\mathrm{d}}
\newcommand{\wt}{\widetilde}

\newcommand{\trans}{^{\mathsf T}}
\DeclareMathOperator{\ESS}{ESS}
\DeclareMathOperator{\CESS}{CESS}
\DeclareMathOperator{\PG}{PG}
\DeclareMathOperator{\GammaDist}{Gamma}
\DeclareMathOperator{\BetaDist}{Beta}
\newcommand{\appref}[1]{\hyperref[#1]{Appendix~\ref*{#1}}}
\newcolumntype{L}[1]{>{\raggedright\arraybackslash}p{#1}}

\crefname{algorithm}{algorithm}{algorithms}
\Crefname{algorithm}{Algorithm}{Algorithms}
\crefname{equation}{Eq.}{Eqs.}
\Crefname{equation}{Eq.}{Eqs.}

\makeatletter
\providecommand{\theHALG@line}{}
\renewcommand{\theHALG@line}{\thealgorithm.\arabic{ALG@line}}
\makeatother

\definecolor{titleaccent}{RGB}{0,0,0}
\definecolor{affilgray}{RGB}{0,0,0}

\newcommand{\rev}[1]{{#1}}
\newenvironment{revision}{\begingroup}{\endgroup}

\title{Online Bayesian Inference for a Dirichlet Process Mixture of Binary Logit Models}
\author{Hajime Watanabe \and Charisma F.\ Choudhury \and Stephane Hess}
\date{}

\makeatletter
\renewcommand{\maketitle}{%
  \begin{center}
    \vspace*{0.6em}
    {\fontsize{18}{22}\selectfont\bfseries\color{titleaccent}\@title\par}
    \vspace{0.5em}
    {\large
    Hajime Watanabe$^{1}$ \quad
    Charisma F.\ Choudhury$^{2}$ \quad
    Stephane Hess$^{2}$\par}
    \vspace{0.8em}
    {\small\color{affilgray}
    $^{1}$Department of Civil Engineering, The University of Tokyo, 7-3-1 Hongo, Bunkyo-ku, Tokyo 113-8656, Japan\par}
    \vspace{0.35em}
    {\small\color{affilgray}
    $^{2}$Choice Modelling Centre and Institute for Transport Studies, University of Leeds, 34--40 University Road, Leeds LS2 9JT, United Kingdom\par}
    \vspace{0.75em}
    {\small
    \href{mailto:hwatanabe@bin.t.u-tokyo.ac.jp}{hwatanabe@bin.t.u-tokyo.ac.jp}
    \quad
    \href{mailto:C.F.Choudhury@leeds.ac.uk}{C.F.Choudhury@leeds.ac.uk}
    \quad
    \href{mailto:S.Hess@leeds.ac.uk}{S.Hess@leeds.ac.uk}\par}
    \vspace{0.75em}
    {\color{titleaccent}\rule{0.72\textwidth}{0.5pt}\par}
    \vspace{0.3em}
  \end{center}
}
\makeatother

\begin{document}

\maketitle

\begin{abstract}
\noindent

\textcolor{black}{Flexible travel behaviour models for preference heterogeneity have largely remained within the batch-estimation paradigm. The growing availability of continuously collected behavioural data, for example from smartphone applications, highlights the need for methods that update posterior inference on such heterogeneity as new observations arrive.}
This study proposes an online Bayesian inference method for a Dirichlet process mixture of binary logit models. The main challenge is to update the posterior as observations arrive while allowing the latent cluster structure to expand and historical cluster assignments to be revised. We develop a sequential Monte Carlo sampler using a sequence of P\'olya--Gamma-augmented partial-data posteriors, local particle-weight corrections, and recursively maintained sufficient statistics. Each observation's contribution can be added, removed, or replaced in these summaries, supporting both sequential coefficient updating and rejuvenation that revisits historical cluster assignments. The simulation study shows that observation-by-observation sequential updating yields final posterior summaries close to those obtained by batch Markov chain Monte Carlo. Sensitivity analyses further demonstrate trade-offs between posterior approximation accuracy and computational cost as the number of particles and rejuvenation settings are varied. An illustrative empirical application shows how the method tracks posterior evolution and produces observation-specific provisional posterior summaries with low latency as real-world trip records are processed sequentially. The proposed approach provides a methodological foundation for online inference in travel behaviour models with parameter heterogeneity and for future extensions to richer choice structures, low-latency decision support, and online anomaly screening.

\end{abstract}


\newpage
\section{Introduction}
\label{sec:introduction}

It has become increasingly common in transportation research for trip-level behavioural data to be accumulated sequentially through smartphone applications, Mobility as a Service, location-based logs, and continuous usage histories \citep{andersson2018promoting}. Updating predictions of travel choices as new observations arrive could support timely information provision or interventions aimed at congestion avoidance, mode shift, or departure-time change \citep{zhu2020personalized,xie2024real}. Because behavioural responses may vary across trips and choice situations, these predictions should account for preference heterogeneity. Such applications therefore call for methods that can learn about this heterogeneity from newly arriving observations and update choice predictions with low latency \citep{zhu2020online}.

The field of travel behaviour analysis has already accumulated a rich body of theory, models, and empirical findings for representing such preference heterogeneity in discrete choice \citep{train2009discrete,hess2024handbook}. For example, through latent class models, mixed logit models, and other mixture models, extensive research has long examined how differences in sensitivities across individuals and latent cluster structures can be interpreted and incorporated into policy analysis (e.g., \citep{bhat2002unified,krueger2020dirichlet}). A comprehensive review of finite-mixture (latent-class) modelling in transportation is provided by \citet{kim2023finite}. However, many of these flexible approaches to modelling heterogeneity still rely on static batch estimation. A direct application of a conventional batch estimator to streaming data involves appending each new observation to the existing dataset and rerunning estimation on the enlarged sample. Repeating this procedure at every arrival can become costly as the observation history grows. Sequential inference instead aims to build on existing estimates and update them as new observations arrive. Although the transportation field offers a rich set of tools for modelling heterogeneity, the theoretical and computational basis for using them in online inference remains limited \citep{danaf2019online,van2022choice}.

Moving beyond this batch-estimation paradigm first requires clarifying the requirements for modelling preference heterogeneity as behavioural observations arrive sequentially. In particular, the level of complexity required to represent heterogeneity at any given stage may not be known in advance. In an online inference setting, this makes it difficult to commit to a fixed specification, such as a finite mixture with a pre-specified number of clusters, and then compare multiple alternatives through repeated trial and error. Instead, the modelling framework should be able both to revise the relative importance of existing clusters and to introduce new ones as new observations reveal previously unobserved patterns of heterogeneity. From this perspective, Dirichlet process (DP) mixture models provide an appealing basis for online learning of heterogeneity under streaming data, because they allow both the reweighting of existing clusters and the creation of new ones within a unified sequential updating framework, without fixing the number of clusters a priori \citep{tank2015streaming,huynh2017streaming}.

This study develops an online Bayesian inference method for a Dirichlet process mixture of binary logit models (DP-BL). \textcolor{black}{The model represents preference heterogeneity across observations (inter-observation heterogeneity).} We adopt the sequential Monte Carlo (SMC) samplers framework to update the posterior as observations arrive. The proposed algorithm uses P\'olya--Gamma (PG) data augmentation to represent the information required for coefficient updates through compact summaries. An observation's contribution can be added to, removed from, or replaced in these summaries without reconstructing them from the full observation history. The algorithm also revisits historical cluster assignments to improve exploration of the current posterior. The simulation study evaluates posterior approximation accuracy and computational cost. An illustrative empirical application tracks posterior evolution as trip records are processed and measures the time required to produce posterior summaries that could support subsequent choice predictions.

An important clarification is that online updating in this paper does not turn the travel behaviour model into a dynamic model, such as a state-space model \citep{lachaab2006modeling} or a hidden Markov model \citep{xiong2017dynamic}. The DP-BL model remains a static choice model, and newly arriving observations update the posterior distribution over its parameters and the assignments of observations to latent clusters. Changes in posterior summaries should therefore be interpreted as sequential learning about preference heterogeneity, not as evidence that travellers' preferences evolve dynamically. Unlike these dynamic formulations, the present model does not require a transition equation specifying how preferences or latent states evolve from one observation to the next. This allows choice predictions to be updated as observations accumulate while preserving the conventional interpretation of coefficients and preference heterogeneity in static travel behaviour models.

\newpage
\section{Literature review}
\label{sec:literature}
\subsection{Sequential inference based on data-tempering SMC}
\label{sec:lit:smc}
\begingroup
\setstretch{1.03}
For online inference in DP-BL, the objective is to update the posterior distribution over behavioural coefficients and latent clusters as observations arrive. This requires representing uncertainty about both coefficient values and the number and composition of clusters, rather than retaining only point estimates or a single clustering. The choice of inference method should therefore be assessed in terms of both its ability to represent this uncertainty and the computational cost of sequential updating.

Several approaches could be considered for this task, including recursive optimisation, streaming approximate Bayesian methods, particle-based posterior updating, and \mbox{offline--online} updating schemes. Several studies in travel behaviour research have examined how individual-level preference information can be updated as new choices or trip histories become available. \citet{danaf2019online} periodically re-estimate their hierarchical model offline, while updating stored individual-level preferences in real time as new choices are observed. \citet{zhu2020online} similarly estimate preference patterns shared across individuals offline and update each individual's weights on these patterns during the online stage. Relatedly, \citet{xie2020personalized} use accumulated personal trip histories to update the individual-specific random-parameter distribution within a specified logit-mixture model; their focus is history-conditioned personalisation rather than recursively updating the model as a whole. These studies demonstrate that preference information can be incorporated in real time or from accumulating histories, but they do not provide a general observation-by-observation scheme for maintaining posterior uncertainty over a nonparametric mixture structure whose occupied components are learned sequentially.

Recursive optimisation-based methods \citep{cappe2009line}, such as stochastic gradient updates and online maximum-likelihood estimation, can deliver fast point updates, but they do not naturally preserve uncertainty in the full posterior distribution, multimodality, or uncertainty regarding latent clusters. Likewise, streaming approximate Bayesian methods, such as online variational inference and related filtering-type approximations \citep{huynh2016streaming}, can provide usable sequential updates. However, assumptions used to approximate the posterior, such as Gaussian coefficient distributions or independence between parameters, may limit how well dependence, multimodality, and discrete latent structure are represented \citep{blei2017variational}. These limitations are relevant here because uncertainty about the latent cluster structure is part of the inferential target.

Online inference has also been developed for dynamic logistic regression, allowing coefficients and model probabilities to change over time \citep{mccormick2012dynamic}. \citet{jiang2023dynamic} develop a logistic state-space model with both time-varying and time-invariant coefficients, updated using data accumulated over prespecified intervals. For static generalised linear models, \citet{ghosh2025online} develop online Bayesian model averaging using new data batches and summaries of previous batches. Their focus is uncertainty about covariate selection, whereas the present study concerns sequential posterior learning about coefficients and latent cluster structure under a static DP-mixture specification.

SMC provides a suitable framework for sequential posterior updating in DP-BL. Each particle represents one possible set of coefficient values and latent cluster assignments. Its weight determines its contribution to posterior summaries. Together, the weighted particles represent uncertainty about the coefficients and latent cluster structure. As observations arrive, SMC updates the particles and their weights, allowing posterior summaries and choice predictions to be revised sequentially \citep{ChopinPapaspiliopoulos2020,hinne2025introduction}. Although SMC originated in particle filtering for dynamic state-space models, the broader SMC samplers framework has made it possible to perform SMC-based sequential inference for sequences of posterior distributions beyond the classical state-space models \citep{DelMoralDoucetJasra2006,DaiHengJacobWhiteley2022}. Here, because the present problem concerns sequential posterior updating for static parameters rather than latent state filtering, we use the term \textit{static SMC} to denote SMC targeting a sequence of posterior distributions over static parameters. Within the SMC samplers framework, two representative ways of constructing the target sequence are \textit{temperature tempering} and \textit{data tempering} \citep{ChopinPapaspiliopoulos2020}. The former constructs an artificial sequence of intermediate distributions by gradually varying a tempering parameter, and is mainly used to stabilise posterior simulation for a complex target under a fixed dataset, that is, in a batch-inference setting \citep{neal2001annealed}. By contrast, the latter introduces observations one by one, or in small batches, and defines the target at each stage as the posterior distribution conditional on the observations available up to that stage \citep{chopin2002sequential}. For this reason, data tempering corresponds more directly to the idea of online posterior updating; at the same time, it is also often used as a bridge construction for reaching a final posterior distribution stably by introducing a fixed static dataset gradually \citep{fearnhead2004particle,griffin2017sequential,burda2023hamiltonian}. Moreover, in practice, temperature tempering can be incorporated within each stage of a data-tempering update in order to improve numerical stability and particle diversity \citep{ChopinPapaspiliopoulos2020}.

Within the broader lineage of data-tempering-based static SMC algorithms, a variety of sequential methods have been discussed and proposed for approximating posterior distributions while introducing data gradually, ranging from logistic regression to Bayesian nonparametric mixture models and, more recently, to DP-based multinomial discrete choice models. For example, \cite{DaiHengJacobWhiteley2022}, in their overview of the general SMC samplers framework, present as an illustrative example the SMC samplers construction of \cite{chopin2002sequential} along a partial-data posterior path for a binary logit model, and visualise how the posterior distribution and predictive performance evolve as observations are added. Building on the same general static-SMC idea as \cite{chopin2002sequential}, \cite{fearnhead2004particle}, \cite{ulker2010sequential}, and \cite{griffin2017sequential} developed sequential inference algorithms for DP mixture models. Within this line of work, \cite{burda2023hamiltonian} constructed an algorithm for inference in a DP-based nonparametric mixed multinomial logit model under a data-tempering framework. Although the construction of \citet{burda2023hamiltonian} can in principle be continued as new data arrive, its primary objective is efficient posterior exploration for a fixed dataset rather than low-latency online inference. The present study focuses on the latter computational objective.

\endgroup
\newpage
\subsection{Design requirements for online inference and the role of P\'olya--Gamma data augmentation}
\label{sec:lit:requirements}
For online inference in DP mixture choice models used in travel behaviour analysis, the key question is whether such inference can be implemented in a genuinely sequential way, with local computational updates whenever possible, and at lower cost than repeating batch estimation whenever a new observation arrives. The following three design requirements motivate the proposed algorithm. First, the posterior sequence must accommodate new latent clusters as observations arrive \citep{DelMoralDoucetJasra2006,ChopinPapaspiliopoulos2020,DaiHengJacobWhiteley2022}. Within a data-tempering SMC samplers framework, the particle weights should be updated using the new observation and the information already retained in each particle, without recalculating the likelihood of all previously processed observations \citep{DelMoralDoucetJasra2006,DaiHengJacobWhiteley2022}; see also \citet{ulker2010sequential} for a DP-mixture implementation. Secondly, local recursive updates are highly desirable. For online inference, the contribution of each new observation should ideally be incorporated into recursively updated low-dimensional summaries. This allows within-particle coefficient updates to use stored summaries without rescanning the full observation history. This makes it easier to decouple the per-observation update cost from the length of the history. This ideal is reflected in the broader literature on sequential Bayesian computation, including methods such as particle learning \citep{carvalho2010particle} and the sequential updating and greedy search algorithm \citep{wang2011fast}, which improve computational efficiency by updating stored summaries as new observations arrive. \rev{Thirdly, particle updates should improve posterior exploration at a manageable computational cost. Each particle has a weight that determines its contribution to posterior summaries. As observations accumulate, these weights can become concentrated in a few particles. To address this concentration, resampling draws a new, equally weighted particle population according to the existing weights, but can produce identical copies \citep{kantas2015particle}. These copies should then be updated to explore different parameter values and cluster assignments without changing the posterior being approximated. Such updates are known as rejuvenation \citep{griffin2017sequential}.} In practice, a limited number of Markov chain Monte Carlo (MCMC) updates can be used for this purpose \citep{DelMoralDoucetJasra2006,kantas2015particle,ChopinPapaspiliopoulos2020,DaiHengJacobWhiteley2022}. Taken together, these considerations suggest that, for SMC to be operationally viable for online inference in complex mixture choice models, such as DP-mixture logit models, one needs
\begin{enumerate*}[label=(\roman*)]
\item a locally updated sequence of data-tempered targets, as embodied in \citet{burda2023hamiltonian},
\item local recursive updates based on sufficient statistics or analogous low-dimensional summaries that avoid repeated recomputation from scratch, and
\item rejuvenation steps that improve posterior exploration at a manageable computational cost.
\end{enumerate*}

In the DP-BL model considered in this study, one of the main obstacles to achieving (ii) and (iii) is the non-conjugacy of the likelihood. In conjugate settings, such as Gaussian DP mixture models, updates of cluster-specific parameters can be implemented via local sequential updates of sufficient statistics, and rejuvenation can also be carried out efficiently using Gibbs sampling. Our aim is to retain such computational conveniences even under the non-conjugate binary logit likelihood. To this end, PG data augmentation provides a way to recover conjugacy-like computational structure by converting the non-conjugate binary logit likelihood into a conditionally Gaussian form \citep{PolsonScottWindle2013}. The basic idea is to augment the target distribution by introducing an auxiliary PG variable for each binary logit observation. Such data augmentation is standard in MCMC for constructing conditionally conjugate updates \citep{tanner1987calculation}, and has been widely used in Bayesian inference for discrete choice models \citep{watanabe2023bayesian,watanabe2024bayesian,watanabe2026sample,oyama2024hierarchical}. PG data augmentation can also be used within the SMC samplers framework. The resulting joint posterior includes both the model parameters and PG variables, while retaining the original posterior as the marginal distribution of the model parameters. The PG variables are omitted when reporting posterior summaries for the model parameters \citep{DelMoralDoucetJasra2006,DaiHengJacobWhiteley2022}. In DP-BL, the conditionally Gaussian form allows coefficient updates to use stored low-dimensional matrices and vectors rather than repeatedly processing the full observation history. These summaries can be updated both when new observations arrive and when historical cluster assignments are revised, supporting sequential updating and particle rejuvenation.

\subsection{Contribution of this study}
\label{sec:lit:contribution}

The central contribution of this study is an online inference algorithm that rapidly incorporates each newly arriving behavioural observation into the posterior approximation and promptly returns updated posterior summaries. This supports sequential learning and timely choice prediction in applications where observations arrive in rapid succession. To achieve this, we develop a static SMC algorithm for DP-BL by combining the SMC samplers framework for data-tempered posterior sequences \citep{DelMoralDoucetJasra2006,burda2023hamiltonian}, existing sequential inference methods for DP mixtures \citep{ulker2010sequential,griffin2017sequential}, and the PG data augmentation identity \citep{PolsonScottWindle2013}. The algorithm targets the PG-augmented posterior based on all observations processed so far and applies local particle-weight corrections as each new observation is incorporated. It maintains sufficient statistics by adding, removing, or replacing observation contributions, without reconstructing them from the full observation history. The same statistics are used when historical cluster assignments are revised during rejuvenation. Observation-specific provisional summaries can be returned after the new observation has been incorporated, without waiting for historical rejuvenation to finish. The simulation study evaluates posterior approximation accuracy against batch MCMC and examines computational trade-offs across tuning settings. The illustrative empirical application examines sequential posterior learning and the latency of observation-specific provisional output using real-world trip records.

\newpage
\section{P\'olya--Gamma-augmented SMC for a Dirichlet process mixture of binary logit models}
\label{sec:method}

\subsection{Overview and posterior inference}
\label{sec:method:overview}

This subsection introduces the notation and basic concepts used to represent, update, and summarise the DP-BL posterior in the proposed SMC algorithm.

We consider posterior inference for DP-BL given a dataset
\[
\D_{1:n} = \{\D_i\}_{i=1}^n, \qquad \D_i = (y_i, x_i, z_i),
\]
where $y_i \in \{0,1\}$ is a binary outcome and $x_i \in \R^o$, $z_i \in \R^q$ are covariates. We formulate inference sequentially: after the first $i$ observations, the algorithm targets the corresponding partial-data posterior. For notational convenience, we write $\D_{1:n}$ for the first $n$ observations currently under analysis; this indexing does not restrict the method to a fixed sample size, and the same recursion applies when additional observations arrive online.

This posterior is defined over observation-level cluster assignments and an unknown, a priori unbounded set of distinct latent coefficient vectors. Let $\rev{C_i}$ denote the set of occupied clusters after processing observations $1{:}i$, and let
\[
\rev{K_i \coloneqq \lvert C_i \rvert}
\]
denote its size. The state consists of (i) cluster-specific coefficients $\beta = \{\beta_h\}_{h \in \rev{C_i}}$, (ii) the shared coefficients $\gamma$ (optional), (iii) cluster assignments $\zeta_{1:i} = (\zeta_1,\dots,\zeta_i)$, and (iv) the DP concentration parameter $\alpha$. Define the partial-data posterior after processing $i$ observations as
\begin{equation*}
\pi_i\bigl(\beta,\gamma,\zeta_{1:i},\alpha \given \D_{1:i}\bigr), \qquad i=1,2,\dots.
\end{equation*}
To obtain a tractable sequential update scheme, we adopt PG data augmentation \citep{PolsonScottWindle2013} and target the resulting augmented posterior sequence
\begin{equation}
\pi_i\bigl(\beta,\gamma,\zeta_{1:i},\omega_{1:i},\alpha \given \D_{1:i}\bigr),
\label{eq:augmented-posterior}
\end{equation}
where $\omega_{1:i} = (\omega_1,\dots,\omega_i)$ are PG auxiliary variables that render $(\beta,\gamma)$ conditionally Gaussian given $(\zeta_{1:i},\omega_{1:i})$. For these conditional coefficient updates, we maintain sufficient statistics: matrices and vectors that sum the outcome and covariate contributions under the current cluster assignments and PG variables. They can be updated by adding each new observation's contribution or by removing and replacing a historical contribution when its cluster assignment or PG variable changes. Using these summaries, the coefficients can be updated without recalculating these sums over the full observation history. Let
\[
\Theta_i \coloneqq (\beta,\gamma,\zeta_{1:i},\omega_{1:i},\alpha)
\]
denote the particle state at stage $i$. The SMC sampler approximates the target distribution $\pi_i(\Theta_i \given \D_{1:i})$ with a weighted particle system $\{\Theta_i^{(m)}, W_i^{(m)}\}_{m=1}^M$, where $W_i^{(m)} \ge 0$ and $\sum_{m=1}^M W_i^{(m)} = 1$. Each particle represents one possible value of $\Theta_i$, and its normalised weight determines its contribution to posterior expectations. Together, the weighted particles provide an empirical approximation of the posterior distribution, thereby representing posterior uncertainty rather than a single estimate. Particle weights are updated through incremental reweighting, which corrects for the change in the target distribution induced by the newly incorporated observation $\D_{i}$, and are then normalised across particles to obtain $\{W_i^{(m)}\}_{m=1}^M$.

In implementation, each particle stores the complete state $\Theta_i$ together with the recursively updated sufficient statistics required for fast updates of $(\beta,\gamma)$. For the static parameters $(\beta,\gamma,\alpha)$, only their current values at stage $i$ are retained; their values at earlier algorithmic stages are not stored. By contrast, the observation-indexed latent variables $(\zeta_{1:i},\omega_{1:i})$ are retained for all processed observations.\footnote{\rev{This implementation is not constant-memory: storage for the retained latent history increases approximately in proportion to both the number of particles and the number of processed observations. Additional storage is required for cluster-specific quantities and temporary copies created during resampling. Memory diagnostics for the baseline simulation are reported in Appendix~F.}} Retaining the current values of $(\beta,\gamma)$ facilitates efficient predictive evaluation and weight calculation. Because the sufficient statistics are deterministic functions of $\D_{1:i}$ and $(\zeta_{1:i},\omega_{1:i})$, they are suppressed from $\Theta_i$ for notational simplicity.

\begin{revision}
We combine resampling with a subsequent MCMC rejuvenation move. As observations are incorporated, particle weights may become concentrated on only a few particles, leaving most particles with negligible weight; this phenomenon is known as weight degeneracy. A standard diagnostic of weight degeneracy is the effective sample size (ESS), which decreases as the weights become more concentrated and can be interpreted as the approximate number of particles contributing meaningfully to the posterior approximation. When the ESS falls below a threshold, SMC samplers resample: low-weight particles are discarded and high-weight particles are replicated, redirecting computational effort towards states with greater posterior support. In implementation, each resampled particle is copied as a complete computational record: $\Theta_i$ and all associated stored quantities, including the sufficient statistics and cluster counts, are inherited from the same selected parent particle. Although resampling alleviates weight degeneracy, replicating high-weight states reduces the number of distinct states represented by the particle system, a phenomenon commonly referred to as \emph{particle impoverishment}. When many particles are duplicates, the nominal particle budget $M$ represents fewer regions of posterior support and fewer alternative allocation structures. This inefficient use of the particle population can increase Monte Carlo approximation error and weaken exploration of the current posterior. We therefore follow resampling with a $\pi_i$-invariant MCMC rejuvenation move, which allows replicated particles to move towards distinct posterior states without changing the target distribution or requiring further reweighting.
\end{revision}

We also include an on-demand safeguard against abrupt particle-weight concentration. Before applying the usual one-step weight update, we compute a diagnostic ESS from the provisional weights implied by the new observation to assess whether such concentration would occur. If the diagnostic indicates a sharp drop, the observation is incorporated gradually through the stabilising update described in \appref{app:tempering}; otherwise, it is incorporated through the standard one-step update. We refer to these as the \emph{stabilising route} and the \emph{standard route}, respectively. This diagnostic only selects between these two computational routes and does not alter the posterior target.

Finally, once the weighted particle system $\{\Theta_i^{(m)}, W_i^{(m)}\}_{m=1}^M$ is available, any posterior summary of interest can be obtained via Monte Carlo integration. Specifically, for any function $f$ under $\pi_i$,
\begin{equation}
\E_{\pi_i}\bigl[f(\Theta_i)\bigr] \approx \sum_{m=1}^M W_i^{(m)} f\bigl(\Theta_i^{(m)}\bigr).
\label{eq:mc-summary}
\end{equation}
For cluster-specific coefficients, each particle contains the occupied set $\{\beta_h^{(m)}:h\in\rev{C_i^{(m)}}\}$. Unless otherwise stated, the marginal posterior summaries of $\beta$ reported below are constructed by pooling these occupied-cluster coefficients across particles and assigning $\beta_h^{(m)}$ weight $W_i^{(m)}n_{h,i}^{(m)}/i$, where $n_{h,i}^{(m)}=\sum_{\ell=1}^{i}\I\{\zeta_\ell^{(m)}=h\}$. This cluster-size weighting yields the posterior distribution of the latent coefficient associated with an observation selected uniformly from the first $i$ processed observations. It differs from an equal-weighted distribution over occupied clusters, the observation-specific posterior introduced in \Cref{sec:method:provisional}, and the predictive coefficient distribution for a future observation, which also allows assignment to a newly instantiated cluster.

The remainder of this section is organised as follows. \Cref{sec:method:dpbl} describes the DP-BL model and highlights its key differences from a standard binary logit model. \Cref{sec:method:alloc} details the DP-mixture mechanism, including posterior updates for the cluster assignments and the concentration parameter $\alpha$. \Cref{sec:method:pg} applies PG data augmentation and derives closed-form recursive updates of the sufficient statistics for the coefficients $(\beta,\gamma)$. In \Crefrange{sec:method:alloc}{sec:method:pg}, the same within-particle update rules are applied separately to every particle. We therefore omit the particle superscript $(m)$ for notational simplicity; unless stated otherwise, each displayed update is understood to be carried out for every particle. \Cref{sec:method:smc} presents the remaining SMC-specific techniques, including the incremental weight update, diagnostic ESS, resampling, and rejuvenation move. Finally, \Cref{sec:method:practical} discusses practical considerations, including tuning parameters and posterior outputs used in sequential applications, and clarifies the roles of the simulation and illustrative empirical application. The complete sequence of operations is summarised in \Cref{alg:complete-smc} in \appref{app:algorithms}.

\clearpage

\subsection{A Dirichlet process mixture of binary logit models}
\label{sec:method:dpbl}

Consider a stream of observations indexed by $i=1,2,\dots$. For any finite $n$, we denote the first $n$ observations by $\D_{1:n}$. For each observation, let $y_i \in \{0,1\}$ denote a binary choice outcome. Let $u_i$ denote the latent utility difference between the two alternatives for observation $i$, and we observe $y_i = 1$ if $u_i > 0$ and $y_i = 0$ otherwise. Under a DP-BL specification, we introduce latent coefficient clusters indexed by $h \in \{1,2,\dots\}$. Conditional on the cluster assignment, the utility difference takes the form
\begin{equation}
u_i = x_i\trans \beta_h + z_i\trans \gamma + \varepsilon_i,
\label{eq:utility}
\end{equation}
where $x_i \in \R^o$ and $z_i \in \R^q$ are observed covariates (including a constant in either block but not both), $\beta_h \in \R^o$ is a cluster-specific coefficient vector that varies across clusters, and $\gamma \in \R^q$ is a coefficient vector shared across clusters. The error term $\varepsilon_i$ is assumed to follow the standard logistic distribution, which yields the binary logit choice probability
\begin{equation}
p\bigl(y_i \given \zeta_i = h, \beta_h, \gamma, x_i, z_i\bigr)
=
\frac{\exp\bigl(y_i(x_i\trans \beta_h + z_i\trans \gamma)\bigr)}{1 + \exp(x_i\trans \beta_h + z_i\trans \gamma)}.
\label{eq:logit-likelihood}
\end{equation}
The key difference from a standard binary logit is that coefficient heterogeneity is represented through a DP mixture over observation-level latent coefficient vectors. Formally, let $\theta_i$ denote the observation-level latent coefficient associated with observation $i$. We assume
\begin{equation}
G \sim \mathrm{DP}(\alpha, G_0),
\qquad
\theta_i \mid G \stackrel{\text{i.i.d.}}{\sim} G,
\qquad
\theta_i = \beta_{\zeta_i},
\label{eq:dp-prior}
\end{equation}
where $G_0$ is the base measure and $\alpha > 0$ is the concentration parameter controlling the expected number of occupied clusters. The discreteness of $G$ induces ties among the latent coefficients $\theta_i$; these ties define the cluster assignments $\zeta_{1:n} = (\zeta_1,\dots,\zeta_n)$ and the distinct cluster-specific coefficients $\{\beta_h\}$. Equivalently, under the Chinese restaurant process (CRP) representation used below, whenever a new cluster is instantiated, its coefficient $\beta_h$ is drawn from $G_0$.

This specification differs from the DP mixture-of-normals formulations used by \citet{burda2008bayesian} and \citet{krueger2020dirichlet}, in which coefficients vary continuously within mixture components. Here, the DP prior is placed directly on the distribution of coefficient vectors, and observations assigned to the same cluster share the same $\beta_h$. Thus, heterogeneity is represented through a discrete mixing distribution over cluster-specific coefficients, without an additional layer of within-cluster coefficient variation. Both formulations share the same underlying aim of flexibly representing coefficient heterogeneity without fixing the number of mixture components in advance.

\textcolor{black}{The latent coefficients $\theta_i$ and cluster assignments $\zeta_i$ are defined at the observation level, so the model represents inter-observation heterogeneity. When each individual contributes only a single observation, this corresponds to inter-person heterogeneity. When all observations come from the same individual, it corresponds to intra-person heterogeneity. When repeated observations are pooled across individuals, the present specification focuses on the overall heterogeneity across observations. Explicitly separating its inter-person and intra-person components would require an additional individual-level structure.}

\subsection{Cluster assignments and concentration-parameter updates}
\label{sec:method:alloc}

Let $\rev{C_{i-1}}$ denote the set of occupied clusters after processing observations $1{:}(i-1)$, and let $\rev{K_{i-1} \coloneqq \lvert C_{i-1} \rvert}$. For each occupied cluster $h \in \rev{C_{i-1}}$, let
\begin{equation*}
n_{h,i-1} \coloneqq \sum_{r=1}^{i-1} \I\{\zeta_r = h\},
\end{equation*}
be its current size, where $\I\{\cdot\}$ denotes the indicator function, which takes the value $1$ when the condition inside the braces is true. Under the CRP representation of the DP prior, the prior predictive allocation probabilities for the next label $\zeta_i$ are
\begin{equation}
p(\zeta_i = h \given \zeta_{1:i-1}, \alpha)
=
\begin{cases}
\dfrac{n_{h,i-1}}{\alpha + i - 1}, & \text{if } h \in \rev{C_{i-1}}, \\
\dfrac{\alpha}{\alpha + i - 1}, & \text{if } h = \text{new},
\end{cases}
\label{eq:crp}
\end{equation}
where ``new'' denotes instantiating a previously unseen cluster. This ``rich-get-richer'' mechanism implies that larger clusters are a priori more likely to attract new observations, while the creation of a new cluster is controlled by $\alpha$.

In this study we target the joint posterior over $\Theta_i = (\beta,\gamma,\zeta_{1:i},\omega_{1:i},\alpha)$. Accordingly, we update $\zeta_i$ without marginalising out $(\beta_h,\gamma)$. Conditional on the particle's current parameter values $(\beta,\gamma)$, the likelihood contribution $L_i(h)$ for assigning observation $i$ to an existing cluster $h \in \rev{C_{i-1}}$ is the binary logit probability in Eq.~\eqref{eq:logit-likelihood}. The assignment of $\zeta_i$ is proportional to the CRP prior mass in Eq.~\eqref{eq:crp} multiplied by that likelihood contribution. Writing $\D_i = (y_i,x_i,z_i)$, we have
\begin{equation}
p(\zeta_i = h \given \Theta_{i-1}, \D_i)
\propto
p(\zeta_i = h \given \zeta_{1:i-1}, \alpha) \, L_i(h).
\label{eq:alloc-unnormalised}
\end{equation}

In non-conjugate DP mixtures, the ``new'' option involves an intractable prior-predictive integral. We thus adopt the auxiliary-parameter method of \citet{Neal2000}, which avoids the intractable prior-predictive integral by introducing auxiliary candidates drawn from $G_0$. Specifically, at each assimilation step $i$, we introduce $J$ independent auxiliary candidate parameters for a potential new cluster:
\[
\wt{\beta}_{\mathrm{new}}^{(1:J)} \coloneqq \bigl(\wt{\beta}_{\mathrm{new}}^{(1)},\dots,\wt{\beta}_{\mathrm{new}}^{(J)}\bigr),
\qquad
\wt{\beta}_{\mathrm{new}}^{(j)} \sim G_0,
\quad j \in \{1,\dots,J\},
\]
and conceptually split the ``new'' mass $\alpha$ into $J$ equally weighted auxiliary labels $\mathrm{new}_1,\dots,\mathrm{new}_J$, each with prior mass $\alpha/J$. For each auxiliary new candidate $h = \mathrm{new}_j$, the prior predictive allocation probability is
\begin{equation*}
p(\zeta_i = \mathrm{new}_j \given \zeta_{1:i-1}, \alpha)
=
\frac{\alpha/J}{\alpha + i - 1},
\end{equation*}
and the likelihood contribution is
\begin{equation*}
L_i(\mathrm{new}_j)
\coloneqq
\frac{\exp\bigl(y_i(x_i\trans \wt{\beta}_{\mathrm{new}}^{(j)} + z_i\trans \gamma)\bigr)}{1 + \exp(x_i\trans \wt{\beta}_{\mathrm{new}}^{(j)} + z_i\trans \gamma)}.
\end{equation*}
Let the candidate label set be
\[
\Hset_i = \rev{C_{i-1}} \cup \{\mathrm{new}_1,\dots,\mathrm{new}_J\}.
\]
Normalising Eq.~\eqref{eq:alloc-unnormalised} over $h \in \Hset_i$ yields a proper discrete distribution from which we sample $\zeta_i$. We define the corresponding allocation normalising constant as
\begin{equation}
Z_i \coloneqq \sum_{h \in \Hset_i} p(\zeta_i = h \given \zeta_{1:i-1}, \alpha) \, L_i(h).
\label{eq:Zi}
\end{equation}
Note that $Z_i$ is computed at the allocation step, using the current particle state and auxiliary candidates before the subsequent Gibbs updates of $\alpha$, $\omega_i$, and $(\beta,\gamma)$. This allocation normalising constant is stored within each particle and used in the weight update described in \Cref{sec:method:weight}. \rev{Because the auxiliary candidate parameters are random draws from $G_0$, $Z_i$ is random conditional on the previous particle state and the newly arriving observation. This auxiliary randomness is included in the extended-state SMC construction derived in \appref{app:weight}. Conditional on the particle state and the realised auxiliary candidate block, $Z_i$ sums over every candidate label and is therefore unchanged by the subsequently realised draw of $\zeta_i$. The same label-invariance property is used by the stabilising update in \appref{app:tempering}.} The assignment probability for each cluster is then given by Eq.~\eqref{eq:alloc-unnormalised} divided by $Z_i$, as follows:
\begin{equation}
p\bigl(\zeta_i = h \given \Theta_{i-1}, \wt{\beta}_{\mathrm{new}}^{(1:J)}, \D_i\bigr)
=
\frac{p(\zeta_i = h \given \zeta_{1:i-1}, \alpha) \, L_i(h)}{Z_i},
\qquad h \in \Hset_i.
\label{eq:alloc-normalised}
\end{equation}
We then sample $\zeta_i$ from the discrete distribution defined by Eq.~\eqref{eq:alloc-normalised}. The realised label and auxiliary candidate block are retained until the diagnostic ESS has selected the update route. Under the standard route, if an auxiliary label $\zeta_i = \mathrm{new}_j$ is selected, we instantiate a new cluster $h^\ast$, set $\beta_{h^\ast} \coloneqq \wt{\beta}_{\mathrm{new}}^{(j)}$, initialise its cluster-specific sufficient statistics at zero, and update the occupied set and cluster counts before adding observation $i$'s contribution. Under the stabilising route, the realised label may be refreshed along the tempered path, so instantiation is deferred until the final allocation has been obtained in \appref{app:tempering}.

We place a Gamma prior on the concentration parameter,
\begin{equation*}
\alpha \sim \GammaDist(a_0,b_0),
\end{equation*}
with shape $a_0 > 0$ and rate $b_0 > 0$. Conditional on the current partition of the first $i$ observations, which is summarised by the number of occupied clusters $K_i$, a standard auxiliary-variable update yields a simple Gibbs step for $\alpha$ \citep{EscobarWest1995}. Introduce
\begin{equation}
\lambda \given \alpha, i \sim \BetaDist(\alpha + 1, i),
\label{eq:escobar-west-lambda}
\end{equation}
and define the mixing probability
\begin{equation*}
p_\lambda = \frac{a_0 + K_i - 1}{(a_0 + K_i - 1) + i\bigl(b_0 - \log \lambda\bigr)}.
\end{equation*}
Then the conditional posterior of $\alpha$ is a two-component Gamma mixture:
\begin{equation}
\begin{aligned}
\alpha \given \lambda, K_i
\sim{}&
p_\lambda \,
\GammaDist\bigl(a_0 + K_i,\, b_0 - \log \lambda\bigr) \\
&+
(1-p_\lambda) \,
\GammaDist\bigl(a_0 + K_i - 1,\, b_0 - \log \lambda\bigr).
\end{aligned}
\label{eq:alpha-mixture}
\end{equation}
This update depends on the data only through the current number of occupied clusters $K_i$, and is therefore inexpensive to embed within sequential algorithms. In our sequential setting, Eqs.~\eqref{eq:escobar-west-lambda}--\eqref{eq:alpha-mixture} can be applied after processing observation $i$ (and, when rejuvenation is invoked, within the rejuvenation move; see \Cref{sec:method:smc}), allowing $\alpha$ to be updated in accordance with the current number of occupied clusters.

\subsection{P\'olya--Gamma data augmentation and recursive sufficient-statistics updates}
\label{sec:method:pg}

Following \citet{PolsonScottWindle2013}, this subsection applies the PG representation to the binary logit likelihood (conditional on $\zeta_i = h$), which yields conditionally Gaussian updates for the coefficient blocks $\{\beta_h\}$ and $\gamma$. The key outcome is a set of closed-form sufficient statistics that can be updated recursively, enabling efficient within-particle updates. As stated in \Cref{sec:method:overview}, we suppress the particle superscript $(m)$; all steps below are understood to be applied within each particle.

\subsubsection{P\'olya--Gamma data augmentation and Gaussian kernel}
\label{sec:method:pg-kernel}

For observation $i$, assigned to cluster $h = \zeta_i$, define the systematic term
\begin{equation*}
\eta_{hi} \coloneqq x_i\trans \beta_h + z_i\trans \gamma,
\end{equation*}
and let $\delta_i \coloneqq y_i - \tfrac{1}{2}$. Using the PG identity \citep{PolsonScottWindle2013}, we can write the binary logit likelihood contribution as
\begin{equation}
\frac{\exp(\eta_{hi})^{y_i}}{1 + \exp(\eta_{hi})}
=
\frac{1}{2}
\int_0^\infty
\exp\!\left(\delta_i \eta_{hi} - \frac{\omega_i}{2}\eta_{hi}^2\right)
\, p(\omega_i \given 1,0)
\, \dd \omega_i,
\label{eq:pg-identity}
\end{equation}
where $p(\omega_i \given 1,0)$ is the $\PG(1,0)$ density. Introducing the auxiliary variables $\omega_i$ yields an augmented likelihood whose kernel is quadratic in $\eta_{hi}$, and hence admits Gaussian-form updates for the coefficients conditional on $\omega_i$. To make this explicit, note that
\begin{align}
\exp\!\left(\delta_i \eta_{hi} - \frac{\omega_i}{2}\eta_{hi}^2\right)
&= \exp\!\left\{-\frac{\omega_i}{2}\left(\frac{\delta_i}{\omega_i} - \eta_{hi}\right)^2\right\}
   \exp\!\left(\frac{\delta_i^2}{2\omega_i}\right) \\
&\propto \exp\!\left\{-\frac{\omega_i}{2}\left(\frac{\delta_i}{\omega_i} - \eta_{hi}\right)^2\right\},
\label{eq:gaussian-kernel}
\end{align}
where $\propto$ denotes equality up to multiplicative factors that do not depend on the parameter block being updated (here, $\{\beta_h\}$ and $\gamma$). Because $\eta_{hi} = x_i\trans \beta_h + z_i\trans \gamma$, Eq.~\eqref{eq:gaussian-kernel} can be viewed as a Gaussian regression kernel with pseudo-outcome $\delta_i/\omega_i$ and variance $1/\omega_i$.

\subsubsection{Conjugate Gaussian priors and recursive sufficient statistics}
\label{sec:method:recursive-stats}

Under the PG representation in Eq.~\eqref{eq:gaussian-kernel}, the binary logit likelihood can be viewed as a conditionally Gaussian regression model. Hence, conditional on the current auxiliary variables $\omega_{1:i}$ and allocations $\zeta_{1:i}$, the full conditional distributions of the coefficient blocks $\beta_h$ and $\gamma$ follow from standard Gaussian posterior algebra. In particular, the required posterior moments depend only on a small set of weighted cross-products and linear terms, which can be updated recursively and carried forward. \appref{app:posterior-derivation} provides the corresponding derivation.

Let $\delta_r \coloneqq y_r - \tfrac{1}{2}$ for $r = 1,\dots,i$. After processing observations $1{:}i$, define the following quantities for the cluster-specific coefficients $\beta_h$ of each occupied cluster $h$:
\begin{align}
    S_{xx,h}^{(i)} &\coloneqq \sum_{r=1}^i \I\{\zeta_r = h\} \, \omega_r x_r x_r\trans, \label{eq:cluster-aggregates-first} \\
s_{x,h}^{(i)} &\coloneqq \sum_{r=1}^i \I\{\zeta_r = h\} \, x_r \delta_r, \\
    S_{xz,h}^{(i)} &\coloneqq \sum_{r=1}^i \I\{\zeta_r = h\} \, \omega_r x_r z_r\trans. \label{eq:cluster-aggregates-last}
\end{align}
Similarly, for the shared coefficients $\gamma$, define
\begin{align}
    S_{zz}^{(i)} &\coloneqq \sum_{r=1}^i \omega_r z_r z_r\trans, \label{eq:shared-aggregates-first} \\
    s_z^{(i)} &\coloneqq \sum_{r=1}^i z_r \delta_r. \label{eq:shared-aggregates-last}
\end{align}
These quantities collect exactly the weighted quadratic and linear terms needed for the Gaussian posterior updates below.

We assume independent Gaussian priors for each coefficient block:
\begin{equation*}
\beta_h \sim \mathcal{N}(\mu_\beta, \Sigma_\beta),
\qquad
\gamma \sim \mathcal{N}(\mu_\gamma, \Sigma_\gamma),
\end{equation*}
with prior precision matrices $\Lambda_\beta \coloneqq \Sigma_\beta^{-1}$ and $\Lambda_\gamma \coloneqq \Sigma_\gamma^{-1}$. The DP base measure for the cluster-specific coefficients is therefore
\[
G_0 = \mathcal{N}(\mu_\beta, \Sigma_\beta).
\]

For $\beta_h$, conditional on $\gamma$ and the current values of $(\omega_{1:i},\zeta_{1:i})$, only observations with $\zeta_r=h$ contribute. Equivalently, Eq.~\eqref{eq:gaussian-kernel} can be read as a weighted Gaussian regression for $\beta_h$ with pseudo-outcome $\delta_r/\omega_r - z_r\trans \gamma$ and variance $1/\omega_r$. Applying standard Gaussian posterior algebra (\appref{app:posterior-derivation}), the conditional posterior of $\beta_h$ is
\begin{equation}
\beta_h \given \Theta_{i,-\beta}, \D_{1:i}
\sim
\mathcal{N}\!\left(\bigl(A_{\beta,h}^{(i)}\bigr)^{-1} b_{\beta,h}^{(i)},\, \bigl(A_{\beta,h}^{(i)}\bigr)^{-1}\right),
\label{eq:beta-posterior}
\end{equation}
where $\Theta_{i,-\beta} \coloneqq (\gamma,\zeta_{1:i},\omega_{1:i},\alpha)$ and
\begin{equation}
A_{\beta,h}^{(i)} = \Lambda_\beta + S_{xx,h}^{(i)},
\qquad
b_{\beta,h}^{(i)} = \Lambda_\beta \mu_\beta + s_{x,h}^{(i)} - S_{xz,h}^{(i)} \gamma.
\label{eq:beta-posterior-moments}
\end{equation}

Likewise, conditional on $\beta = \{\beta_h\}_{h \in \rev{C_i}}$, Eq.~\eqref{eq:gaussian-kernel} can be read as a weighted Gaussian regression for $\gamma$ with pseudo-outcome $\delta_r/\omega_r - x_r\trans \beta_{\zeta_r}$ and variance $1/\omega_r$. Standard Gaussian posterior algebra then gives
\begin{equation}
\gamma \given \Theta_{i,-\gamma}, \D_{1:i}
\sim
\mathcal{N}\!\left(\bigl(A_\gamma^{(i)}\bigr)^{-1} b_\gamma^{(i)},\, \bigl(A_\gamma^{(i)}\bigr)^{-1}\right),
\label{eq:gamma-posterior}
\end{equation}
where $\Theta_{i,-\gamma} \coloneqq (\beta,\zeta_{1:i},\omega_{1:i},\alpha)$ and
\begin{equation}
A_\gamma^{(i)} = \Lambda_\gamma + S_{zz}^{(i)},
\qquad
b_\gamma^{(i)} = \Lambda_\gamma \mu_\gamma + s_z^{(i)} - \sum_{h \in \rev{C_i}} \bigl(S_{xz,h}^{(i)}\bigr)\trans \beta_h.
\label{eq:gamma-posterior-moments}
\end{equation}

Upon arrival of observation $i$ with assignment $h = \zeta_i$, the cluster-specific aggregates are updated as
\begin{align}
S_{xx,h}^{(i)} &= S_{xx,h}^{(i-1)} + \omega_i x_i x_i\trans, \label{eq:cluster-recursions-first} \\
s_{x,h}^{(i)} &= s_{x,h}^{(i-1)} + x_i \delta_i, \\
S_{xz,h}^{(i)} &= S_{xz,h}^{(i-1)} + \omega_i x_i z_i\trans. \label{eq:cluster-recursions-last}
\end{align}
Similarly,
\begin{align}
S_{zz}^{(i)} &= S_{zz}^{(i-1)} + \omega_i z_i z_i\trans, \label{eq:shared-recursions-first} \\
s_z^{(i)} &= s_z^{(i-1)} + z_i \delta_i. \label{eq:shared-recursions-last}
\end{align}
Thus, the sufficient statistics of $(\beta_h,\gamma)$ can be updated recursively. Note that only the aggregated quantities $(S_{xx,h}^{(i)}, s_{x,h}^{(i)}, S_{xz,h}^{(i)})$ for the currently assigned cluster $h = \zeta_i$ are updated for $\beta$, whereas $(S_{zz}^{(i)}, s_z^{(i)})$ are updated at every observation since $\gamma$ is common across clusters.

PG data augmentation also yields a simple full conditional distribution for $\omega_i$ \citep{PolsonScottWindle2013}. Conditional on the current coefficients and assignment $\zeta_i = h$,
\begin{equation}
\omega_i \given \Theta_{i,-\omega}, \D_i \sim \PG(1,\eta_{hi}),
\label{eq:omega-full-conditional}
\end{equation}
where $\Theta_{i,-\omega} \coloneqq (\beta,\gamma,\zeta_{1:i},\alpha)$ and $\eta_{hi} = x_i\trans \beta_h + z_i\trans \gamma$. We sample $\omega_i$ from Eq.~\eqref{eq:omega-full-conditional} and \rev{update the sufficient statistics via Eqs.~\eqref{eq:cluster-recursions-first}--\eqref{eq:shared-recursions-last}}. We then perform a Gibbs update for the coefficients $(\beta,\gamma)$: for each occupied cluster $h$, sample $\beta_h$ from Eq.~\eqref{eq:beta-posterior} conditional on $\gamma$, and then sample $\gamma$ from Eq.~\eqref{eq:gamma-posterior} conditional on $\beta$.

\subsection{Incremental weight update, resampling, and rejuvenation move}
\label{sec:method:smc}

\subsubsection{Weight update and resampling}
\label{sec:method:weight}

Again, we approximate the joint PG-augmented posterior sequence $\{\pi_i\}_{i=1}^n$ by a weighted particle system $\{\Theta_i^{(m)}, W_i^{(m)}\}_{m=1}^M$, where $\Theta_i \coloneqq (\beta,\gamma,\zeta_{1:i},\omega_{1:i},\alpha)$.

After the allocation normalising constants $Z_i^{(m)}$ have been computed, we evaluate the diagnostic ESS using the normalised candidate weights proportional to $\{W_{i-1}^{(m)}Z_i^{(m)}\}_{m=1}^{M}$. If this diagnostic indicates a sharp drop, we use the stabilising update in \appref{app:tempering}, which incorporates the incremental weights gradually through a sequence of tempered reweighting steps \citep{ChopinPapaspiliopoulos2020}. Otherwise, we follow the standard route described below.

On the standard route, we complete the remaining within-particle updates described in \Crefrange{sec:method:alloc}{sec:method:pg} and then update the particle weights to account for the new observation $\D_i$. As in sequential importance sampling, we define the stage-$i$ unnormalised weights by
\begin{equation*}
\wt{W}_i^{(m)} = W_{i-1}^{(m)} \cdot w_i^{(m)},
\qquad m = 1,\dots,M.
\end{equation*}
Here, $\wt{W}_i^{(m)}$ denotes the unnormalised stage-$i$ weight and $w_i^{(m)}$ denotes the incremental weight. In our algorithm, the incremental weight simplifies to the particle-specific allocation normalising constant $Z_i^{(m)}$ computed and stored at the cluster-allocation step in Eq.~\eqref{eq:Zi}. The subsequent Gibbs updates of $\alpha$, $\omega_i$, and $(\beta,\gamma)$ are exact conditional updates under the augmented target and therefore introduce no additional importance-weight factor (see \appref{app:weight} for details):
\begin{equation*}
w_i^{(m)} = Z_i^{(m)},
\qquad m = 1,\dots,M.
\end{equation*}
Consequently, the weight update reduces to
\begin{equation}
\wt{W}_i^{(m)} = W_{i-1}^{(m)} \cdot Z_i^{(m)},
\qquad m = 1,\dots,M.
\label{eq:weight-update}
\end{equation}
This is a one-step weight update in the sense that the particle-specific incremental weight $Z_i^{(m)}$ associated with observation $\D_i$ is incorporated into the particle weight in a single reweighting operation.
Normalised weights are obtained by
\begin{equation*}
W_i^{(m)} = \frac{\wt{W}_i^{(m)}}{\sum_{\ell=1}^M \wt{W}_i^{(\ell)}},
\qquad m = 1,\dots,M.
\end{equation*}
\rev{Sequential importance weighting can progressively concentrate particle weights, leaving most weights negligible while a few particles dominate.} We monitor this weight degeneracy using the effective sample size,
\begin{equation}
\ESS_i \coloneqq \frac{1}{\sum_{m=1}^M \bigl(W_i^{(m)}\bigr)^2}.
\label{eq:ess}
\end{equation}
If $\ESS_i$ falls below a pre-specified resampling threshold $\rho M$ (e.g., $\rho = 0.5$), we resample particles according to $\{W_i^{(m)}\}_{m=1}^M$ and reset weights to
\begin{equation*}
W_i^{(m)} \leftarrow \frac{1}{M},
\qquad m = 1,\dots,M.
\end{equation*}
We use a low-variance scheme such as systematic or stratified resampling. \rev{Because resampling replicates high-weight particle states, we follow it with the rejuvenation move (mutation) described below to improve exploration of the current posterior.}

\subsubsection{Rejuvenation move}
\label{sec:method:rejuvenation}

\rev{To improve exploration of the high-dimensional augmented posterior after resampling, we apply an MCMC rejuvenation move.} For each particle, we use a Markov transition kernel that leaves the augmented posterior $\pi_i(\Theta_i \given \D_{1:i})$ invariant. We implement this kernel as one or more sweeps of blocked Gibbs updates. \rev{At each rejuvenation trigger and independently for each particle $m$, let $R_i^{(m)} \subseteq \{1,\dots,i\}$ denote the set of observation indices updated within a sweep. The set is sampled uniformly without replacement, is retained across all sweeps in that rejuvenation invocation, and has size $|R_i^{(m)}|=\lceil\kappa i\rceil$. When $\kappa=1$, $R_i^{(m)}=\{1,\dots,i\}$.} Let $T_{\mathrm{move}}$ denote the number of rejuvenation sweeps applied whenever the move step is invoked. \rev{In the proposed procedure, $T_{\mathrm{move}}\ge1$; setting $T_{\mathrm{move}}=0$ disables historical rejuvenation and is used only for the simulation ablation. The routine conditional updates performed during assimilation remain active in this setting.} The corresponding update sequence is summarised in \Cref{alg:rejuvenation} in \appref{app:algorithms}.

Specifically, within each particle \rev{$m$}, for each sweep $t=1,\dots,T_{\mathrm{move}}$, and for each observation $r \in \rev{R_i^{(m)}}$, perform the following steps. Let $h_{\mathrm{old}} = \zeta_r$. Before removing observation $r$, record whether its current cluster is a singleton. If $n_{h_{\mathrm{old}},i}=1$, also store its current coefficient value as $\beta_{\mathrm{old}}^\star=\beta_{h_{\mathrm{old}}}$ so that returning observation $r$ to the same singleton cluster remains possible after the temporary removal step. First, remove observation $r$'s contribution from the sufficient statistics and cluster size:
\begin{align*}
S_{xx,h_{\mathrm{old}}} &\leftarrow S_{xx,h_{\mathrm{old}}} - \omega_r x_r x_r\trans, \\
s_{x,h_{\mathrm{old}}} &\leftarrow s_{x,h_{\mathrm{old}}} - x_r \delta_r, \\
S_{xz,h_{\mathrm{old}}} &\leftarrow S_{xz,h_{\mathrm{old}}} - \omega_r x_r z_r\trans,
\end{align*}
and
\begin{align*}
S_{zz} &\leftarrow S_{zz} - \omega_r z_r z_r\trans, \\
s_z &\leftarrow s_z - z_r \delta_r, \\
n_{h_{\mathrm{old}},i} &\leftarrow n_{h_{\mathrm{old}},i} - 1.
\end{align*}
If $n_{h_{\mathrm{old}},i} = 0$, the cluster becomes empty and is removed from the occupied cluster set $\rev{C_i}$, and we delete its associated parameter $\beta_{h_{\mathrm{old}}}$ and aggregated quantities $(S_{xx,h_{\mathrm{old}}}, s_{x,h_{\mathrm{old}}}, S_{xz,h_{\mathrm{old}}})$.

After removing the contribution of observation $r$, we jointly refresh $(\zeta_r, \omega_r)$. First, before resampling $\zeta_r$, we construct the auxiliary candidate set following the auxiliary-parameter update of \citet{Neal2000}. If the original cluster of observation $r$ was not a singleton, we draw $J$ auxiliary candidate parameters $\wt{\beta}_{\mathrm{new}}^{(1:J)}$ independently from $G_0$, exactly as in \Cref{sec:method:alloc}. If the original cluster was a singleton, we retain its previous coefficient value as one of the auxiliary candidates, say $\wt{\beta}_{\mathrm{new}}^{(1)}=\beta_{\mathrm{old}}^\star$, and, if $J>1$, draw the remaining $J-1$ candidates independently from $G_0$. We then treat observation $r$ as if it were a newly arrived one and resample $\zeta_r$ using the same allocation rule as in \Cref{sec:method:alloc}, with $\omega_r$ marginalised out. If an auxiliary new label $\zeta_r = \mathrm{new}_j$ is selected, we instantiate a new cluster $h^\ast$, set $\beta_{h^\ast} \coloneqq \wt{\beta}_{\mathrm{new}}^{(j)}$, and update the occupied cluster set accordingly before sampling $\omega_r$. We then resample $\omega_r$ from the full conditional distribution as in Eq.~\eqref{eq:omega-full-conditional}. This two-step update is a valid blocked Gibbs move on the augmented target: we first draw $\zeta_r$ from its conditional distribution with $\omega_r$ integrated out, using the standard singleton-retaining auxiliary construction for non-conjugate DP mixtures, and then draw $\omega_r$ from its full conditional given the updated label.

Subsequently, let $h_{\mathrm{new}} = \zeta_r$ and add the contribution of observation $r$ to the aggregates and increment the new cluster size:
\begin{align*}
S_{xx,h_{\mathrm{new}}} &\leftarrow S_{xx,h_{\mathrm{new}}} + \omega_r x_r x_r\trans, \\
s_{x,h_{\mathrm{new}}} &\leftarrow s_{x,h_{\mathrm{new}}} + x_r \delta_r, \\
S_{xz,h_{\mathrm{new}}} &\leftarrow S_{xz,h_{\mathrm{new}}} + \omega_r x_r z_r\trans,
\end{align*}
and
\begin{align*}
S_{zz} &\leftarrow S_{zz} + \omega_r z_r z_r\trans, \\
s_z &\leftarrow s_z + z_r \delta_r, \\
n_{h_{\mathrm{new}},i} &\leftarrow n_{h_{\mathrm{new}},i} + 1.
\end{align*}

After completing the updates for all $r \in \rev{R_i^{(m)}}$ within a given sweep, we refresh the current values of $(\alpha, \{\beta_h\}_{h \in \rev{C_i}}, \gamma)$. Specifically, we draw $\alpha$ using Eqs.~\eqref{eq:escobar-west-lambda}--\eqref{eq:alpha-mixture}, and sample $(\{\beta_h\}_{h \in \rev{C_i}},\gamma)$ from their conditional posterior distributions given the updated sufficient statistics. Repeating this blocked Gibbs sweep multiple times can improve mixing, although at an approximately linear increase in computational cost.

These recursively maintained sufficient statistics support both observation assimilation and historical rejuvenation. New observations contribute additive updates, whereas revisited observations require removal of their previous contributions and addition of those computed from the refreshed allocations and PG variables, with clusters created or removed as needed. The coefficient updates can therefore use the maintained summaries rather than repeatedly reconstructing them from the full observation history.

\subsection{Practical considerations}
\label{sec:method:practical}
\subsubsection{Tuning parameters}
\label{sec:method:tuning}

The tuning parameters considered here are algorithmic choices for the proposed SMC inference procedure, rather than behavioural parameters or assumptions of the DP-BL model. \rev{The main choices varied in this study are the number of particles $M$, the number $J$ of auxiliary new-cluster candidates, the resampling threshold $\rho$, the history fraction $\kappa$, and the number of rejuvenation sweeps $T_{\mathrm{move}}$.} The optional stabilising update additionally uses a conditional effective sample size (CESS) threshold fraction $\mu$, which is held fixed in the reported analyses. Together, these choices control accuracy, robustness to degeneracy, and computational cost. Hence, prior to live deployment, it is good practice to conduct short \emph{pilot runs} offline using an initial portion of the stream (or representative historical data). Such pilot runs provide a straightforward way to set sensible defaults under the intended time budget and to check that posterior summaries and algorithmic diagnostics \rev{such as the ESS trajectory and resampling frequency} behave stably.

\begin{revision}
The number of particles $M$ primarily controls Monte Carlo variability. In practice, $M$ should be chosen large enough that posterior summaries and the ESS trajectory are stable in pilot runs. The number of auxiliary candidates $J$ controls how many candidate parameter values represent the ``new cluster'' option within each allocation update. For any fixed $J\geq1$, these candidates form part of the auxiliary-variable construction and do not change the posterior target. Increasing $J$ can improve proposal efficiency and exploration of the new-cluster option \citep{Neal2000,griffin2017sequential}. It also increases the cost of generating and evaluating auxiliary candidates, although its effect on total runtime depends on the other updates. A small to moderate value is therefore often sufficient unless new clusters are expected to appear frequently.
\end{revision}

The resampling threshold $\rho$ should balance weight stability against unnecessary loss of diversity. A conventional default such as $\rho = 0.5$ is often reasonable, but smaller values reduce resampling frequency at the cost of allowing more severe weight concentration. In all analyses reported in this study, a ``sharp drop'' is operationally defined as a decrease of at least $50\%$ in the diagnostic ESS relative to the ESS of the particle system before incorporating $\D_i$. The diagnostic ESS is computed from the normalised candidate weights proportional to $\{W_{i-1}^{(m)}Z_i^{(m)}\}_{m=1}^M$, before any resampling. When the optional stabilising update is invoked, the CESS threshold fraction $\mu$, defined formally in \appref{app:tempering}, controls the intermediate temperature increments, with values close to one limiting incremental-weight variation. The resampling threshold $\rho$ separately determines whether resampling is required based on the ESS after reweighting.

\begin{revision}
The rejuvenation design is governed jointly by the particle-specific updated index sets $R_i^{(m)}$ and the number of sweeps $T_{\mathrm{move}}$. The parameterisation $|R_i^{(m)}|=\lceil\kappa i\rceil$, with $\kappa\in(0,1]$, specifies the fraction of processed observations revisited within each rejuvenation invocation. The set $R_i^{(m)}$ is used only in historical rejuvenation, which is invoked after main resampling on the standard route or as the final step of the stabilising update. Ordinary assimilation still updates the stage-$i$ posterior when every new observation arrives, irrespective of the value of $\kappa$. Consequently, the baseline choice $\kappa=1$ remains an online observation-by-observation procedure rather than repeated batch re-estimation. Choosing $\kappa<1$ reduces the number of historical allocation updates within a sweep and its computational cost, at the possible cost of weaker rejuvenation. The parameter $T_{\mathrm{move}}$ controls the number of such sweeps whenever the move is invoked. Larger values can improve exploration of the current posterior, but increase computational cost approximately linearly. The simulation study therefore examines the accuracy--cost implications of partial-history rejuvenation and separately evaluates the inferential and computational contribution of $T_{\mathrm{move}}$, including a no-move ablation ($T_{\mathrm{move}}=0$).
\end{revision}

\subsubsection{Provisional and fully updated posteriors}
\label{sec:method:provisional}

For interpretive purposes, it is useful to distinguish between a \emph{provisional posterior} and a \emph{fully updated posterior} at stage $i$. The provisional posterior is the normalised weighted particle approximation available after incorporating the new observation but before the final resampling--rejuvenation stage on the standard route, or the final historical rejuvenation move on the stabilising route. On the latter route, the tempering sequence has already been completed.

More precisely, the provisional-output point is immediately after Step~18 of \Cref{alg:complete-smc} on the standard route, or immediately before Step~20 of \Cref{alg:tempering} on the stabilising route. By contrast, the fully updated posterior denotes the particle approximation obtained after all main resampling and rejuvenation steps invoked at stage $i$ have been completed. Both approximate the same stage-$i$ partial-data posterior target, but they correspond to different points in the within-stage computational workflow. If the resampling--rejuvenation step is not triggered on the standard route, the provisional and fully updated particle approximations coincide.

For the newly incorporated observation $i$, its observation-specific provisional posterior of the assigned cluster-specific coefficient $\beta_{\zeta_i}$ is represented by the weighted sample
\[
\left\{
\left(
\beta_{\zeta_i^{(m)}}^{(m)},
 W_i^{(m)}
\right)
\right\}_{m=1}^{M},
\]
where all quantities in each pair are evaluated at the provisional-output point defined above. Each particle therefore contributes exactly one coefficient: the coefficient of the cluster to which observation $i$ is assigned in that particle. No additional cluster-size weight is applied. This representation incorporates uncertainty in both the focal observation's cluster assignment and its assigned cluster-specific coefficient. Because $\D_i=(y_i,x_i,z_i)$ has already been assimilated, it is a post-assimilation posterior rather than a one-step-ahead predictive distribution available before $y_i$ is observed.

The provisional posterior is particularly relevant in latency-sensitive online applications because an observation-specific summary becomes available immediately after the new observation has been incorporated. Choice predictions for subsequent observations can be computed from the updated particle approximation by averaging over parameter uncertainty and possible cluster assignments, as in \Cref{sec:simulation:base}. In the present implementation, the fully updated particle approximation is passed to the next observation, while provisional summaries can be returned earlier.

\subsubsection{Role of the simulation and illustrative empirical application}
\label{sec:method:evaluation-roadmap}

The simulation study and illustrative empirical application in \Cref{sec:simulation,sec:case} use these practical considerations in different ways. The simulation study in \Cref{sec:simulation} provides a controlled setting in which the accuracy and computational behaviour of the proposed SMC algorithm can be examined against a known data-generating process and a batch MCMC benchmark. It therefore examines posterior fidelity and sensitivity to key tuning parameters, alongside predictive checks. The illustrative empirical application in \Cref{sec:case}, by contrast, focuses on the online outputs that motivate the proposed framework in applied behavioural settings. In particular, it illustrates how fully updated posteriors evolve over the observation stream, how provisional posteriors can be produced for newly arriving observations, and how the associated computational latency can be evaluated.

\begin{revision}
The replication designs also serve different purposes. In \Cref{sec:simulation}, we generate 20 datasets using distinct data-generation seeds and apply each SMC tuning configuration once to each dataset. Results are averaged across datasets to assess performance over repeated data realisations. In \Cref{sec:case}, we instead keep the empirical dataset fixed and repeat each SMC configuration with 20 algorithmic seeds. These repetitions allow us to assess variability due to algorithmic randomness.
\end{revision}

\newpage
\section{Simulation}
\label{sec:simulation}

We study the finite-sample behaviour of the proposed online algorithm using synthetic streams generated from a known data-generating process (DGP). The design follows common practice in simulation studies for DP mixture choice models by using a deliberately challenging heterogeneity distribution with both multimodality and skewness \citep{krueger2020dirichlet}. Our primary goal is to characterise the algorithmic accuracy--cost trade-off under streaming updates and to examine the tuning guidance in \Cref{sec:method:tuning} under a controlled DGP. \rev{The analysis uses 20 independently generated datasets, with results averaged across datasets.}

\subsection{Data-generating process and prior}
\label{sec:simulation:dgp}

We generate a stream of observations $\mathcal{D}_{1:n}=\{(y_i,x_i,z_i)\}_{i=1}^{n}$ with $n=3{,}000$. Again, let $o$ denote the number of covariates whose coefficients vary across latent clusters and let $q$ denote the number of covariates with shared coefficients. In this simulation, we set $(o,q)=(2,1)$. The first component of $x_i$ is fixed to one (a \emph{constant}), yielding $x_i=(1,\tilde x_i)\trans$ with $\tilde x_i\in\mathbb{R}$.

For each $i$,
\begin{align}
\Pr(y_i=1\mid x_i,z_i,\beta_i,\gamma)
&=
\frac{\exp(x_i\trans\beta_i + z_i\,\gamma)}{1+\exp(x_i\trans\beta_i + z_i\,\gamma)},
\label{eq:sim:logit}
\end{align}
and $y_i\mid x_i,z_i,\beta_i,\gamma \sim \mathrm{Bernoulli}(\Pr(y_i=1\mid \cdot))$.

\rev{Throughout this subsection, the second argument of a univariate Gaussian distribution denotes its variance, whereas the second argument of a multivariate Gaussian distribution denotes its covariance matrix.} We draw covariates independently as
\begin{align}
\qquad \tilde x_i \sim \mathcal{N}(0,3),\qquad z_i \sim \mathcal{N}(0,3).
\end{align}
The shared coefficient is fixed at $\gamma=-0.50$ throughout. To generate multimodal and skewed coefficient heterogeneity, we draw $\beta_i$ from a three-component mixture of skew-normal--logistic (SNL) distributions (as in \citealp{krueger2020dirichlet}). For a scalar random variable $u$, we write $u\sim \mathrm{SNL}(\xi,\sigma,\nu)$ if it has density
\begin{align}
f(u\mid\xi,\sigma,\nu)
=
2\,\phi(u\mid\xi,\sigma)\,
\frac{1}{1+\exp\{-\nu(u-\xi)\}},
\label{eq:sim:snl}
\end{align}
where $\phi(\cdot\mid\xi,\sigma)$ is the $\mathcal{N}(\xi,\sigma^2)$ density. We draw a latent component label $c_i\in\{1,2,3\}$ with probabilities $(0.50,0.25,0.25)$. Conditional on $c_i=k$, we draw $\beta_i=(\beta_{i,1},\beta_{i,2})\trans$ independently across coordinates as follows:
\begin{align*}
c_i = 1:\qquad
&\beta_{i,1} \sim \mathrm{SNL}(0.0, 3.0, 0.0),
&&\beta_{i,2} \sim \mathrm{SNL}(0.0, 3.0, 0.0), \\
c_i = 2:\qquad
&\beta_{i,1} \sim \mathrm{SNL}(-1.0, 2.0, 10.0),
&&\beta_{i,2} \sim \mathrm{SNL}(-1.0, 2.0, -5.0), \\
c_i = 3:\qquad
&\beta_{i,1} \sim \mathrm{SNL}(1.0, 2.0, -5.0),
&&\beta_{i,2} \sim \mathrm{SNL}(1.0, 2.0, 10.0).
\end{align*}
In the reported summaries, $\beta_1$ and $\beta_2$ denote the intercept and slope components, respectively.

For estimation, we set the base measure $G_0$ for the cluster-specific coefficients to be Gaussian:
\[
G_0 = \mathcal{N}\!\left(0, 10 I_2\right).
\]
The prior of the shared coefficient follows
\begin{align}
\gamma \sim \mathcal{N}(0,10).
\end{align}
For the DP concentration parameter, we set
\begin{align}
\alpha \sim \mathrm{Gamma}(a_0,b_0),\qquad (a_0,b_0)=(2,1),
\end{align}
where the Gamma distribution is parameterised by shape $a_0$ and rate $b_0$. The Gaussian coefficient priors are zero-centred and have variance 10, while the Gamma prior for $\alpha$, under the shape--rate parameterisation, has mean 2 and standard deviation $\sqrt{2}$.

\subsection{Experimental design and evaluation}
\label{sec:simulation:base}

Following the tuning discussion in \Cref{sec:method:tuning}, we define a baseline configuration and then vary one tuning parameter at a time, keeping all remaining settings fixed at their baseline values. We use systematic resampling throughout. In all runs, the on-demand stabilising update in \appref{app:tempering} is enabled.

The baseline settings are:
\begin{align*}
M = 1{,}000,\quad
J = 5,\quad
\rho = 0.5,\quad
\mu = 0.9,\quad
\kappa = 1,\quad
T_{\mathrm{move}} = 5.
\end{align*}
\begin{revision}
The sensitivity grid for the tuning parameters varied in this section is:
\begin{align*}
M &\in \{500,1{,}000,5{,}000\},\\
J &\in \{1,5,10\},\\
\rho &\in \{0.3,0.5,0.8\},\\
T_{\mathrm{move}} &\in \{0,1,5,10\},\\
\kappa &\in \{0.25,0.5,0.75,1\}.
\end{align*}
The one-at-a-time sensitivity analysis is therefore carried out over $(M,J,\rho,T_{\mathrm{move}},\kappa)$, while $\mu$ is fixed at 0.9 and the stabilising update is enabled in all scenarios.
\end{revision}

To check one-step-ahead predictive performance during sequential updating, we use a prequential log predictive score (LPS) at each arrival. Let $\pi_{i-1}$ denote the target posterior after processing $\mathcal{D}_{1:i-1}$ and let $\{(\Theta_{i-1}^{(m)},W_{i-1}^{(m)})\}_{m=1}^M$ be its particle approximation. Before assimilating observation $i$, we compute
\begin{align*}
\mathrm{LPS}_i
&\coloneqq
\log p(y_i\mid x_i,z_i,\mathcal{D}_{1:i-1})
\;\approx\;
\log\!\left(
\sum_{m=1}^{M} W_{i-1}^{(m)}\,
Z_i^{(m)}
\right).
\end{align*}
\begin{revision}
Here $Z_i^{(m)}$ is the particle-specific allocation normalising constant introduced in Eq.~\eqref{eq:Zi}, evaluated for observation $i$ under particle $m$ before the weight update. It is computed using the same auxiliary-candidate construction as the allocation step and sums over both existing clusters and the sampled auxiliary new-cluster candidates. For each replication, we summarise online predictive performance by averaging the final 500 one-step-ahead scores $\mathrm{LPS}_i$; we refer to this average as the \emph{Last-500 LPS}. The Last-500 LPS is used only to check whether gross predictive deterioration occurs, rather than as the primary measure of posterior fidelity.
\end{revision}

To complement this online criterion, we also report an out-of-sample \emph{Holdout LPS}. Specifically, after generating an estimation sample of size $n=3{,}000$, we generate, for each replication, an independent holdout sample of size $n_{\mathrm{hold}}=500$ from the same DGP and evaluate the posterior predictive density for the holdout outcomes using the final posterior approximation. Let $\mathcal{D}^{\mathrm{hold}}=\{(y_j^{\mathrm{hold}},x_j^{\mathrm{hold}},z_j^{\mathrm{hold}})\}_{j=1}^{n_{\mathrm{hold}}}$ denote the holdout sample. For a given fitted model, we define
\begin{align*}
\mathrm{Holdout\;LPS}
&\coloneqq
\frac{1}{n_{\mathrm{hold}}}
\sum_{j=1}^{n_{\mathrm{hold}}}
\log p\!\left(y_j^{\mathrm{hold}}
\mid
x_j^{\mathrm{hold}},z_j^{\mathrm{hold}},\mathcal{D}_{1:n}\right)
\\
&\approx
\frac{1}{n_{\mathrm{hold}}}
\sum_{j=1}^{n_{\mathrm{hold}}}
\log\!\left(
\sum_{m=1}^{M}
W_n^{(m)}
Z_{j,\mathrm{hold}}^{(m)}
\right).
\end{align*}
Thus, the holdout predictive density
$
p\!\left(
y_j^{\mathrm{hold}}
\mid
x_j^{\mathrm{hold}},z_j^{\mathrm{hold}},\mathcal{D}_{1:n}
\right)
$
is approximated by averaging the holdout analogue $Z_{j,\mathrm{hold}}^{(m)}$ over the final weighted particle approximation. The holdout analogue $Z_{j,\mathrm{hold}}^{(m)}$ is computed in the same way as $Z_i^{(m)}$, but using the final particle state $\Theta_n^{(m)}$ and the holdout observation $j$; it is used only for evaluation and does not update the particle system. Hence, the prequential score $\mathrm{LPS}_i$ measures sequential one-step-ahead forecasting performance during online updating, whereas the Holdout LPS measures out-of-sample predictive accuracy based on the final fitted posterior.

\begin{revision}
The Wasserstein-1 distance measures the minimum average distance over which probability mass must be moved to transform one distribution into another; smaller values indicate closer distributions. For replication $r$, let $F_{r,j}^{\mathrm{SMC}}$ and $F_{r,j}^{\mathrm{MCMC}}$ denote the marginal posterior distributions of $\beta_j$ obtained by the proposed SMC algorithm and batch MCMC, respectively. We summarise posterior fidelity for $(\beta_1,\beta_2)$ by the mean marginal Wasserstein-1 distance
\pagebreak[3]
\begin{align*}
\overline{W}_{1,r}
&\coloneqq
\frac{1}{2}\sum_{j=1}^{2}
W_1\!\left(F_{r,j}^{\mathrm{SMC}},F_{r,j}^{\mathrm{MCMC}}\right).
\end{align*}
Scenario summaries report the mean (standard deviation) of $\overline{W}_{1,r}$ across 20 replications.
\end{revision}

\subsection{Computational setup and posterior summarisation}
\label{sec:simulation:computation}

All SMC experiments were run on a laptop equipped with an Intel Core i7-10875H CPU at 2.30 GHz (8 cores, 16 threads). \rev{All SMC runs in the simulation study used 14 Julia threads for parallel computation.}

For benchmarking, we also estimated the same DP-BL by batch MCMC. The MCMC algorithm combines Algorithm 8 of \citet{Neal2000} for non-conjugate DP mixture allocation updates with PG data augmentation and Gibbs updates for the binary logit component following \citet{PolsonScottWindle2013}. The prior specification was kept identical to that used for SMC. The reported batch MCMC results are based on $52{,}000$ iterations per chain, with the first $2{,}000$ discarded as burn-in, and three independent chains were run for each replication. Here, \(\widehat{R}\) denotes the split-chain potential scale reduction factor, with values close to one indicating good agreement across chains. Convergence was assessed using split-\(\widehat{R}\) based on the number of occupied clusters, \(\alpha\), \(\gamma\), and posterior means of \(\beta_1\) and \(\beta_2\); all replications satisfied the comparison criterion \(\widehat{R}<1.05\), with a maximum of 1.0480. For Holdout LPS only, a systematic subset of 1,000 retained draws per chain was used; posterior summaries, convergence diagnostics, and Wasserstein-1 distances use all 50,000 retained draws per chain.

Unless otherwise stated, the SMC posterior summaries and posterior density plots in this section are based on the fully updated posterior approximation after all updates at the final observation, including resampling and rejuvenation when triggered. The displayed density curves are replication-averaged summaries over the 20 independent replications.
\begin{revision}
All simulation runtimes are defined as the sum of per-observation algorithm-processing times. Auxiliary diagnostic calculations, plotting, and final file output are excluded, whereas ESS calculations required by the algorithm are included.
\end{revision}
To visualise the posterior distributions, we smooth the weighted posterior with a Gaussian kernel. For each one-dimensional marginal, we first construct a weighted histogram and then convolve it with a Gaussian kernel, with the bandwidth chosen by a Silverman-type rule applied to the weighted coefficient samples. The two-dimensional density plots in \Cref{fig:truth_posterior} are constructed similarly by smoothing weighted bivariate histograms. \rev{Smoothing is used only for visualisation. The Wasserstein-1 distances are calculated directly from the unsmoothed weighted posterior samples of SMC and batch MCMC.}

\newpage
\subsection{Results and discussion}
\label{sec:simres:base}

\begin{revision}
\Cref{tab:scenario_lps_clusters} shows only very small differences in predictive scores across the examined SMC tuning scenarios, despite the substantial differences in computational cost reported in \Cref{tab:scenario_runtime_resampling}. We interpret these scores only as indicating that no gross predictive deterioration is detected under the examined scenarios, not as evidence that the posterior approximations are equivalent. \textcolor{black}{Posterior distributions with different shapes can yield similar choice probabilities, which may explain why the LPS varies little across tuning settings. This echoes the broader point raised by \citet{kim2023finite} in their discussion of the Rashomon effect: similar predictive performance need not imply similar representations of heterogeneity.} Posterior fidelity is assessed separately through coefficient distributions, parameter summaries, occupied-cluster counts, and the mean marginal Wasserstein-1 distance to batch MCMC defined in \Cref{sec:simulation:base}.
\end{revision}

\rev{Tables~\ref{tab:scenario_lps_clusters}--\ref{tab:scenario_beta_gamma_alpha} report all 13 SMC configurations, including no rejuvenation ($T_{\mathrm{move}}=0$) and partial-history rejuvenation ($\kappa=0.25,0.50,0.75$).} The baseline SMC setting yields 27.94 occupied clusters on average, while batch MCMC yields 28.29. The full-history configurations with $\kappa=1$ and $T_{\mathrm{move}}\geq1$ yield mean cluster counts between 27.52 and 28.55. The no-move condition ($T_{\mathrm{move}}=0$) yields 26.32, whereas partial-history rejuvenation with $\kappa=0.25$ yields 30.57. Both show greater variation across replications, with standard deviations of 5.52 and 4.74, respectively, compared with 2.71 for the baseline.

\begin{figure}[htbp]
    \centering
    \includegraphics[width=1.0\textwidth]{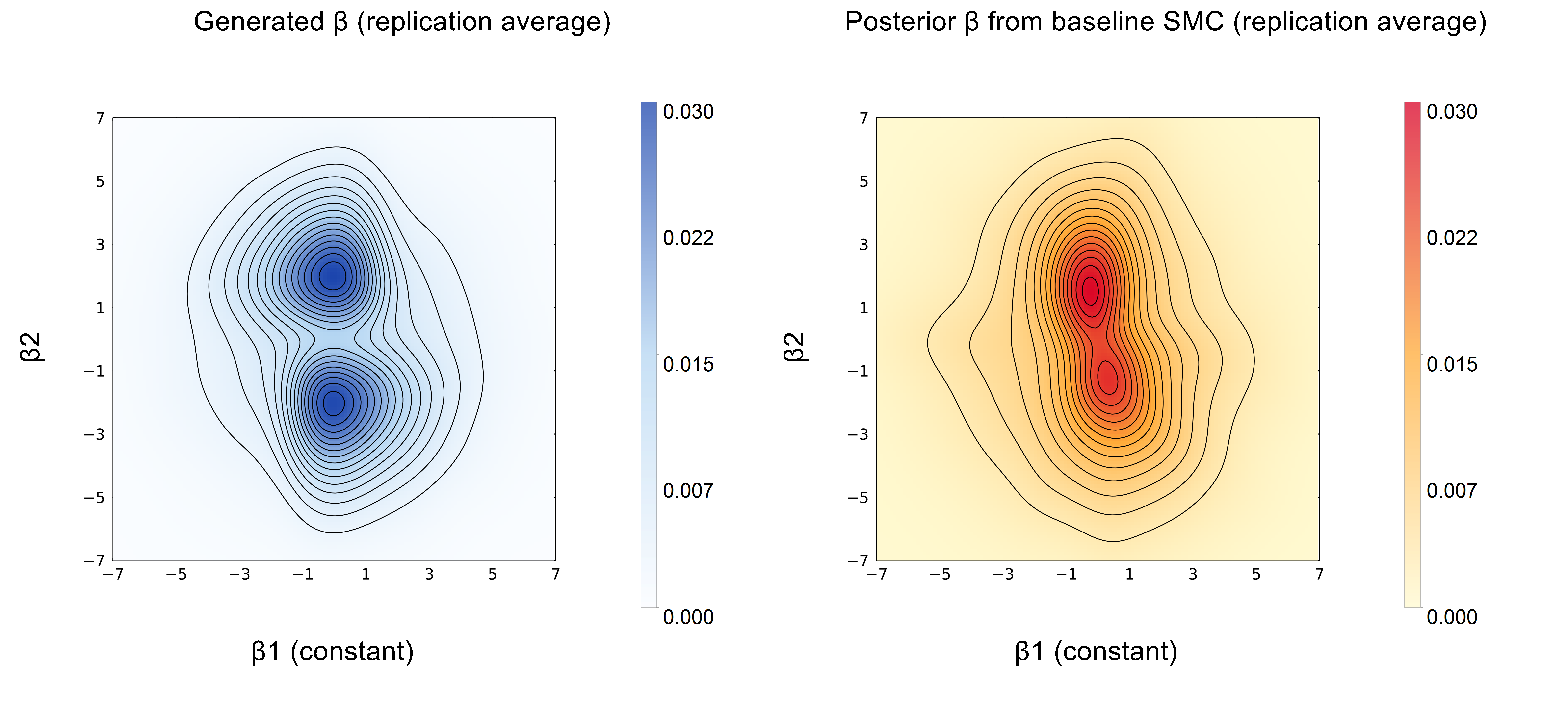}
    \caption{Generated distribution and posterior density of $\beta$ under the baseline SMC setting.}
    \label{fig:truth_posterior}
\end{figure}

\begin{figure}[htbp]
    \centering
    \includegraphics[width=0.9\textwidth]{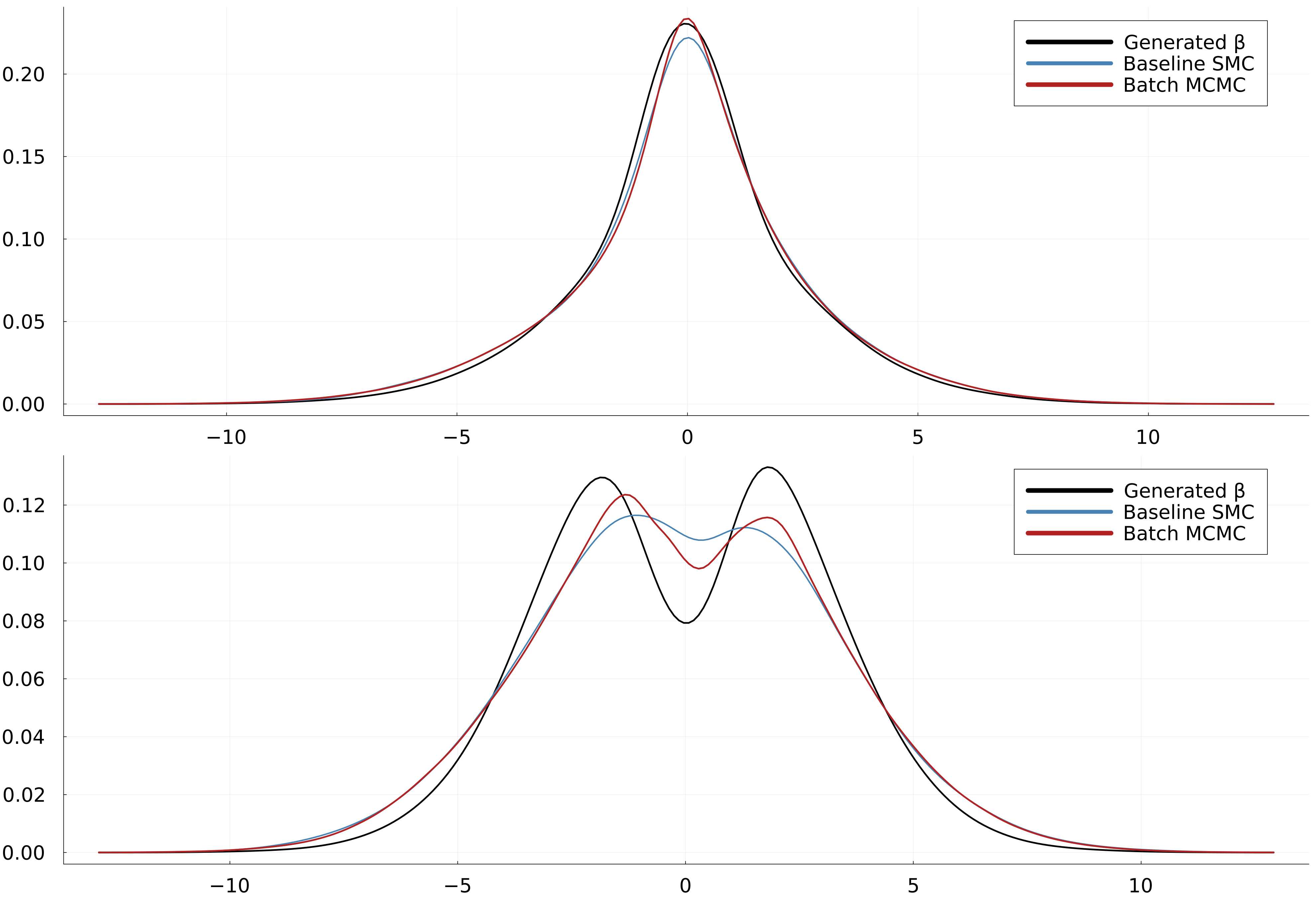}
    \caption{Generated distributions of $\beta$ and posterior densities under the baseline SMC setting and batch MCMC. Upper: $\beta_1$ (intercept). Lower: $\beta_2$.
}
    \label{fig:marginal}
\end{figure}

\Cref{fig:truth_posterior} shows that the baseline SMC captures the existence and approximate locations of the two main modes. However, it does not fully reproduce finer geometric features of the true SNL mixture, such as the precise degree of skewness, the relative peak heights, and the sharpness of the valley between the two modes. \Cref{fig:marginal} indicates that, even under batch MCMC, the posterior distribution does not fully recover the true SNL mixture at this sample size, especially for \(\beta_2\), for which the peak heights and the depth of the valley are still not fully recovered. This suggests that part of the remaining discrepancy reflects limited information at this sample size, rather than a failure of SMC alone. Visually, however, batch MCMC yields a somewhat sharper and more refined representation of the \(\beta_2\) marginal than the baseline SMC, as shown in \Cref{fig:marginal}. \rev{Hence, batch MCMC is useful as a benchmark for posterior shape.} \appref{app:marginal-comparisons} provides replication-level comparisons of the marginal posterior densities from baseline SMC and batch MCMC.

\begin{figure}[htbp]
    \centering
    \includegraphics[width=1.0\textwidth]{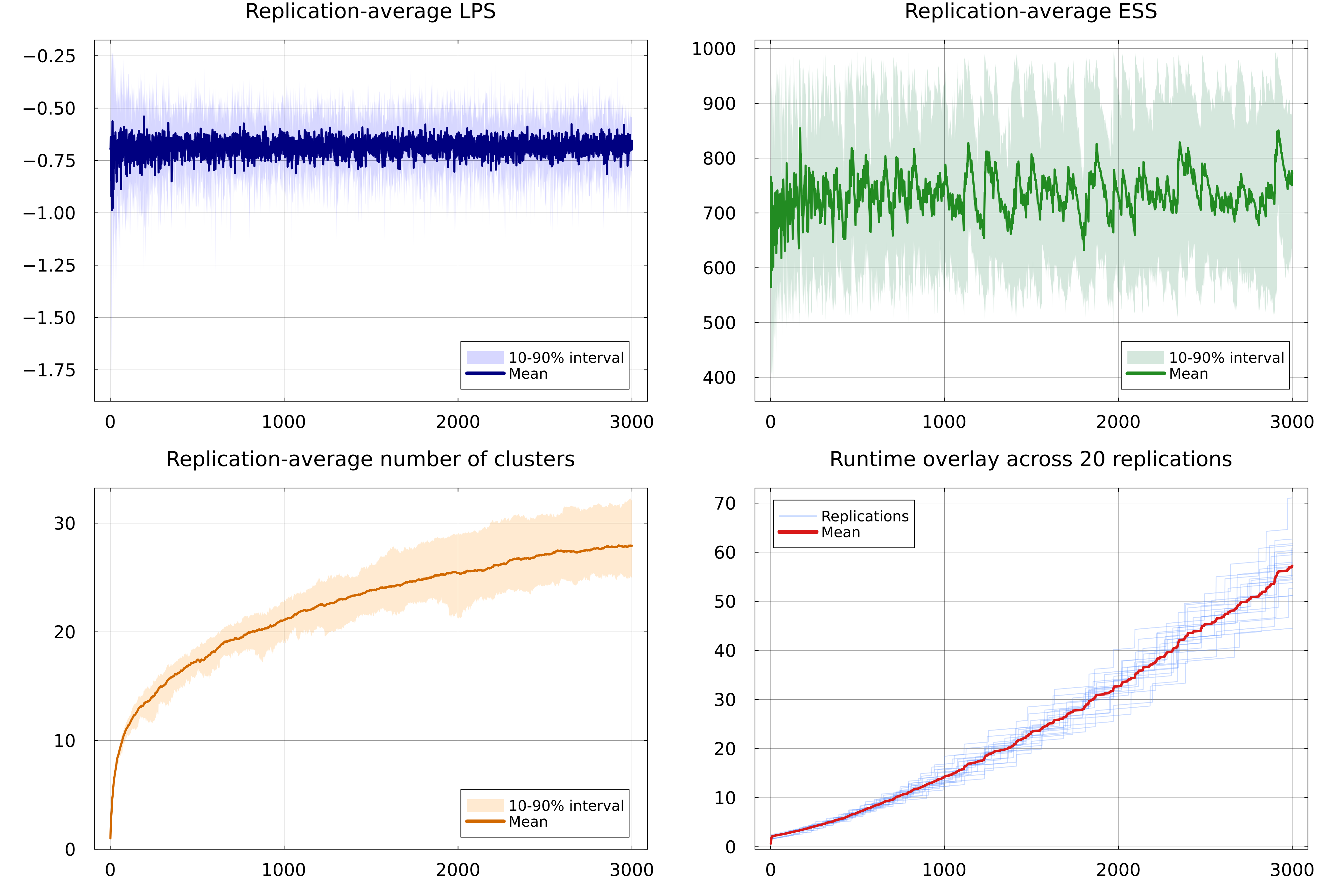}
    \caption{Replication-averaged diagnostics for the baseline SMC setting: LPS, ESS, number of occupied clusters, and \rev{cumulative algorithm-processing time} (seconds) across 20 replications, shown as functions of the sequential observation index. ESS is recorded after reweighting and before main resampling on the standard route, and after the tempering sequence (including any intermediate resampling) on the stabilising route.
}
    \label{fig:diagnostics}
\end{figure}

\Cref{fig:diagnostics} provides informative diagnostics for the data-tempering SMC run under the baseline setting. First, the LPS stabilises fairly early along the sample path. Second, the average number of occupied clusters increases gradually as more observations are processed. Third, the replication-averaged ESS mostly lies between about 600 and 800, although it is somewhat lower during the earliest phase of the run. Finally, the runtime plot shows that the vertical jumps in cumulative processing time are associated with the rejuvenation step. \rev{Under the baseline setting $\kappa=1$, rejuvenation revisits the full history \(R_i^{(m)}=\{1,\dots,i\}\). As this set grows with \(i\), rejuvenation becomes more costly later in the run.}

\begin{figure}[htbp]
    \centering
    \includegraphics[width=1.0\textwidth]{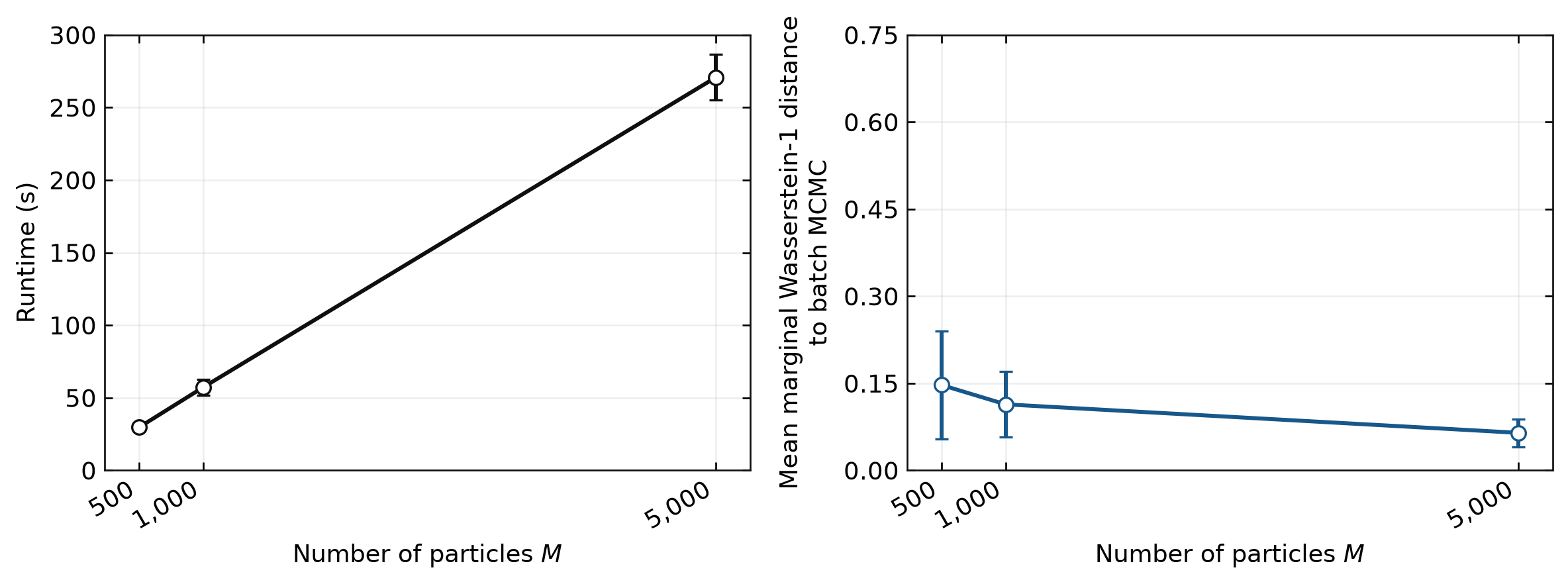}
    \caption{\rev{Runtime and posterior fidelity for different numbers of particles $M$. Points show means across 20 matched replications and error bars show standard deviations across replications.}}
    \label{fig:M_tradeoff}
\end{figure}

\begin{figure}[htbp]
    \centering
    \includegraphics[width=1.0\textwidth]{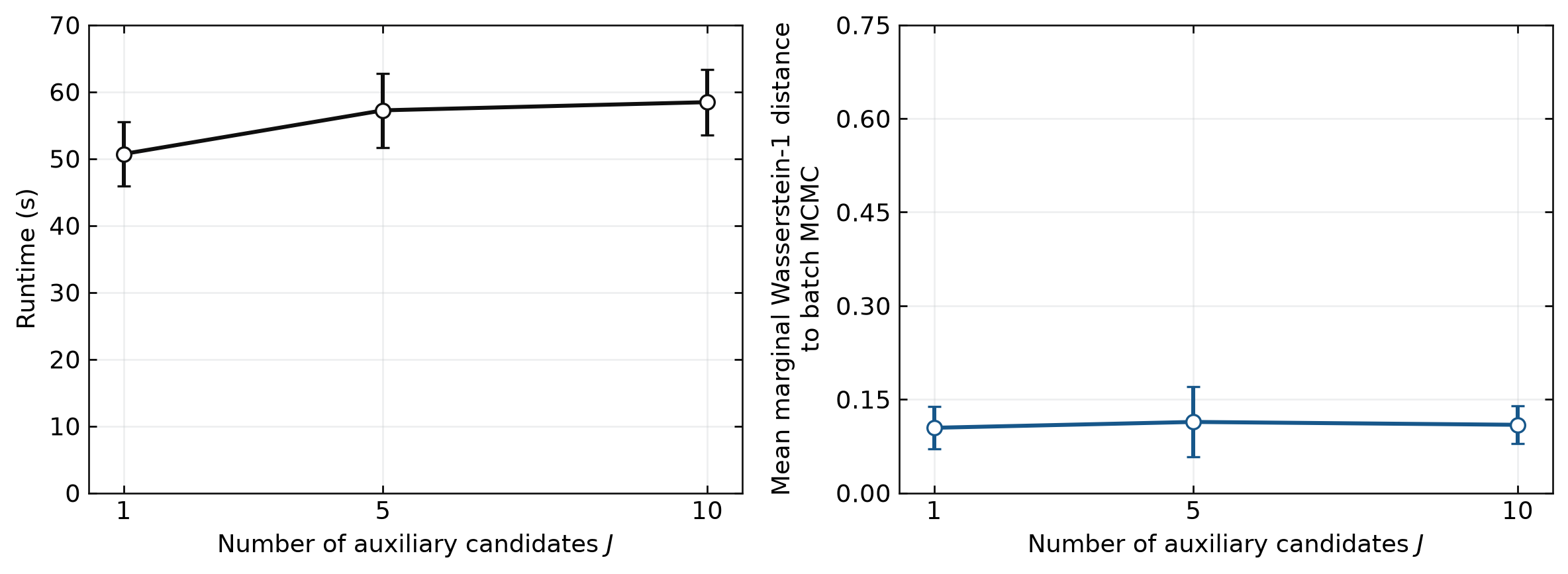}
    \caption{\rev{Runtime and posterior fidelity across numbers of auxiliary candidates $J$. Points show means across 20 matched replications and error bars show standard deviations across replications.}}
    \label{fig:J_tradeoff}
\end{figure}

\begin{figure}[htbp]
    \centering
    \includegraphics[width=1.0\textwidth]{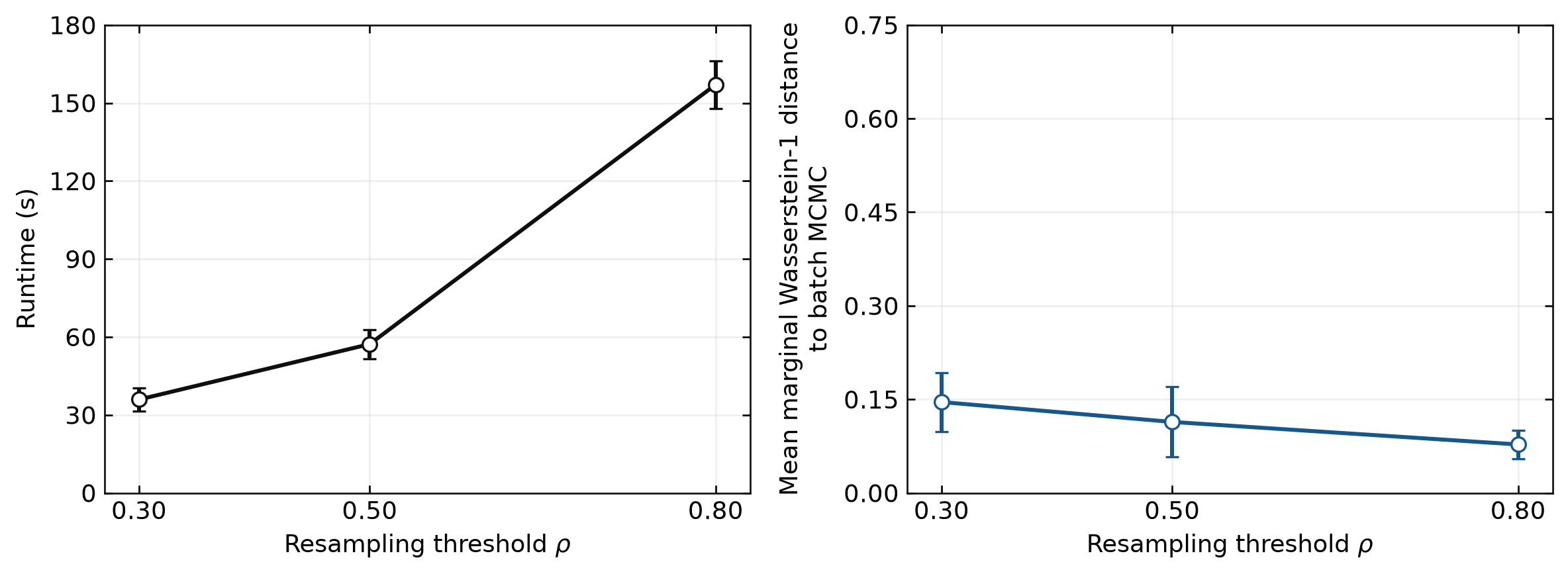}
    \caption{\rev{Runtime and posterior fidelity across resampling thresholds $\rho$. Points show means across 20 matched replications and error bars show standard deviations across replications.}}
    \label{fig:rho_tradeoff}
\end{figure}

\begin{figure}[htbp]
    \centering
    \includegraphics[width=1.0\textwidth]{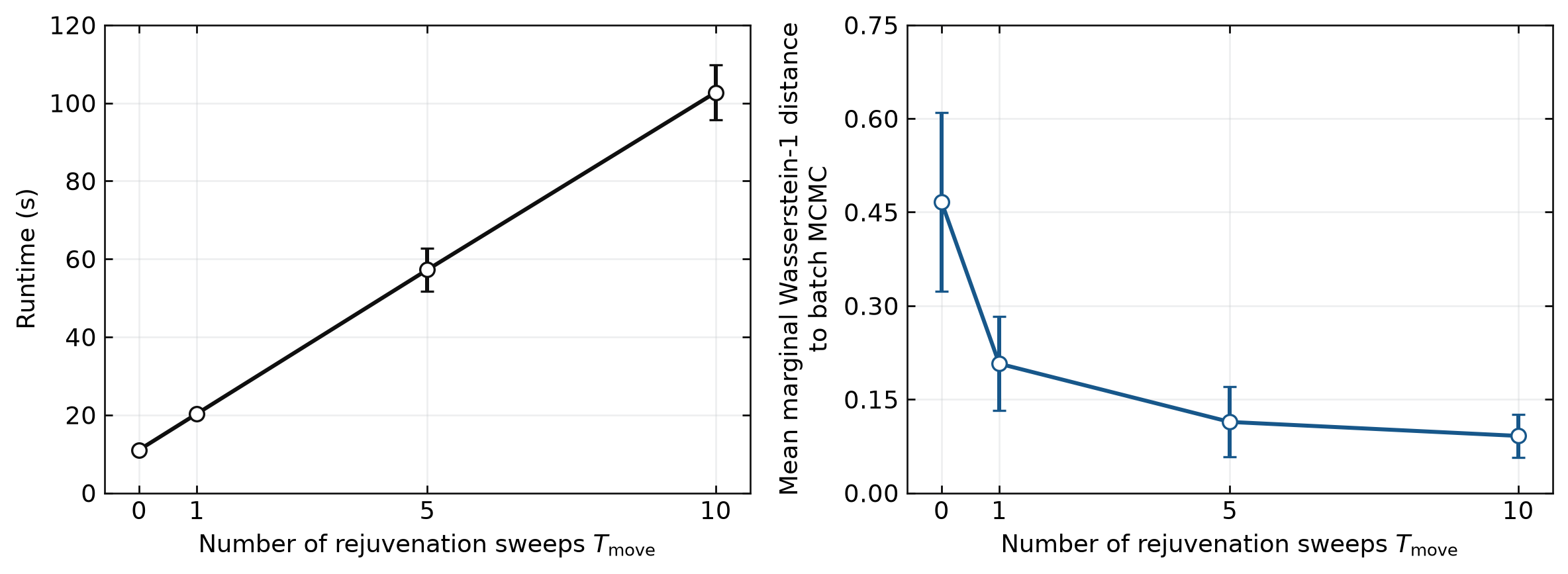}
    \caption{\rev{Runtime and posterior fidelity across rejuvenation sweep counts $T_{\mathrm{move}}$, including the no-move ablation at $T_{\mathrm{move}}=0$. Points show means across 20 matched replications and error bars show standard deviations across replications.}}
    \label{fig:T_tradeoff}
\end{figure}

\begin{figure}[htbp]
    \centering
    \includegraphics[width=1.0\textwidth]{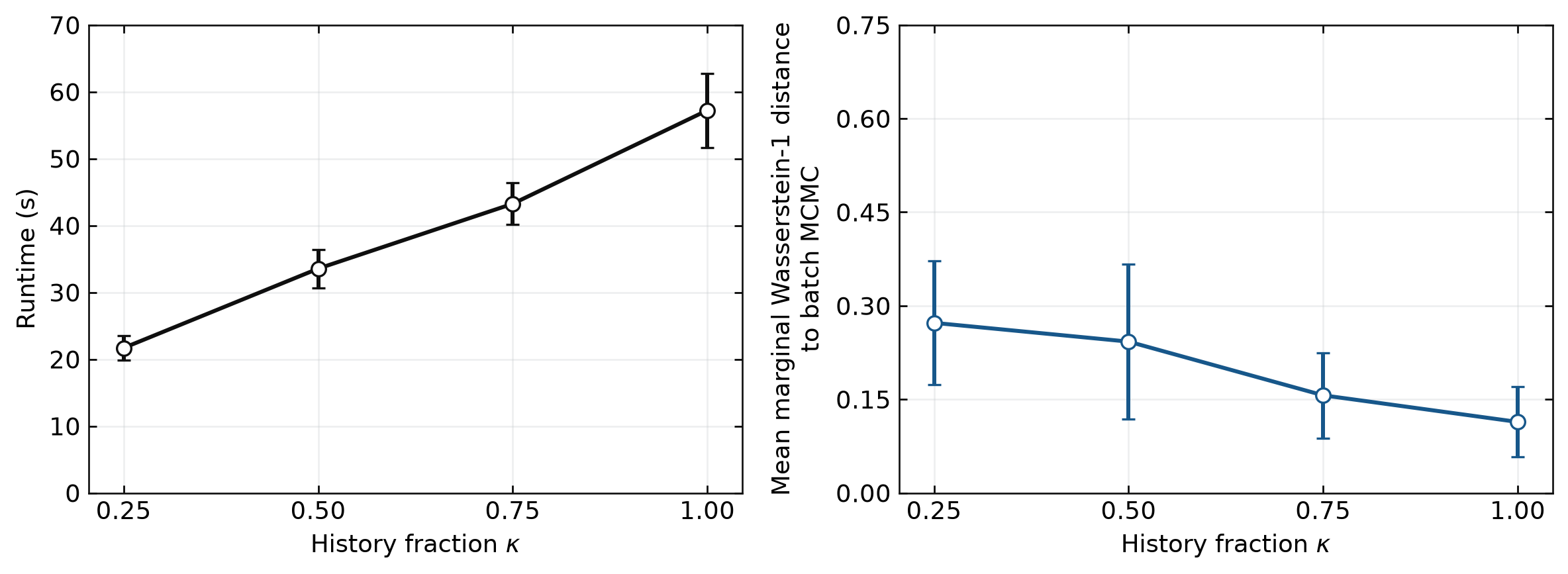}
    \caption{\rev{Runtime and posterior fidelity across history fractions $\kappa$ used for rejuvenation. Points show means across 20 matched replications and error bars show standard deviations across replications.}}
    \label{fig:kappa_tradeoff}
\end{figure}

\begin{revision}
\Cref{fig:M_tradeoff,fig:J_tradeoff,fig:rho_tradeoff,fig:T_tradeoff,fig:kappa_tradeoff} compare runtime and mean marginal Wasserstein-1 distance for each tuning parameter, using the same 20 simulated datasets. Complementary posterior-density comparisons are shown in \appref{app:marginal-comparisons}.

Increasing $M$ improves mean posterior fidelity at a substantial computational cost (\Cref{fig:M_tradeoff}). For $M=500$, $1{,}000$, and $5{,}000$, the mean marginal Wasserstein-1 distances are 0.147, 0.114, and 0.065, while mean runtimes are 29.68, 57.23, and 270.69~s, respectively. This pattern is consistent with reduced Monte Carlo approximation error as the number of particles increases, although variation remains across replications. Whether this improvement justifies the additional computational cost depends on the application. In practice, $M$ may therefore be selected to achieve an acceptable balance between posterior accuracy and processing time.

Increasing $J$ does not produce a corresponding improvement in posterior fidelity (\Cref{fig:J_tradeoff}). The mean distances are 0.104, 0.114, and 0.109 for $J=1$, $5$, and $10$, with mean runtimes of 50.71, 57.23, and 58.46~s. These differences in distance are small relative to the variation across replications. Thus, larger auxiliary candidate blocks do not consistently improve the final approximation in the present experiment. These results suggest that fine-tuning $J$ may be less critical within the examined range. In the absence of application-specific reasons for a different choice, $J=5$ may serve as a reasonable starting value.

The parameter $\rho$ affects both computational cost and posterior fidelity (\Cref{fig:rho_tradeoff}). For $\rho=0.3$, $0.5$, and $0.8$, mean runtimes are 35.93, 57.23, and 157.06~s, while the mean distances are 0.146, 0.114, and 0.078. Larger thresholds trigger main resampling and its associated rejuvenation more often, as reported in \Cref{tab:scenario_runtime_resampling}. In this experiment, mean posterior fidelity improves at each increase in the threshold. However, increasing $\rho$ from 0.5 to 0.8 reduces the mean distance by approximately one-third while requiring about 2.7 times the runtime. The additional resampling and rejuvenation should therefore be judged against this computational cost rather than maximised irrespective of cost.

\Cref{fig:T_tradeoff} provides empirical support for rejuvenation by showing that it reduces the mean marginal Wasserstein-1 distance to batch MCMC relative to the no-move condition ($T_{\mathrm{move}}=0$). Specifically, the mean distances across the same 20 simulated datasets are 0.466, 0.207, 0.114, and 0.091 for $T_{\mathrm{move}}=0$, $1$, $5$, and $10$, respectively; the corresponding mean runtimes are 11.03, 20.28, 57.23, and 102.71~s. The first sweep provides a substantial improvement over $T_{\mathrm{move}}=0$, while increasing from five to ten sweeps yields a smaller additional reduction in mean distance. Across the matched replications, the mean accumulated time spent in the rejuvenation kernel is 9.48~s with one sweep and 46.35~s with five sweeps, explaining most of the increase in total runtime relative to the no-move condition ($T_{\mathrm{move}}=0$). The substantial improvement from $T_{\mathrm{move}}=0$ to $1$ supports using at least one rejuvenation sweep as a practical starting point, although additional sweeps may be needed to achieve the required posterior accuracy.

\Cref{fig:kappa_tradeoff} summarises the accuracy--cost trade-off in the partial-history experiment. Reducing $\kappa$ from 1 to 0.5 lowers mean runtime from 57.23~s to 33.55~s, while increasing the mean marginal Wasserstein-1 distance from 0.114 to 0.242. At $\kappa=0.25$, runtime falls further to 21.69~s, but the distance increases to 0.273 and the mean number of occupied clusters rises to 30.57, compared with 28.29 under batch MCMC. At $\kappa=0.5$, the mean cluster count remains close to the batch benchmark, but this does not imply an equally accurate coefficient distribution. Hence, partial-history rejuvenation can reduce cost substantially, but revisiting too small a fraction of past observations reduces posterior accuracy in the present experiment. \textcolor{black}{A comparison of \Cref{fig:T_tradeoff,fig:kappa_tradeoff} shows that a single full-history sweep achieves a lower mean Wasserstein-1 distance than five sweeps over one-quarter of the history, at a similar computational cost. Under the simulation design examined here, this suggests reducing the number of sweeps before reducing the history fraction when computation time is limited.}

Taken together, these experiments provide practical guidance for tuning the proposed SMC algorithm to the posterior accuracy and processing-time requirements of an online application.
\end{revision}

\begin{table}[htbp]
\centering
\caption{Scenario-wise predictive fit and final clustering.}
\label{tab:scenario_lps_clusters}
\small
\setlength{\tabcolsep}{5pt}
\begin{tabular}{ll ccc}
\toprule
Scenario & Setting & \makecell[b]{Last-500\\LPS} & \makecell[b]{Holdout\\LPS} & \makecell[b]{Final number\\of clusters} \\
\midrule
baseline & Baseline & $-0.6821$ $(0.0059)$ & $-0.6798$ $(0.0066)$ & $27.94$ $(2.71)$ \\
M500 & $M=500$ & $-0.6821$ $(0.0061)$ & $-0.6798$ $(0.0066)$ & $28.55$ $(2.49)$ \\
M5000 & $M=5000$ & $-0.6821$ $(0.0059)$ & $-0.6798$ $(0.0066)$ & $28.26$ $(1.65)$ \\
J1 & $J=1$ & $-0.6821$ $(0.0059)$ & $-0.6798$ $(0.0067)$ & $27.67$ $(2.58)$ \\
J10 & $J=10$ & $-0.6821$ $(0.0059)$ & $-0.6799$ $(0.0067)$ & $27.96$ $(2.31)$ \\
rho03 & $\rho=0.3$ & $-0.6822$ $(0.0059)$ & $-0.6798$ $(0.0066)$ & $28.46$ $(3.07)$ \\
rho08 & $\rho=0.8$ & $-0.6821$ $(0.0059)$ & $-0.6798$ $(0.0066)$ & $27.98$ $(2.22)$ \\
T0 & $T_{\mathrm{move}}=0$ & $-0.6822$ $(0.0062)$ & $-0.6800$ $(0.0066)$ & $26.32$ $(5.52)$ \\
T1 & $T_{\mathrm{move}}=1$ & $-0.6821$ $(0.0060)$ & $-0.6798$ $(0.0065)$ & $27.52$ $(3.56)$ \\
T10 & $T_{\mathrm{move}}=10$ & $-0.6821$ $(0.0059)$ & $-0.6798$ $(0.0066)$ & $28.09$ $(1.81)$ \\
kappa025 & $\kappa=0.25$ & $-0.6821$ $(0.0060)$ & $-0.6798$ $(0.0067)$ & $30.57$ $(4.74)$ \\
kappa050 & $\kappa=0.5$ & $-0.6822$ $(0.0059)$ & $-0.6801$ $(0.0067)$ & $28.44$ $(3.72)$ \\
kappa075 & $\kappa=0.75$ & $-0.6821$ $(0.0059)$ & $-0.6798$ $(0.0065)$ & $27.95$ $(3.04)$ \\
\midrule
Batch MCMC &  & --- & $-0.6798$ $(0.0066)$ & $28.29$ $(1.65)$ \\
\bottomrule
\end{tabular}

\par\smallskip
{\footnotesize Entries are averages across 20 replications. Standard deviations across replications are reported in parentheses. \rev{The baseline includes $T_{\mathrm{move}}=5$ and $\kappa=1$; other rows change only the indicated setting. Final cluster counts are posterior means.} For batch MCMC, the final number of clusters is reported as the pooled posterior mean of the number of occupied clusters across the three post-burn-in chains, averaged across replications.}
\end{table}

\begin{table}[htbp]
\centering
\caption{Scenario-wise runtime, stabilising updates, and resampling.}
\label{tab:scenario_runtime_resampling}
\footnotesize
\setlength{\tabcolsep}{4pt}
\begin{tabular}{l rrrcc}
\toprule
Scenario &
\makecell[b]{Runtime\\(s)} &
\makecell[b]{Stabilising-\\update count} &
\makecell[b]{Resampling\\count} &
\makecell[b]{Processing time\\ at resampling\\ trigger (s)} &
\makecell[b]{Processing time at\\last resampling\\trigger (s)} \\
\midrule
baseline & $57.23$ $(5.53)$ & $0.95$ $(0.60)$ & $32.25$ $(2.29)$ & $1.46$ $(0.12)$ & $5.89$ $(0.51)$ \\
M500 & $29.68$ $(2.27)$ & $0.80$ $(0.70)$ & $31.85$ $(2.18)$ & $0.75$ $(0.06)$ & $2.95$ $(0.18)$ \\
M5000 & $270.69$ $(15.73)$ & $0.90$ $(0.72)$ & $32.70$ $(1.63)$ & $6.99$ $(0.44)$ & $27.78$ $(1.51)$ \\
J1 & $50.71$ $(4.84)$ & $1.10$ $(0.79)$ & $33.35$ $(2.01)$ & $1.23$ $(0.11)$ & $5.18$ $(0.46)$ \\
J10 & $58.46$ $(4.89)$ & $0.85$ $(0.75)$ & $31.90$ $(2.07)$ & $1.53$ $(0.14)$ & $6.01$ $(0.42)$ \\
rho03 & $35.93$ $(4.39)$ & $0.90$ $(0.85)$ & $17.80$ $(1.88)$ & $1.46$ $(0.17)$ & $5.54$ $(0.70)$ \\
rho08 & $157.06$ $(9.24)$ & $0.90$ $(0.85)$ & $96.05$ $(3.79)$ & $1.54$ $(0.07)$ & $6.09$ $(0.37)$ \\
T0 & $11.03$ $(0.48)$ & $0.95$ $(0.76)$ & $29.75$ $(2.45)$ & $0.03$ $(0.00)$ & $0.02$ $(0.00)$ \\
T1 & $20.28$ $(0.81)$ & $0.90$ $(0.79)$ & $31.95$ $(1.79)$ & $0.32$ $(0.02)$ & $1.19$ $(0.12)$ \\
T10 & $102.71$ $(7.04)$ & $1.05$ $(0.60)$ & $32.35$ $(1.81)$ & $2.87$ $(0.21)$ & $11.39$ $(0.78)$ \\
kappa025 & $21.69$ $(1.87)$ & $0.85$ $(0.75)$ & $31.25$ $(2.15)$ & $0.39$ $(0.05)$ & $1.47$ $(0.19)$ \\
kappa050 & $33.55$ $(2.85)$ & $0.85$ $(0.75)$ & $32.25$ $(2.15)$ & $0.74$ $(0.06)$ & $2.90$ $(0.20)$ \\
kappa075 & $43.26$ $(3.10)$ & $0.80$ $(0.62)$ & $32.20$ $(1.58)$ & $1.05$ $(0.08)$ & $4.28$ $(0.40)$ \\
\bottomrule
\end{tabular}

\par\smallskip
{\footnotesize Entries are averages across 20 replications. Standard deviations across replications are reported in parentheses. The fifth column reports the mean algorithm-processing time for observations that trigger main resampling, calculated within each replication and then averaged across replications. The final column reports the algorithm-processing time for the last such observation in each replication, averaged across replications. These times include the complete observation update, including resampling and rejuvenation when enabled; they are not timings of resampling alone.}
\end{table}

\begin{sidewaystable}[htbp]
\centering
\caption{Scenario-wise posterior summaries for $\beta$, $\gamma$, and $\alpha$.}
\label{tab:scenario_beta_gamma_alpha}
\small
\begin{tabular}{lcccc}
\toprule
Scenario & $\beta_1$ & $\beta_2$ & $\gamma$ & $\alpha$ \\
\midrule
baseline & $0.00$ [$-5.43$, $5.20$] & $-0.03$ [$-6.04$, $6.05$] & $-0.52$ [$-0.81$, $-0.30$] & $3.93$ [$1.02$, $8.66$] \\
M500 & $0.01$ [$-5.36$, $5.22$] & $-0.01$ [$-6.05$, $5.93$] & $-0.52$ [$-0.80$, $-0.31$] & $4.00$ [$1.06$, $8.48$] \\
M5000 & $0.00$ [$-5.53$, $5.22$] & $-0.01$ [$-6.12$, $6.14$] & $-0.53$ [$-0.82$, $-0.31$] & $3.96$ [$1.06$, $8.58$] \\
J1 & $0.01$ [$-5.46$, $5.25$] & $0.01$ [$-6.00$, $6.17$] & $-0.52$ [$-0.80$, $-0.31$] & $3.88$ [$1.11$, $8.29$] \\
J10 & $0.00$ [$-5.50$, $5.28$] & $-0.02$ [$-6.05$, $5.99$] & $-0.53$ [$-0.81$, $-0.32$] & $3.91$ [$1.05$, $8.35$] \\
rho03 & $0.00$ [$-5.53$, $5.12$] & $-0.01$ [$-6.22$, $6.13$] & $-0.52$ [$-0.82$, $-0.32$] & $3.99$ [$1.12$, $8.62$] \\
rho08 & $0.00$ [$-5.52$, $5.18$] & $0.00$ [$-6.06$, $6.11$] & $-0.52$ [$-0.81$, $-0.31$] & $3.91$ [$1.05$, $8.50$] \\
T0 & $0.03$ [$-4.64$, $4.84$] & $0.06$ [$-5.29$, $5.71$] & $-0.49$ [$-0.67$, $-0.33$] & $3.67$ [$1.35$, $7.26$] \\
T1 & $0.00$ [$-5.42$, $5.05$] & $-0.05$ [$-6.17$, $6.01$] & $-0.53$ [$-0.78$, $-0.33$] & $3.85$ [$1.12$, $8.58$] \\
T10 & $0.00$ [$-5.55$, $5.27$] & $-0.02$ [$-6.02$, $5.96$] & $-0.52$ [$-0.81$, $-0.31$] & $3.93$ [$0.99$, $8.69$] \\
kappa025 & $0.02$ [$-5.33$, $5.30$] & $-0.01$ [$-5.95$, $5.96$] & $-0.53$ [$-0.80$, $-0.33$] & $4.30$ [$1.47$, $8.25$] \\
kappa050 & $0.03$ [$-5.53$, $5.51$] & $0.00$ [$-6.05$, $6.11$] & $-0.54$ [$-0.81$, $-0.33$] & $3.98$ [$1.16$, $8.30$] \\
kappa075 & $0.01$ [$-5.44$, $5.17$] & $-0.01$ [$-5.95$, $5.95$] & $-0.51$ [$-0.81$, $-0.30$] & $3.91$ [$1.12$, $8.36$] \\
\midrule
Batch MCMC & $0.00$ [$-5.55$, $5.26$] & $-0.01$ [$-6.07$, $6.08$] & $-0.52$ [$-0.82$, $-0.31$] & $3.96$ [$1.06$, $8.60$] \\
\bottomrule
\end{tabular}

\par\smallskip
{\footnotesize Entries are averages across 20 replications. For each SMC scenario, within each replication, the posterior mean and the equal-tail 95\% credible interval are computed from the weighted SMC posterior, and the reported interval endpoints are then averaged across replications. For batch MCMC, within each replication, the pooled 3-chain posterior mean and the equal-tail 95\% credible interval are computed from the pooled post-burn-in draws, and the reported interval endpoints are then averaged across replications.}
\end{sidewaystable}

\clearpage

\newpage
\clearpage
\section{Illustrative empirical application}
\label{sec:case}

\subsection{Data, trip-level model specification, and computational setup}
\label{sec:case:data-model-setup}

This section presents an illustrative empirical application of the proposed sequential inference method using smartphone-based person-trip records collected in the Tsukechi district of Nakatsugawa City, Japan, in October 2024. The purpose of the application is to demonstrate how the proposed algorithm processes real-world trip observations sequentially and to examine its posterior approximation and computational performance, rather than to provide a comprehensive empirical analysis of travel behaviour in the study area. The analysis sample is trip-based and is restricted to trips whose origin lies within the Tsukechi district. We also exclude trips whose recorded purpose is walking/running, as these observations are not directly comparable with destination-oriented trips for which driving a car is a meaningful mode-choice outcome. After these restrictions, the final estimation sample contains 2{,}502 trips from 32 individuals, of which 2{,}293 were made with the respondent driving a car and 209 were made using other modes.

The binary outcome equals one if the respondent was the driver of a car on the trip, and zero otherwise. The shared coefficient block $\gamma$ is included to adjust for observed trip and respondent characteristics and contains an intercept, a weekend dummy, a female dummy, an indicator for age 60 or above, household car ownership, a single-person household dummy, a same-district destination dummy, and trip-purpose dummies. Definitions of the model covariates are reported in \Cref{tab:case-vars}, and descriptive statistics for the observed variables are reported in \Cref{tab:case-desc}. The cluster-specific block contains only the standardised origin--destination (OD) log-distance variable. Log-distance is standardised using the mean and standard deviation calculated from all 2{,}502 trips in the estimation sample. These scaling constants are held fixed throughout sequential processing. Within each posterior state, each trip is assigned to a latent coefficient cluster, and $\beta_h$ represents the distance sensitivity associated with cluster $h$, capturing heterogeneity across trip observations in the association between trip length and the propensity to drive a car. The occupied clusters are therefore interpreted as trip-level latent coefficient clusters learned from the pooled trip records rather than as persistent respondent-level types.

Trips are processed in their observed chronological order, and the resulting posterior path is therefore conditional on the realised arrival sequence of the data. As in the simulation study, the \rev{stabilising update} is triggered whenever the diagnostic ESS falls by at least 50\% relative to the ESS before incorporating $\D_i$. The tuning parameters are set as
\begin{align*}
M = 5{,}000,\quad
J = 5,\quad
\rho = 0.5,\quad
\mu = 0.9,\quad
\kappa = 1,\quad
T_{\mathrm{move}} = 1.
\end{align*}
As in the simulation study in \Cref{sec:simulation}, we use independent zero-centred Gaussian coefficient priors with variance 10 and a Gamma prior for $\alpha$ with shape 2 and rate 1. \rev{The sensitivity analysis for the prior scale of the distance-sensitivity parameter $\beta$ is reported in full in \Cref{sec:case:sensitiv}.} All computations for the empirical application were run on a laptop equipped with an Intel Core i7-10875H CPU at 2.30 GHz (8 cores, 16 threads).

\begin{revision}
All reported SMC configurations in this section use 12 Julia threads for parallel computation and are evaluated on the fixed empirical dataset using the same 20 algorithmic seeds. Parameter summaries are computed from weighted posterior samples pooled across the 20 runs, with each run receiving equal total weight. Density curves show the pointwise mean of the run-specific smoothed densities. Run-level quantities such as runtime, Wasserstein-1 distance, and the posterior mean number of occupied clusters are summarised across runs. Shaded bands in the corresponding figures show the 10th--90th percentiles of the run-specific quantities. These repeated runs quantify finite-particle Monte Carlo variability and do not alter the posterior target at any observation index.
\end{revision}

\begin{table}[htbp]
\centering
\small
\caption{Variable definitions}
\label{tab:case-vars}
\renewcommand{\arraystretch}{1.08}
\begin{tabular}{@{}L{4.5cm}L{10.0cm}@{}}
\toprule
Variable & Definition \\
\midrule
Weekend dummy & Equals 1 if the trip is observed on Saturday or Sunday. \\
Female dummy & Equals 1 if the respondent is female. \\
Age 60+ dummy & Equals 1 if the respondent is aged 60 or above. \\
Household car ownership & Number of cars owned by the respondent's household. \\
Single-person household dummy & Equals 1 if the respondent belongs to a one-person household. \\
Same-district destination dummy & Equals 1 if the trip destination lies within the Tsukechi district. \\
Trip-purpose dummies & Indicators included in the shared coefficient block for commute/school, delivery, farm work, other work, pick-up/drop-off, medical, shopping, eating/leisure, hobby/lesson, and other private purposes. Return home and unknown purposes are reported in the descriptive statistics but are absorbed into the omitted reference category in the reported specification. \\
Standardised log-distance & Standardised $\log(1+\text{OD distance in km})$, where OD distance is the straight-line distance between matched departure-side and arrival-side GPS points. \\
\bottomrule
\end{tabular}
\end{table}

\begin{table}[htbp]
\centering
\small
\caption{Descriptive statistics for the estimation sample}
\label{tab:case-desc}
\renewcommand{\arraystretch}{1.05}
\begin{tabular}{@{}L{9.5cm}r@{}}
\toprule
Variable & Mean (SD) or \% \\
\midrule
OD straight-line distance (km) & 4.24 (10.10) \\
Weekend dummy & 21.8\% \\
Female dummy & 63.7\% \\
Age 60+ dummy & 31.3\% \\
Household car ownership & 2.90 (1.27) \\
Single-person household dummy & 6.3\% \\
Same-district destination dummy & 86.2\% \\
\addlinespace
\multicolumn{2}{@{}l@{}}{\textit{Trip-purpose dummies}} \\
Commute/school & 20.8\% \\
Return home & 29.5\% \\
Delivery & 1.1\% \\
Farm work & 0.4\% \\
Other work & 11.5\% \\
Pick-up/drop-off & 11.3\% \\
Medical & 1.0\% \\
Shopping & 11.1\% \\
Eating/leisure & 2.7\% \\
Hobby/lesson & 1.4\% \\
Other private & 9.0\% \\
Unknown & 0.3\% \\
\bottomrule
\end{tabular}

\vspace{0.35em}
\parbox{0.94\linewidth}{\footnotesize Notes: $N=2{,}502$ trips from 32 individuals. Continuous variables are reported as means, with standard deviations (SDs) in parentheses. Binary variables are reported as percentages. All statistics are calculated across trip records; respondent and household attributes therefore contribute once for each recorded trip.}
\end{table}

\subsection{Results and discussion}
\label{sec:case:results}

\subsubsection{Comparison with batch MCMC}
\label{sec:case:comp_MCMC}

\begin{revision}
We first compare the approximation obtained by the proposed SMC algorithm with batch MCMC for the same full-data posterior. The posterior comparisons use a representative batch MCMC run with 25,000 iterations per chain, including 5,000 burn-in iterations, pooling the retained draws from three chains. Split-$\widehat{R}<1.05$ was satisfied for the observation-weighted mean and positive mass of $\beta$, every component of $\gamma$, $\alpha$, and the number of occupied clusters. This run provides a common reference posterior for assessing the SMC approximations.
Both algorithms were implemented in Julia on the same hardware under a nominal total allocation of 12 Julia threads. The proposed SMC algorithm uses particle-level parallelism, whereas batch MCMC runs three chain processes concurrently, each with four Julia threads.
We repeated the baseline SMC run 20 times on the fixed empirical dataset using the common algorithmic seeds specified above. Mean algorithmic runtime was 216.61~s (SD 8.71~s), excluding auxiliary file output used only for diagnostic and posterior-summary plots.\footnote{\rev{The reported runtime should not be interpreted as evidence of a general speed advantage over batch MCMC. As an additional check, we conducted five MCMC runs with 25,000 iterations per chain, including 1,000 burn-in iterations, using three concurrent chains with four Julia threads each. All monitored quantities satisfied split-$\widehat{R}<1.05$ in every run. The mean sampling runtime was 497.75~s (SD 13.51~s), measured as the maximum sampler-loop elapsed time across the three chains and excluding final file output. The mean Wasserstein-1 distance to the 25,000-iteration reference posterior for $\beta$ was 0.023, compared with 0.091 across the 20 SMC runs with $M=5{,}000$.}}
Across the 20 runs, the stabilising update was invoked 20.20 times on average (SD 0.77), and main resampling was triggered 59.70 times on average (SD 0.92). The mean processing time of the final main-resampling update was 3.06~s (SD 0.17~s), of which the rejuvenation move accounted for 2.92~s on average (SD 0.16~s).
\end{revision}

\begin{figure}[htbp]
    \centering
    \includegraphics[width=0.9\textwidth]{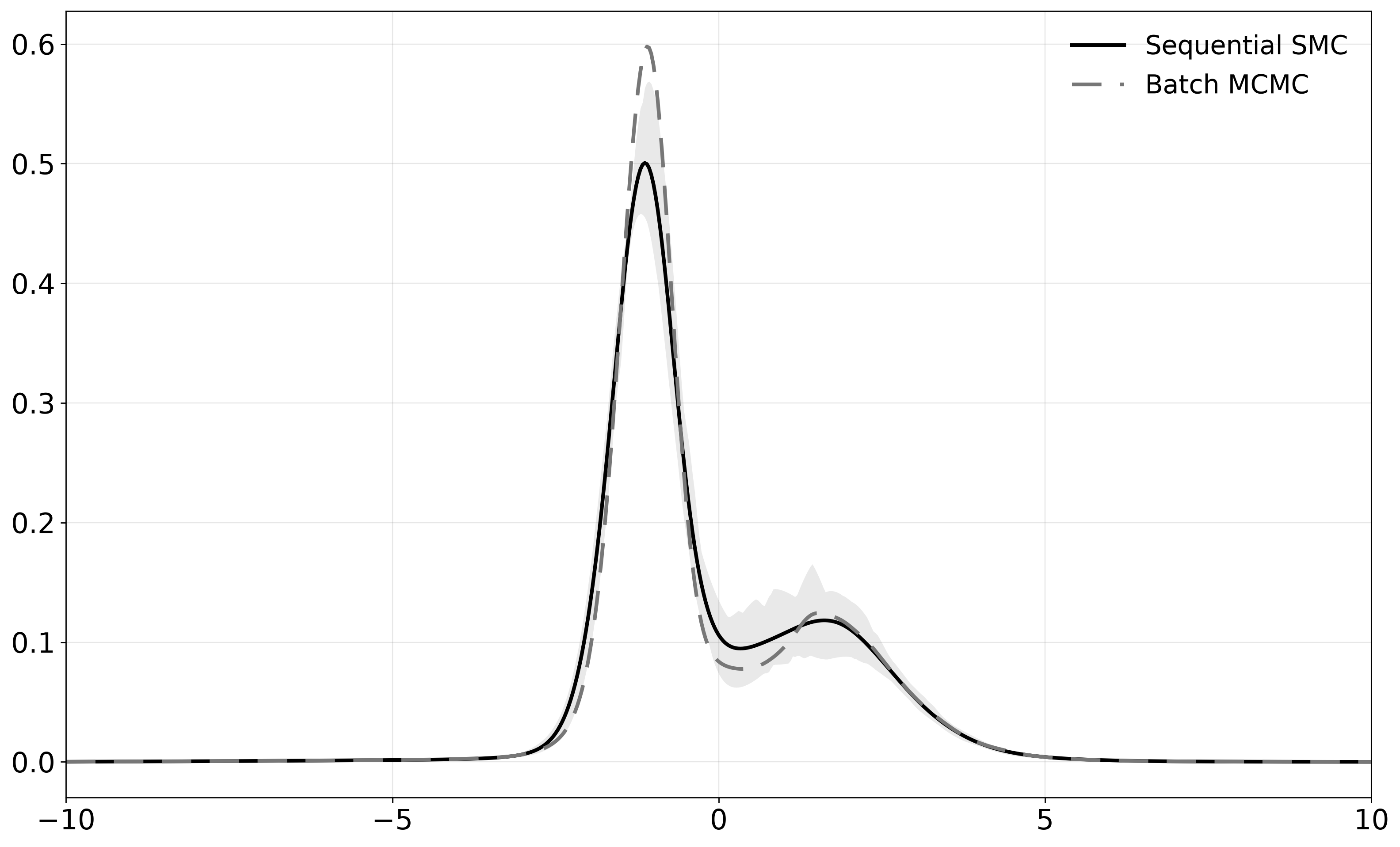}
    \caption{\rev{Final posterior densities of $\beta$ under the proposed SMC algorithm and batch MCMC. The SMC line shows the pointwise mean density across 20 algorithmic seeds for visualising the average posterior shape, and the shaded band shows the 10th--90th percentiles across seeds.}
}
    \label{fig:smc_mcmc_case}
\end{figure}

\begin{revision}
\Cref{fig:smc_mcmc_case} shows that the final posterior density of $\beta$ obtained by the proposed SMC algorithm is close to that obtained by batch MCMC. The posterior means of $\beta$ are $-0.192$ for the pooled SMC approximation across 20 seeds and $-0.193$ under batch MCMC, and the posterior probabilities $\Pr(\beta>0)$ are 0.339 and 0.328, respectively. \rev{Across the 20 individual SMC runs, the mean Wasserstein-1 distance to the batch MCMC posterior is 0.091. The Wasserstein-1 distance calculated for each SMC run is used throughout the empirical application to characterise the fidelity of a typical $M$-particle approximation. The pointwise mean density in \Cref{fig:smc_mcmc_case} is used only to visualise the average posterior shape across seeds.}
\end{revision}

\subsubsection{Sequential posterior learning and provisional output}
\label{sec:case:posteriors}

\Cref{tab:case-posteriors} reports posterior summaries for the fully updated posterior at the final observation. The covariates in $z_i$ are included to adjust for observed trip and respondent characteristics, and their shared coefficients $\gamma$ are reported for completeness. The discussion therefore focuses on $\beta$, the trip-level heterogeneous distance coefficient modelled through the DP mixture.

\begin{figure}[htbp]
    \centering
    \includegraphics[width=0.9\textwidth]{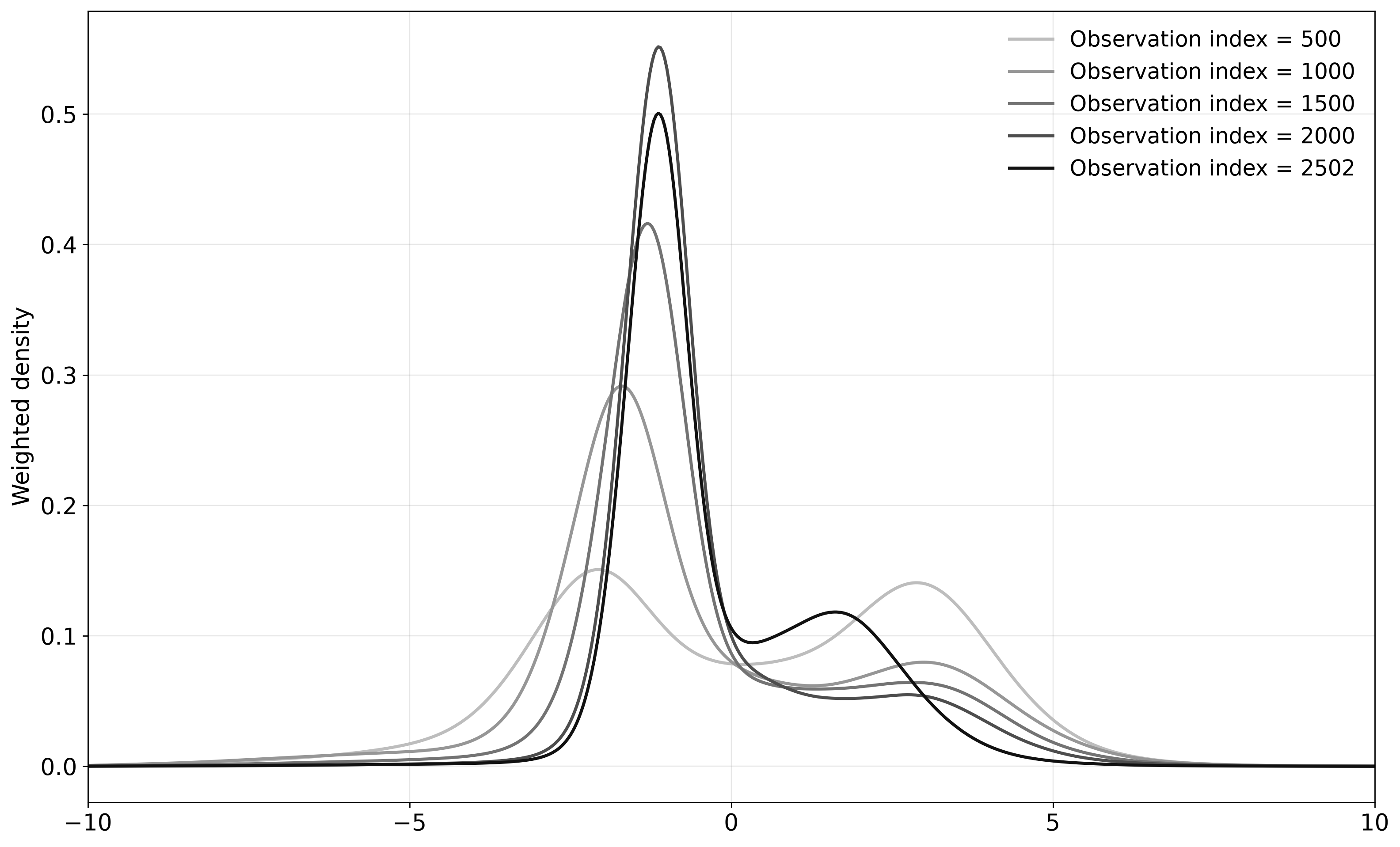}
    \caption{\rev{Evolution of the seed-averaged fully updated posterior of $\beta$ across selected observation indices. Each curve is averaged across 20 algorithmic seeds.}}
    \label{fig:Evo_pos}
\end{figure}

\begin{revision}
For the observation-weighted marginal posterior of the trip-level distance coefficient $\beta$ defined in \Cref{sec:method:overview}, the posterior mean is slightly negative, at $-0.19$, whereas the equal-tail 95\% credible interval, $[-2.10,\,3.39]$, is wide and covers both negative and positive values. The corresponding posterior mass above zero is 0.34. These summaries place greater posterior mass on negative distance sensitivity while retaining non-negligible support for positive values. \Cref{fig:Evo_pos} shows the seed-averaged evolution of this distribution at selected observation indices as new observations are incorporated. In the earlier part of the sequence, the posterior is relatively diffuse, whereas by around the middle to later part of the sequence a more stable shape emerges. The posterior progressively concentrates on a dominant negative mode while retaining non-negligible mass on the positive side.
\end{revision}

\begin{figure}[htbp]
    \centering
    \includegraphics[width=0.9\textwidth]{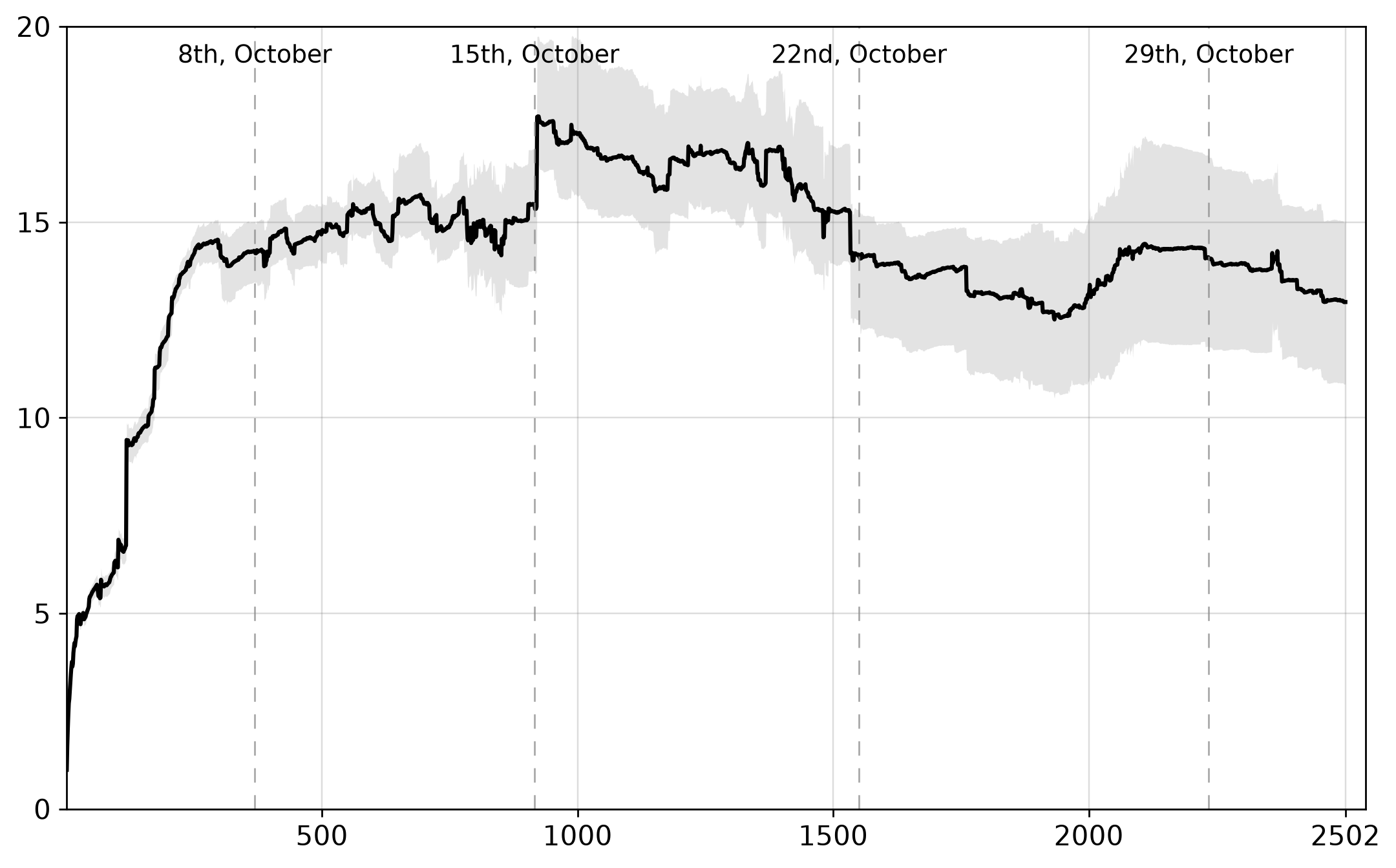}
    \caption{\rev{Posterior mean number of occupied clusters over the sequential observation index under the original data order. The solid line is the mean across 20 algorithmic seeds, and the shaded band shows the 10th--90th percentiles across seeds.}}
    \label{fig:Num_clus}
\end{figure}

\begin{revision}
\rev{\Cref{fig:Num_clus} plots the seed-averaged posterior mean number of occupied trip-level latent coefficient clusters over the sequential observation index under the original observation order. The number rises during the earlier part of the stream, reaches a relatively high level around the middle, and then declines after approximately 1{,}500 observations before ending at 12.96 on average across the 20 algorithmic seeds (SD 1.86). The 10th--90th percentile band shows that the precise trajectory varies across seeds.}
\end{revision}

\begin{figure}[htbp]
    \centering
    \includegraphics[width=1.0\textwidth]{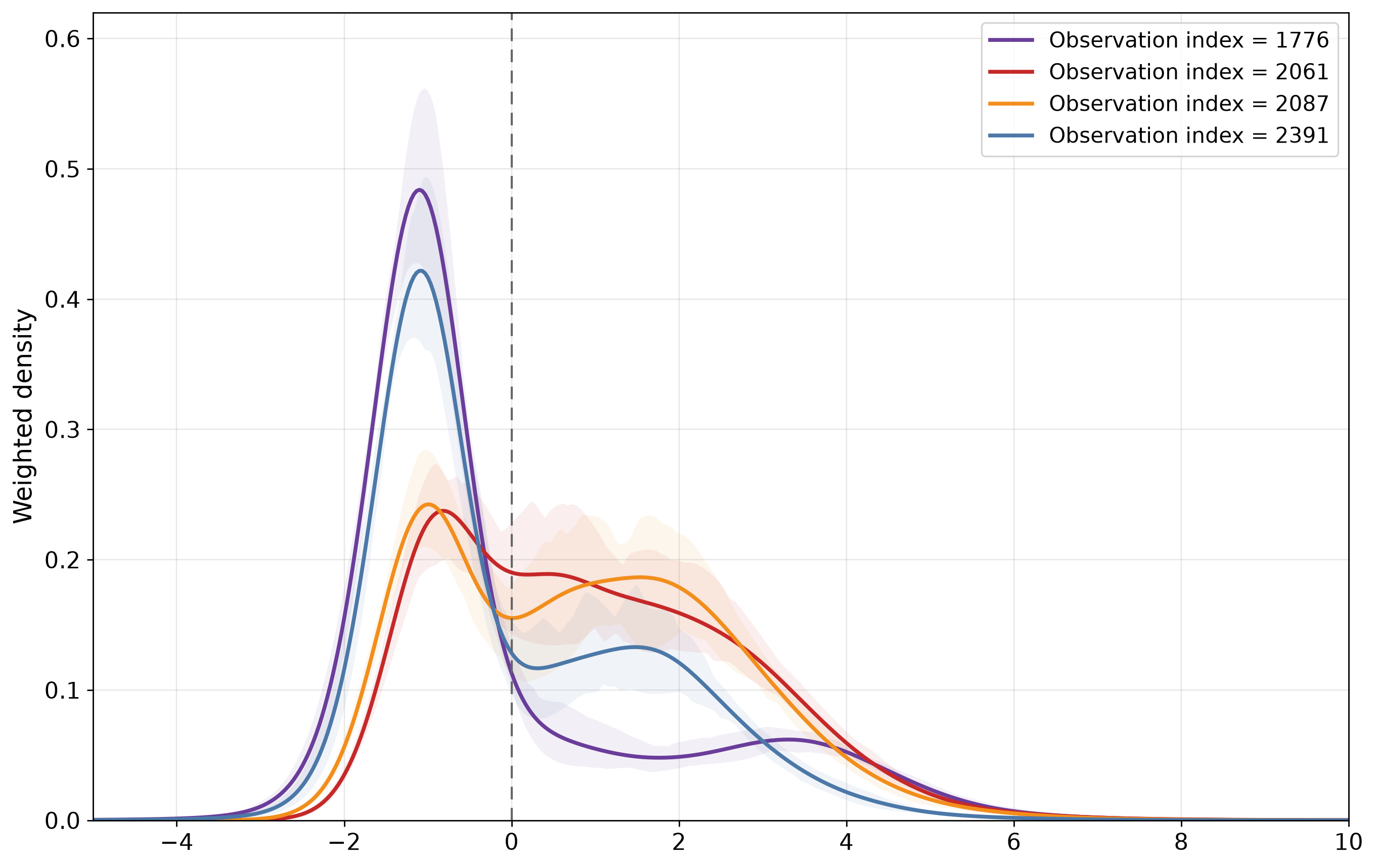}
    \caption{\rev{Seed-averaged observation-specific provisional posterior distributions of the assigned cluster-specific distance coefficient $\beta_{\zeta_i}$ for the last four outside-district trips made with the respondent driving a car for eating/leisure purposes. Lines are means across 20 algorithmic seeds and shaded bands show the 10th--90th percentiles across seeds.}}
    \label{fig:Provisional}
\end{figure}

\begin{revision}
Finally, \Cref{fig:Provisional} presents seed-averaged observation-specific provisional posterior distributions of the assigned cluster-specific distance coefficient $\beta_{\zeta_i}$ for the last four outside-district trips made with the respondent driving a car for eating/leisure purposes. The selected trips correspond to observation indices 1776, 2061, 2087, and 2391 and were made by four different respondents. For each focal trip $i$, the distribution is constructed from the weighted sample $\{(\beta_{\zeta_i^{(m)}}^{(m)},W_i^{(m)})\}_{m=1}^{M}$, as defined in \Cref{sec:method:provisional}, and then averaged across the 20 algorithmic seeds. The corresponding mean values of $\Pr(\beta_{\zeta_i}>0)$ are 0.284, 0.643, 0.628, and 0.396, respectively. A notable feature of the figure is that the four trips yield provisional posteriors with different shapes, and the 10th--90th percentile bands show their variability across seeds. Each curve summarises posterior uncertainty in both the focal trip's cluster assignment and the coefficient associated with that cluster, conditional on the data assimilated through observation $i$. The figure therefore illustrates that the proposed algorithm can produce distinct trip-specific posterior summaries immediately after newly arriving binary choice observations have been incorporated.
\end{revision}

\begin{revision}
Another operationally important result is the computational speed at which the provisional posteriors can be produced. Provisional latency is measured from the start of processing each trip record to the completion of its provisional posterior summaries, before file output and historical rejuvenation. The reported latencies are measured separately in each run and then summarised across seeds; they do not include construction of the seed-averaged density curves. The four selected trips in \Cref{fig:Provisional} followed the standard route in all 20 runs, and their mean provisional latencies were 0.029~s, 0.029~s, 0.031~s, and 0.031~s, respectively. Across the 20 runs, 404 observation updates followed the stabilising route; their median provisional latency was 0.174~s and their maximum was 0.399~s. Thus, the stabilising route generally required more processing time than the standard route, while still producing provisional output within 0.4~s in these runs. These results demonstrate the computational feasibility of producing low-latency observation-specific posterior summaries under both standard and stabilised updates.
\end{revision}

\begin{revision}
\begin{table}[htbp]
\centering
\small
\caption{Posterior summaries of the observation-weighted distribution of $\beta$ and the shared coefficients $\gamma$}
\label{tab:case-posteriors}
\renewcommand{\arraystretch}{1.05}
\begin{tabular}{@{}L{2.0cm}L{6.2cm}r@{ [}r@{, }r@{]}@{}}
\toprule
Parameter & Covariate & \multicolumn{3}{c@{}}{Posterior summary} \\
\midrule
$\beta$ & Standardised log-distance & -0.19 & -2.10 & 3.39 \\
\midrule
$\gamma$ & Intercept & 5.29 & 3.86 & 6.86 \\
 & Weekend dummy & -1.26 & -1.65 & -0.88 \\
 & Female dummy & 0.25 & -0.18 & 0.69 \\
 & Age 60+ dummy & -0.06 & -0.49 & 0.37 \\
 & Household car ownership & -0.06 & -0.26 & 0.15 \\
 & Single-person household dummy & 4.00 & 2.25 & 6.38 \\
 & Same-district destination dummy & -1.78 & -3.17 & -0.54 \\
 & Commute/school & 0.72 & 0.05 & 1.42 \\
 & Delivery & 3.16 & -0.09 & 7.61 \\
 & Other work & -1.24 & -1.88 & -0.61 \\
 & Shopping & -0.13 & -0.77 & 0.52 \\
 & Eating/leisure & -2.04 & -2.90 & -1.19 \\
 & Hobby/lesson & -1.56 & -2.64 & -0.39 \\
 & Medical & -0.91 & -2.47 & 0.80 \\
 & Pick-up/drop-off & 0.98 & 0.14 & 1.94 \\
 & Farm work & -1.66 & -3.47 & 0.31 \\
 & Other private & -1.41 & -2.00 & -0.84 \\
\bottomrule
\end{tabular}

\vspace{0.35em}
\begin{minipage}{0.92\textwidth}
\footnotesize
Notes: Posterior means and equal-tail 95\% credible intervals are computed from weighted samples of the fully updated posteriors at the final observation ($i = 2{,}502$), pooled across 20 algorithmic seeds with equal total weight assigned to each run. The entries for $\beta$ summarise the observation-weighted coefficient distribution defined in \Cref{sec:method:overview}. The entries for $\gamma$ summarise the posterior distributions of coefficients shared across trips and associated with covariates included for adjustment. \textcolor{black}{Respondent-level attributes are constant across each respondent's trips, and the specification includes no respondent-level effect. The credible intervals for their coefficients therefore do not account for within-respondent dependence and may understate uncertainty.}
\end{minipage}
\end{table}
\end{revision}

\newpage
\subsubsection{\rev{Sensitivity to the number of particles, data ordering, and prior scale}}
\label{sec:case:sensitiv}

\rev{We next examine sensitivity to two computational features of the sequential approximation---the number of particles and the order in which observations arrive---and to the prior scale of $\beta$.} We first examine whether the baseline particle size provides an adequate approximation for the empirical dataset. We reran the proposed SMC algorithm with $M=1{,}000$, $2{,}500$, $5{,}000$, and $10{,}000$, keeping the remaining settings fixed. \rev{We then compared the final posterior of $\beta$ obtained by the proposed SMC algorithm with the batch MCMC posterior using the Wasserstein-1 distance introduced in \Cref{sec:simulation:base}.}

\begin{figure}[htbp]
    \centering
    \includegraphics[width=1.0\textwidth]{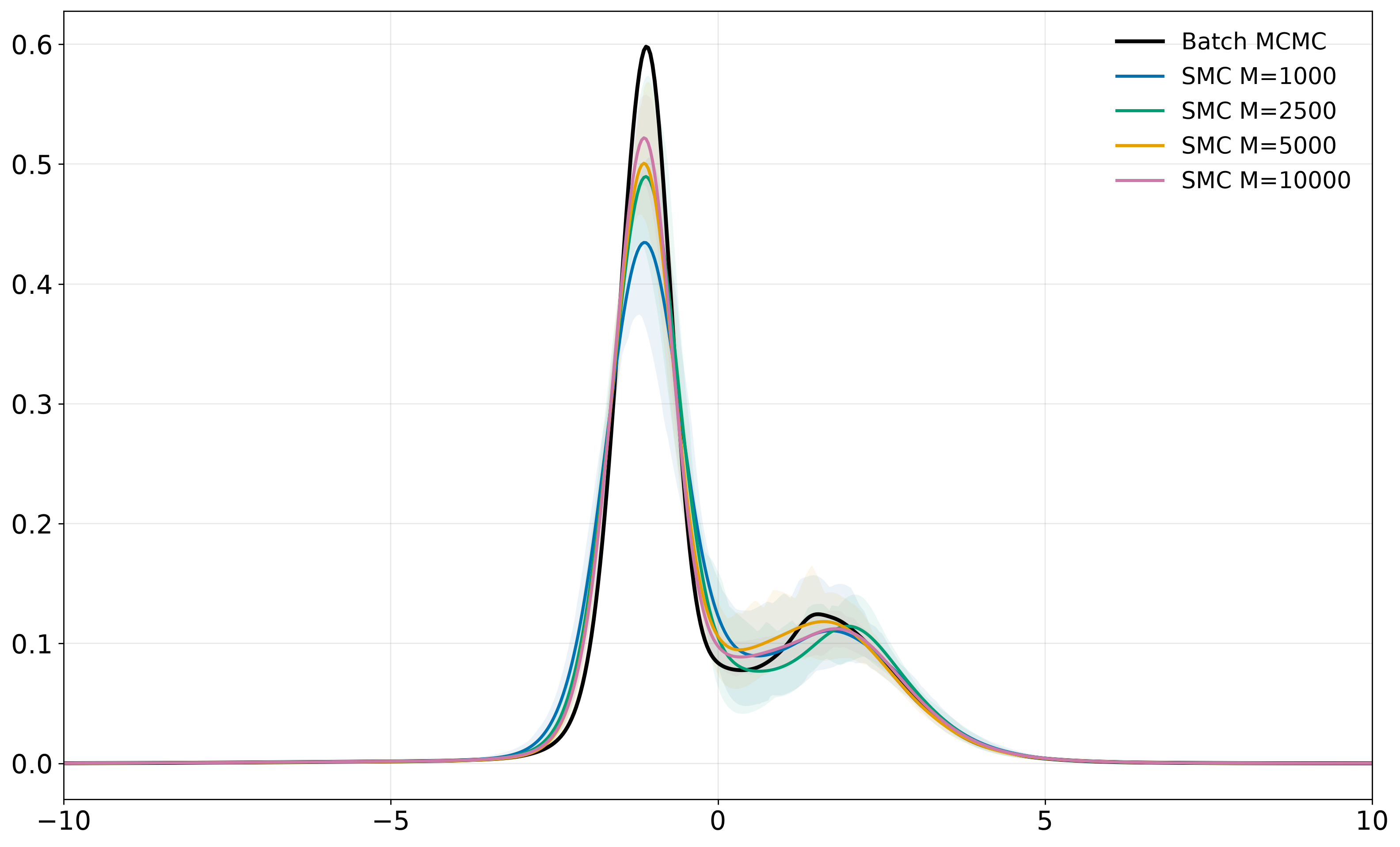}
    \caption{\rev{Sensitivity of the final posterior density of $\beta$ to the number of particles. SMC lines are means across 20 common algorithmic seeds, shaded bands show the 10th--90th percentiles across seeds, and the black line is the batch MCMC reference.}}
    \label{fig:case_M_sensitiv}
\end{figure}

\begin{figure}[htbp]
    \centering
    \includegraphics[width=1.0\textwidth]{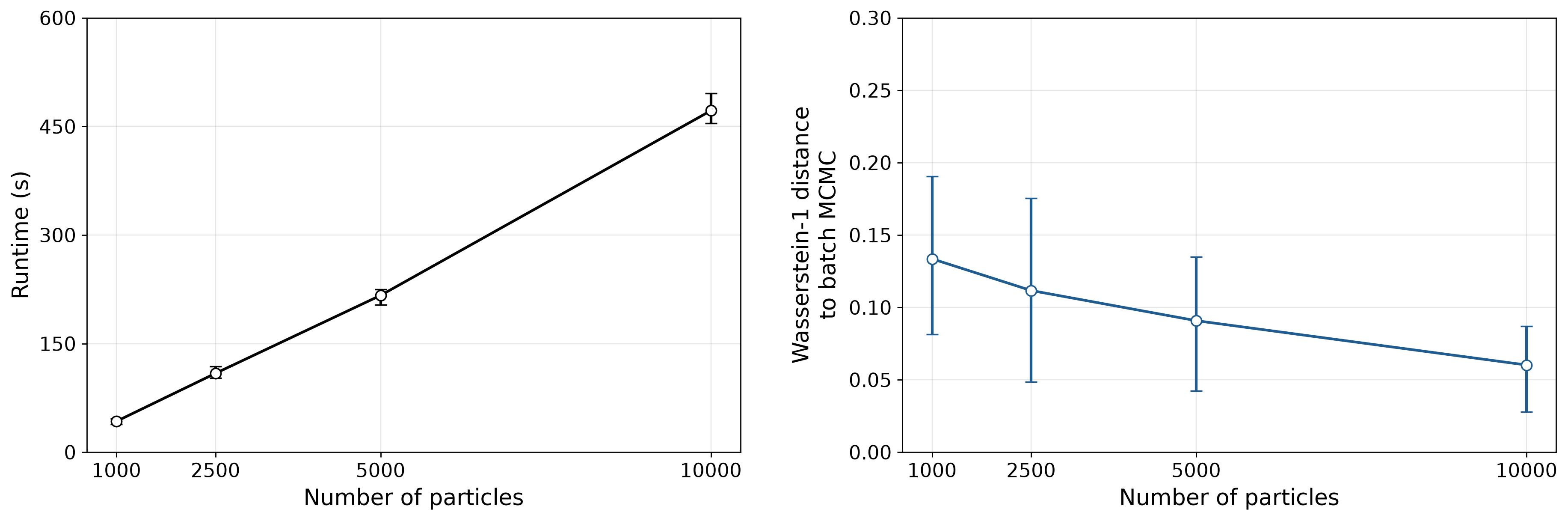}
    \caption{\rev{Runtime and Wasserstein-1 distance from the batch MCMC posterior for different numbers of particles. Points are means across 20 common algorithmic seeds and error bars show the 10th--90th percentiles across seeds.}}
    \label{fig:case_M_tradeoff}
\end{figure}

\begin{revision}
\Cref{fig:case_M_sensitiv,fig:case_M_tradeoff} summarise sensitivity to the number of particles. Consistent with the simulation results, increasing $M$ generally improves the agreement with batch MCMC, but at a clear computational cost. The mean Wasserstein-1 distance from the batch MCMC posterior decreases from 0.134 for $M=1{,}000$ to 0.112 for $M=2{,}500$, 0.091 for $M=5{,}000$, and 0.060 for $M=10{,}000$. At the same time, mean runtime increases from 42.60~s to 108.89~s, 216.61~s, and 471.76~s, respectively. Mean fidelity improves at each increase in $M$, although the 10th--90th percentile ranges overlap across the examined settings. These results highlight the importance of assessing sensitivity to the number of particles for the dataset at hand rather than selecting $M$ on the basis of sample size alone.
\end{revision}

\newpage
\begin{revision}
We also examine how the order of observation arrival affects sequential posterior learning. At intermediate observation indices, different orderings generally involve different subsets of the data and therefore different partial-data posterior targets. Once all observations have been processed, however, the posterior target is the same across orderings. We compare the original chronological order with three shuffled orders, using the baseline number of particles $M=5{,}000$. Each order is evaluated using the same 20 algorithmic seeds.

\Cref{fig:case_orderA_sensitiv} shows the seed-averaged final posterior densities, with 10th--90th percentile bands across seeds. The mean Wasserstein-1 distances to batch MCMC are 0.091, 0.058, 0.045, and 0.045 for the original order and Shuffles A--C, respectively. The final posterior densities are broadly similar in shape across these orders. In contrast, \Cref{fig:case_orderB_sensitiv} shows more marked differences in how the posterior density evolves as observations accumulate. \Cref{fig:case_orderC_sensitiv} likewise shows different intermediate trajectories for the posterior mean number of occupied clusters, although the trajectories end at similar values. Taken together, \Cref{fig:case_orderA_sensitiv,fig:case_orderB_sensitiv,fig:case_orderC_sensitiv} indicate that, in this experiment, intermediate posterior shapes and cluster-count trajectories are more sensitive to the order in which trip records arrive than the final coefficient posterior.
\end{revision}

\begin{revision}
\begin{figure}[htbp]
    \centering
    \includegraphics[width=0.9\textwidth]{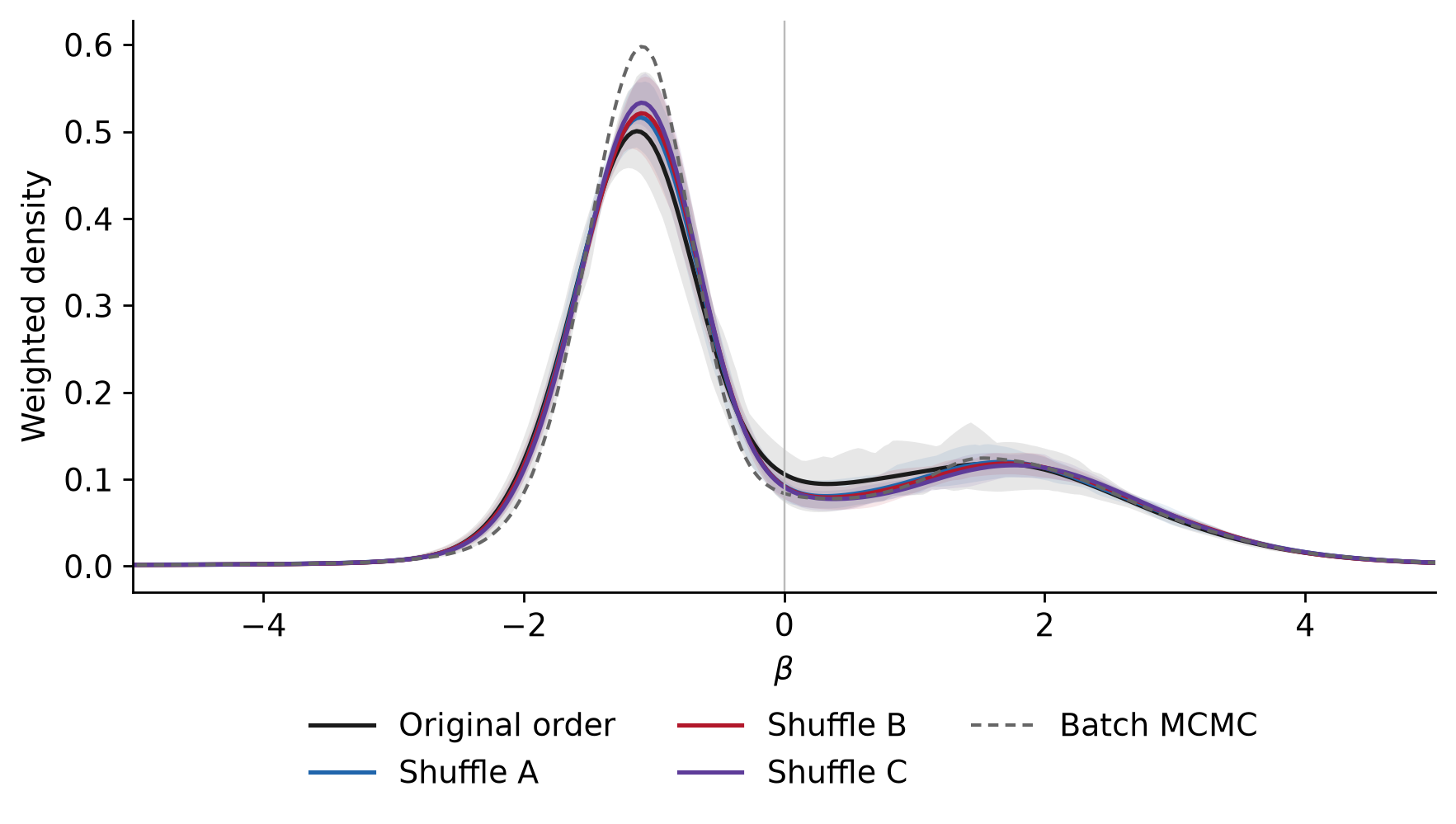}
    \caption{\rev{Sensitivity of the final posterior density of $\beta$ to data ordering. Solid lines are means across 20 common algorithmic seeds, shaded bands show the 10th--90th percentiles across seeds, and the dashed line is the batch MCMC reference.}}
    \label{fig:case_orderA_sensitiv}
\end{figure}
\end{revision}

\begin{revision}
\begin{figure}[htbp]
    \centering
    \includegraphics[width=1.0\textwidth]{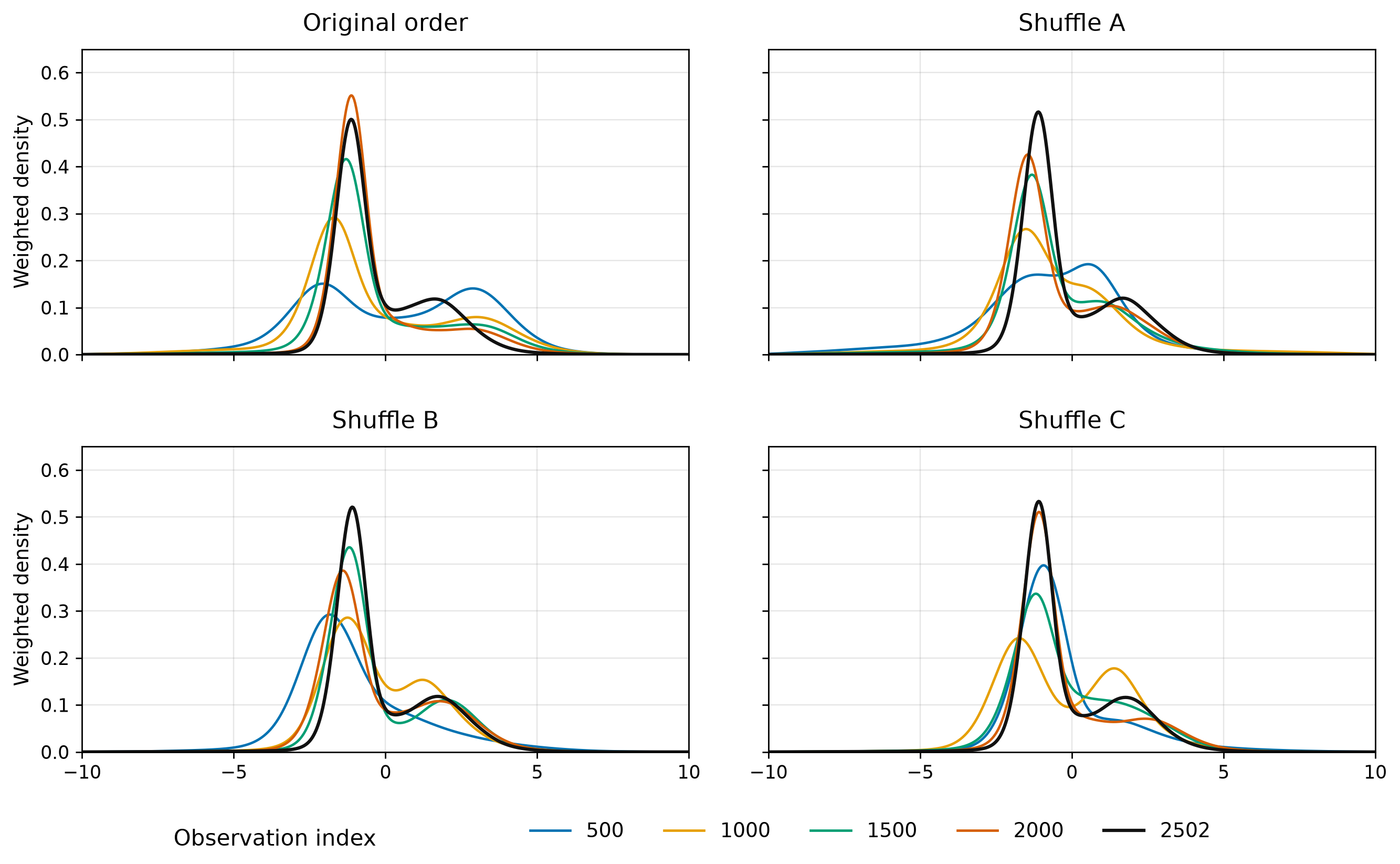}
    \caption{\rev{Evolution of the seed-averaged fully updated posterior density of $\beta$ under different data orderings. Each curve is averaged across 20 common algorithmic seeds.}}
    \label{fig:case_orderB_sensitiv}
\end{figure}
\end{revision}

\begin{revision}
\begin{figure}[htbp]
    \centering
    \includegraphics[width=1.0\textwidth]{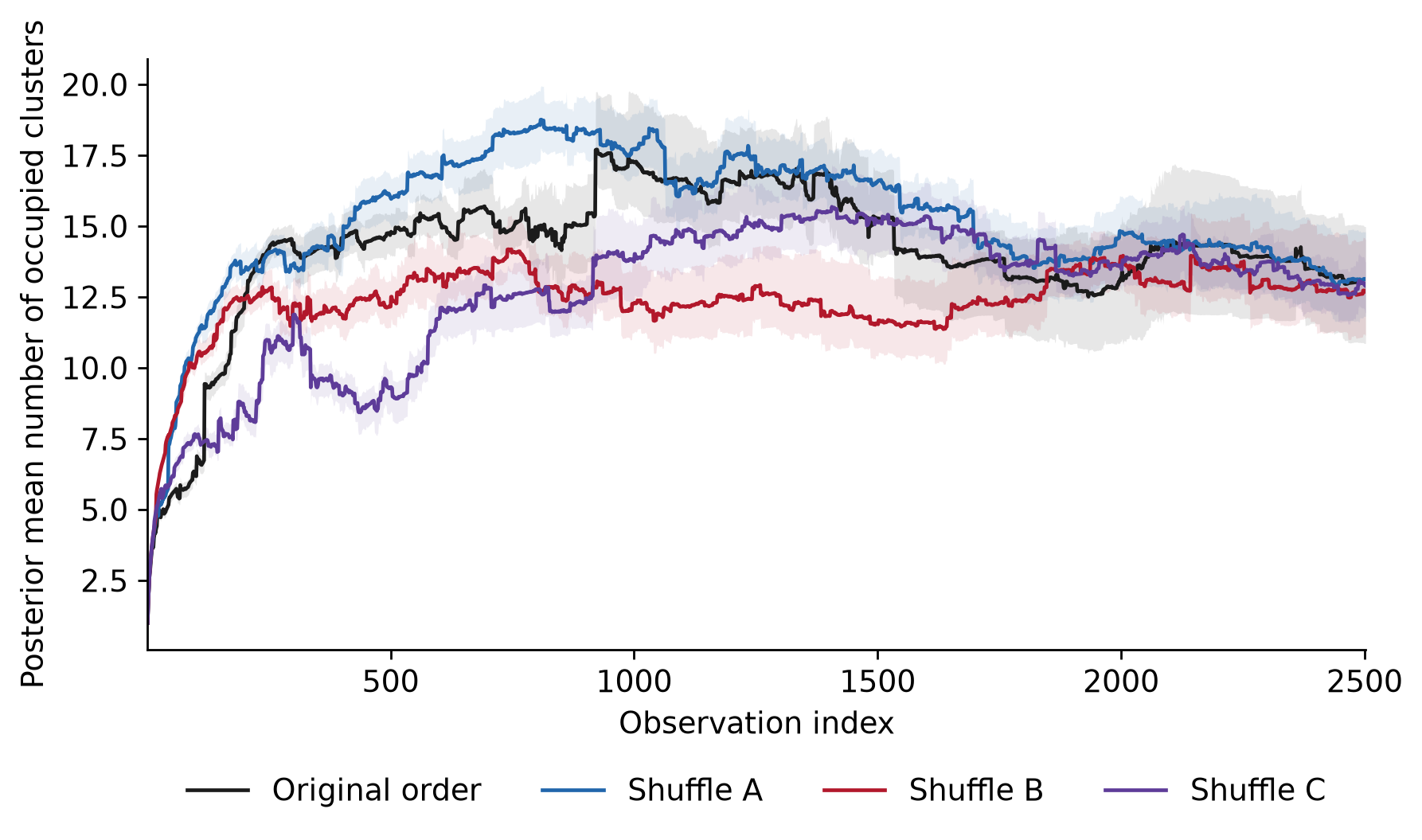}
    \caption{\rev{Evolution of the posterior mean number of occupied clusters under different data orderings. Lines are means across 20 common algorithmic seeds and shaded bands show the 10th--90th percentiles across seeds.}}
    \label{fig:case_orderC_sensitiv}
\end{figure}
\end{revision}

\FloatBarrier

\begin{revision}
We finally assess sensitivity to the prior scale of the cluster-specific coefficient $\beta$. The baseline Gaussian base measure has standard deviation $\sqrt{10}$; alternatives with standard deviations 1 and 10 are evaluated using the same 20 algorithmic seeds, with all remaining settings held fixed. \Cref{fig:beta_pri_den} shows that negative and positive modes remain visible under all three specifications, although their relative mass and spread differ. Under the broadest prior, the negative peak is lower and the tails are more extended. \Cref{fig:beta_pri_panels} further shows that the early posterior densities are more diffuse under this specification, with the modal structure becoming clearer as observations accumulate. Thus, the prior scale influences how concentrated the coefficient posterior is during the earlier stages of online learning. \Cref{fig:beta_pri_clu} shows that the clustering structure is also sensitive to the prior scale: the final posterior mean numbers of occupied clusters are 16.39, 12.96, and 8.30 for prior standard deviations of 1, $\sqrt{10}$, and 10, respectively. One possible explanation is that a broader base measure allocates more probability to coefficient values that fit the observations poorly, reducing the prior-averaged likelihood of a new cluster relative to existing clusters.
\end{revision}

\begin{revision}
\begin{figure}[htbp]
    \centering
    \includegraphics[width=0.9\textwidth]{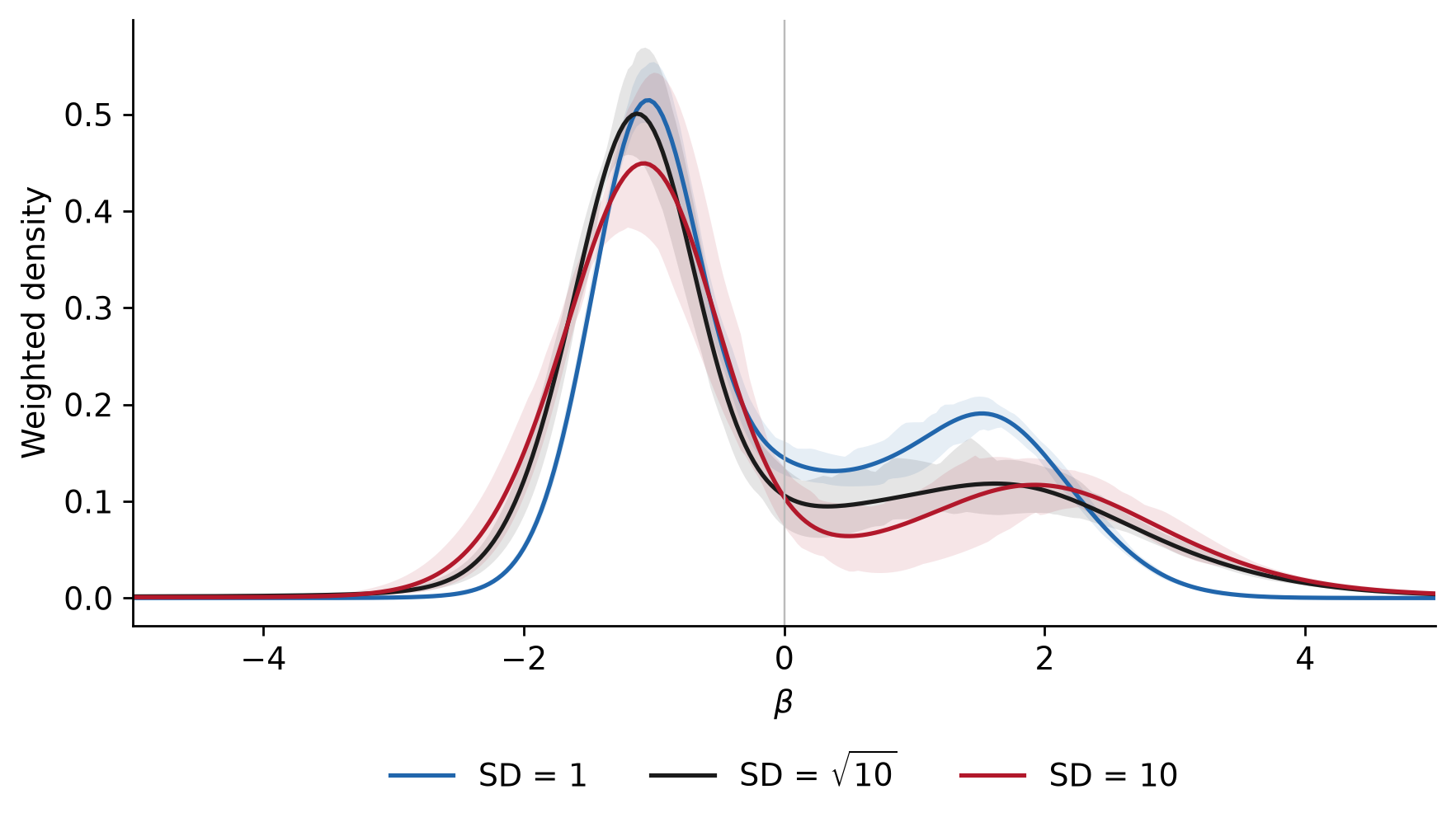}
    \caption{\rev{Sensitivity of the final posterior density of $\beta$ to its prior scale. Lines are means across 20 common algorithmic seeds and shaded bands show the 10th--90th percentiles across seeds.}}
    \label{fig:beta_pri_den}
\end{figure}
\end{revision}

\begin{revision}
\begin{figure}[htbp]
    \centering
    \includegraphics[width=1.0\textwidth]{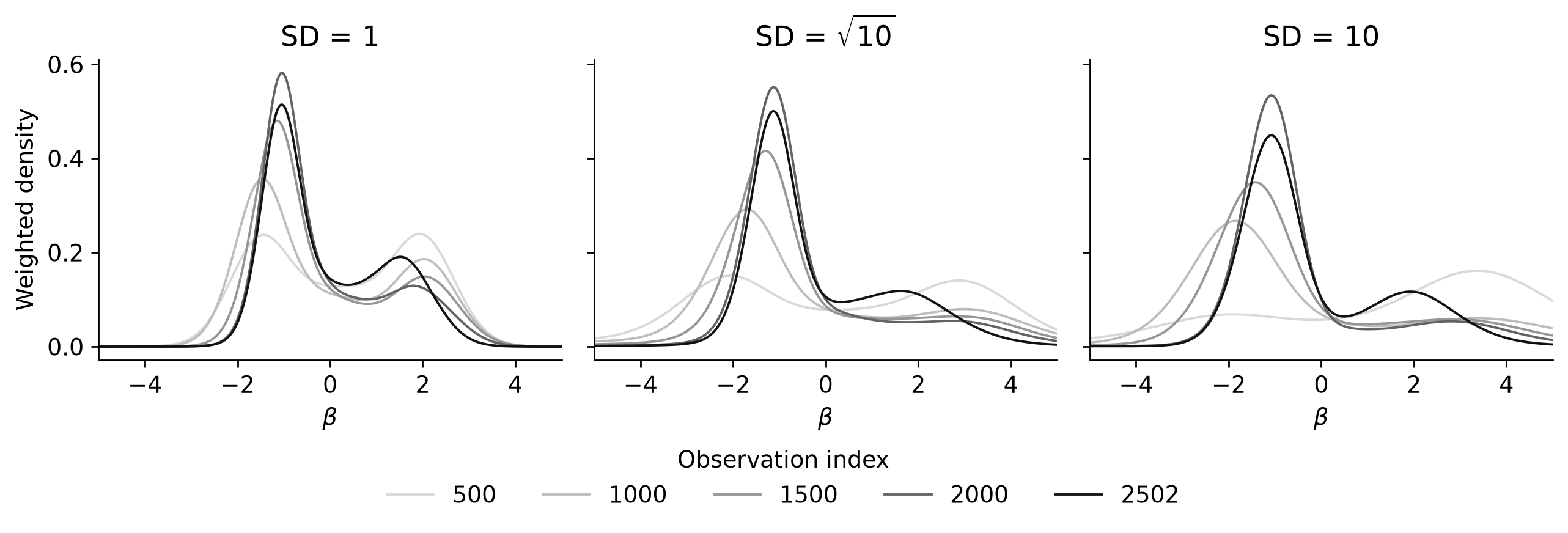}
    \caption{\rev{Evolution of the fully updated posterior density of $\beta$ at selected observation indices under different prior scales. Each curve is averaged across 20 common algorithmic seeds.}}
    \label{fig:beta_pri_panels}
\end{figure}
\end{revision}

\begin{revision}
\begin{figure}[htbp]
    \centering
    \includegraphics[width=1.0\textwidth]{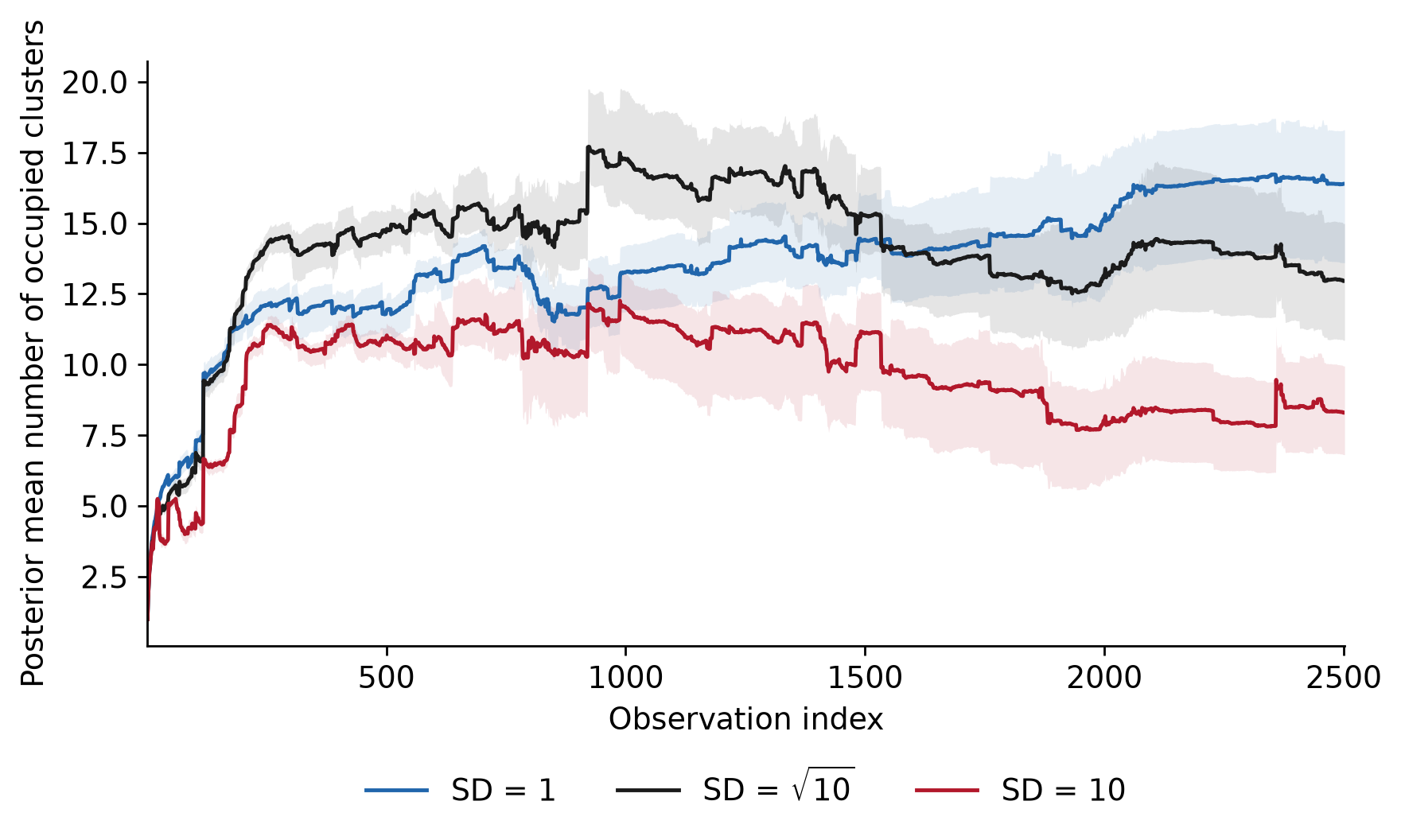}
    \caption{\rev{Posterior mean number of occupied clusters over the sequential observation index under different prior scales for $\beta$. Lines are means across 20 common algorithmic seeds and shaded bands show the 10th--90th percentiles across seeds.}}
    \label{fig:beta_pri_clu}
\end{figure}
\end{revision}

\FloatBarrier

\newpage
\subsubsection{Limitations and future extensions}
\label{sec:case:limit}

\begin{revision}
Several limitations of the illustrative empirical application should be noted. First, the 2,502 trip records consist of repeated observations contributed by only 32 respondents. Given the illustrative purpose of the application and the limited length of the respondent-specific histories, the present analysis pools all trips into a single DP-mixture stream. Conditional on the included covariates and trip-level latent allocations, the specification does not include an additional respondent-level effect or dependence structure for repeated trips. It therefore does not explicitly separate within-person variation from between-person heterogeneity, and the resulting clusters should not be interpreted as persistent personalised traveller types. With substantially longer individual-level streams, a useful future extension would be to conduct person-specific sequential inference for $\beta$, linking posterior learning directly to recurring behavioural patterns within each respondent. Second, the empirical data were processed retrospectively in their observed chronological order rather than being ingested in a live streaming environment. It remains to be examined whether the observed computational performance, posterior stability, and provisional-update latency can be maintained over longer observation periods and under real-time data arrivals.
\end{revision}

The proposed framework also provides a basis for extensions to richer choice structures. Because the proposed algorithm is developed for a binary logit likelihood, a promising extension is a sequential-logit formulation that represents choice among multiple alternatives through a behaviourally specified sequence of conditional binary decisions. The PG-based updates could then be adapted to each stage, although this would require further development of both the behavioural specification and the inference algorithm. A direct extension to a standard multinomial-logit likelihood would require a different augmentation and update design. A separate challenge is to update choice predictions as newly introduced transport modes expand the available choice set, as examined by \citet{chen2026addressing}.

The low-latency observation-specific posterior summaries also create opportunities for future operational extensions. Future applications could use the updated particle approximation to predict subsequent choices, quantify predictive uncertainty, and identify observations that are unlikely under the predictive distribution. Another important direction is to move beyond the present static DP-mixture specification towards dynamic specifications of preference heterogeneity, for example state-space formulations in which latent coefficients evolve over time. Such formulations would require a transition structure and corresponding changes to the sequential propagation and update steps, while the general design principles developed here would remain relevant.

\clearpage
\section{Conclusion}
\label{sec:conclusion}

This study develops an online Bayesian inference algorithm for a Dirichlet process mixture of binary logit models. The algorithm rapidly incorporates newly arriving behavioural observations and makes updated posterior summaries promptly available for subsequent choice predictions. The main methodological challenge is to achieve such sequential updating under a non-conjugate binary logit likelihood while allowing new clusters to be introduced and historical assignments to be revised. Within the sequential Monte Carlo samplers framework, we address this challenge through P\'olya--Gamma data augmentation, local particle-weight corrections, and recursively maintained sufficient statistics. An observation's contribution can be added to, removed from, or replaced in these statistics without reconstructing them from the full observation history. The same computational structure therefore supports both the incorporation of new observations and the revision of historical cluster assignments. The latter is implemented through a rejuvenation step using blocked Gibbs updates that leave the current posterior target unchanged. Under the simulation design examined in this study, sequential updating yields final posterior summaries close to those obtained by batch Markov chain Monte Carlo. Sensitivity analyses further demonstrate trade-offs between posterior approximation accuracy and computational cost as the number of particles and rejuvenation settings are varied. The illustrative empirical application tracks posterior learning about heterogeneous distance sensitivity as trip records are processed sequentially. It also demonstrates low-latency output of observation-specific provisional posterior summaries. Together, these results provide a computational foundation for future extensions to richer choice structures, low-latency decision support, and online anomaly screening.

\begin{appendices}

\counterwithin{equation}{section}
\counterwithin{figure}{section}
\counterwithin{table}{section}
\counterwithin{algorithm}{section}

\renewcommand{\theequation}{\Alph{section}.\arabic{equation}}
\renewcommand{\thefigure}{\Alph{section}\arabic{figure}}
\renewcommand{\thetable}{\Alph{section}\arabic{table}}
\renewcommand{\thealgorithm}{\Alph{section}\arabic{algorithm}}

\newpage
\section{Algorithmic summaries}
\label{app:algorithms}

This appendix summarises the execution order of the proposed sampler and its rejuvenation move. \Cref{alg:complete-smc} gives the complete observation-by-observation workflow, while \Cref{alg:rejuvenation} records the rejuvenation subroutine invoked after resampling in the standard route and as the final step of the stabilising update in \appref{app:tempering}.

In both algorithms, operations within the loops over $m=1,\ldots,M$ can be performed independently across particles. This particle-wise structure offers scope for reducing wall-clock time through parallel processing.

\begin{algorithm}[H]
\caption{PG-augmented SMC for DP-BL}
\label{alg:complete-smc}
\small
\begin{algorithmic}[1]
\State Initialise particles $\{\Theta_0^{(m)},W_0^{(m)}=1/M\}_{m=1}^M$.
\While{a new observation $\D_i$ arrives}
    \For{$m=1$ to $M$}
        \State Sample $\zeta_i^{(m)}$, retain the auxiliary candidates, and store the corresponding $Z_i^{(m)}$.
    \EndFor
    \State Compute the diagnostic ESS using $\{W_{i-1}^{(m)}Z_i^{(m)}\}_{m=1}^M$.
    \If{the diagnostic ESS indicates a sharp drop}
        \State Apply the stabilising update in \appref{app:tempering} and proceed to $\D_{i+1}$.
    \EndIf
    \For{$m=1$ to $M$}
        \State Finalise $\zeta_i^{(m)}$ and update the cluster counts, instantiating a selected new-cluster candidate when required.
        \State Update the current value of $\alpha^{(m)}$.
        \State Sample $\omega_i^{(m)}$.
        \State Update the sufficient statistics of $(\beta^{(m)},\gamma^{(m)})$.
        \State Update the current values of $(\beta^{(m)},\gamma^{(m)})$.
        \State Update the unnormalised weight $\wt W_i^{(m)}$.
    \EndFor
    \State Normalise $\{\wt W_i^{(m)}\}_{m=1}^M$ to obtain $\{W_i^{(m)}\}_{m=1}^M$.
    \State Compute $\ESS_i$ from the normalised weights.
    \If{$\ESS_i<\rho M$}
        \State Resample $\{\Theta_i^{(m)}\}_{m=1}^M$ and reset $W_i^{(m)}=1/M$ for all $m$.
        \State Apply the rejuvenation move in \Cref{alg:rejuvenation}.
    \EndIf
\EndWhile
\end{algorithmic}
\end{algorithm}
\FloatBarrier

\begin{algorithm}[H]
\caption{Rejuvenation move}
\label{alg:rejuvenation}
\small
\begin{algorithmic}[1]
\For{$m = 1$ to $M$}
    \State \rev{Sample $R_i^{(m)}$ uniformly without replacement from $\{1,\ldots,i\}$, with $|R_i^{(m)}|=\lceil\kappa i\rceil$.}
    \For{$t = 1$ to $T_{\mathrm{move}}$}
        \For{$r \in \rev{R_i^{(m)}}$}
            \State If observation $r$ currently occupies a singleton cluster, retain its coefficient.
            \State Remove observation $r$'s contribution, decrement its cluster count, and delete the cluster if it becomes empty.
            \State Construct auxiliary candidates, including the retained singleton coefficient when applicable; sample $\zeta_r^{(m)}$ and instantiate a selected new-cluster candidate when required.
            \State Sample $\omega_r^{(m)}$.
            \State Add observation $r$'s refreshed contribution to the sufficient statistics and increment the assigned cluster count.
        \EndFor
        \State Update the current value of $\alpha^{(m)}$.
        \State Update the current values of $(\beta^{(m)},\gamma^{(m)})$.
    \EndFor
\EndFor
\end{algorithmic}
\end{algorithm}
\FloatBarrier

\newpage
\section{\texorpdfstring{\raggedright Posterior derivation under \mbox{P\'olya--Gamma} data augmentation}{Posterior derivation under P\'olya--Gamma data augmentation}}
\label{app:posterior-derivation}

As shown in Eq.~\eqref{eq:gaussian-kernel}, the binary logit likelihood can be written as a conditionally Gaussian regression model given the auxiliary variables $\omega_i$. Specifically, conditional on $\omega_i$, the contribution of observation $i$ can be expressed in a weighted least-squares form with variance $1/\omega_i$.

For $\beta_h$, conditional on $\gamma$ and $\{\omega_r,\zeta_r\}_{r \le i}$, the model can be viewed as having the pseudo-outcome $\delta_r/\omega_r - z_r\trans \gamma$ with variance $1/\omega_r$. Likewise, for $\gamma$, conditional on $\beta$ and $\{\omega_r,\zeta_r\}_{r \le i}$, the model can be viewed as having the pseudo-outcome $\delta_r/\omega_r - x_r\trans \beta_{\zeta_r}$, with variance $1/\omega_r$. Therefore, by standard Gaussian posterior algebra, the conditional posteriors of $(\beta_h,\gamma)$ can be derived through the following accumulated quantities:
\begin{align}
A_{\beta,h}^{(i)}
&\coloneqq \Lambda_\beta + \sum_{r=1}^i \I\{\zeta_r = h\} \, \omega_r x_r x_r\trans, \\
b_{\beta,h}^{(i)}
&\coloneqq \Lambda_\beta \mu_\beta + \sum_{r=1}^i \I\{\zeta_r = h\} \, \omega_r x_r \left(\frac{\delta_r}{\omega_r} - z_r\trans \gamma\right) \\
&= \Lambda_\beta \mu_\beta + \sum_{r=1}^i \I\{\zeta_r = h\} \, x_r \delta_r - \left(\sum_{r=1}^i \I\{\zeta_r = h\} \, \omega_r x_r z_r\trans\right) \gamma.
\label{eq:pg-beta-accumulation}
\end{align}
Similarly,
\begin{align}
A_\gamma^{(i)}
&\coloneqq \Lambda_\gamma + \sum_{r=1}^i \omega_r z_r z_r\trans, \\
b_\gamma^{(i)}
&\coloneqq \Lambda_\gamma \mu_\gamma + \sum_{r=1}^i \omega_r z_r \left(\frac{\delta_r}{\omega_r} - x_r\trans \beta_{\zeta_r}\right) \\
&= \Lambda_\gamma \mu_\gamma + \sum_{r=1}^i z_r \delta_r - \sum_{r=1}^i (\omega_r z_r x_r\trans) \beta_{\zeta_r}.
\label{eq:pg-gamma-accumulation}
\end{align}
\rev{Substituting the accumulated quantities in Eqs.~\eqref{eq:cluster-aggregates-first}--\eqref{eq:shared-aggregates-last} yields Eqs.~\eqref{eq:beta-posterior}--\eqref{eq:gamma-posterior-moments}.}

\newpage
\section{Incremental weight derivation}
\label{app:weight}
\begingroup
\setlength{\abovedisplayskip}{6pt plus 2pt minus 2pt}
\setlength{\belowdisplayskip}{6pt plus 2pt minus 2pt}
\setlength{\belowdisplayshortskip}{4pt plus 2pt minus 2pt}

This appendix derives the incremental-weight update used in \Cref{sec:method:weight}. In the general SMC samplers framework, the incremental weight is given by the standard importance-sampling ratio on a joint space formed by the previous state and the proposed state, using a forward kernel and an auxiliary backward kernel \citep{DelMoralDoucetJasra2006,ChopinPapaspiliopoulos2020,DaiHengJacobWhiteley2022}. In the present algorithm, we apply this construction only to the cluster-allocation step, because the subsequent updates of $\omega_i$ and $(\beta,\gamma,\alpha)$ are exact full-conditional Gibbs draws under the stage-$i$ target distribution and therefore contribute unit incremental weight. Let
\[
\psi_i \coloneqq \bigl(\zeta_i,\wt{\beta}_{\mathrm{new}}^{(1:J)}\bigr)
\]
denote the variables introduced in the cluster-allocation step at stage $i$. Conditional on the previous particle state $\Theta_{i-1}$ and the new observation $\D_i$, the forward proposal for this step is
\begin{equation}
f_i(\Theta_{i-1},\psi_i)
=
\left[\prod_{j=1}^J G_0\bigl(\wt{\beta}_{\mathrm{new}}^{(j)}\bigr)\right]
\, p\bigl(\zeta_i \given \Theta_{i-1}, \wt{\beta}_{\mathrm{new}}^{(1:J)}, \D_i\bigr),
\label{eq:weight-forward-proposal}
\end{equation}
where the proposal probability on the right-hand side is the discrete assignment probability defined in Eq.~\eqref{eq:alloc-normalised}. To connect this step to the standard SMC samplers formula, consider the extended state $(\Theta_{i-1},\psi_i)$ and choose the backward kernel as its deterministic projection onto $\Theta_{i-1}$. The forward retention and backward projection then contain matching Dirac factors on the ancestor coordinate, which cancel in the joint-space importance ratio \citep{DelMoralDoucetJasra2006,ulker2010sequential}. After cancelling the factors associated with the exact full-conditional Gibbs updates of $\omega_i$ and $(\beta,\gamma,\alpha)$, the remaining non-cancelling part of the stage-$i$ unnormalised target is
\begin{equation*}
\wt{\pi}_i(\Theta_{i-1},\psi_i)
\propto
\wt{\pi}_{i-1}(\Theta_{i-1})
\left[\prod_{j=1}^J G_0\bigl(\wt{\beta}_{\mathrm{new}}^{(j)}\bigr)\right]
p(\zeta_i \given \zeta_{1:i-1},\alpha) \, L_i(\zeta_i),
\end{equation*}
where $p(\zeta_i \given \zeta_{1:i-1},\alpha)$ and $L_i(\zeta_i)$ are defined in \Cref{sec:method:alloc}. Integrating out the unused auxiliary candidates and summing over the auxiliary new-cluster labels recovers the original DP-mixture allocation target, with any selected candidate retained as the coefficient of the new cluster \citep{Neal2000}. Therefore, after cancellation of the matching Dirac factors, the local incremental weight for the cluster-allocation step is
\begin{equation}
w_i = \frac{\wt{\pi}_i(\Theta_{i-1},\psi_i)}{\wt{\pi}_{i-1}(\Theta_{i-1}) \, f_i(\Theta_{i-1},\psi_i)}.
\label{eq:weight-local-ratio}
\end{equation}
Substituting the target and proposal densities gives
\begin{equation*}
w_i
=
\frac{\left[\prod_{j=1}^J G_0\bigl(\wt{\beta}_{\mathrm{new}}^{(j)}\bigr)\right] p(\zeta_i \given \zeta_{1:i-1},\alpha) \, L_i(\zeta_i)}{\left[\prod_{j=1}^J G_0\bigl(\wt{\beta}_{\mathrm{new}}^{(j)}\bigr)\right] p\bigl(\zeta_i \given \Theta_{i-1},\wt{\beta}_{\mathrm{new}}^{(1:J)},\D_i\bigr)}.
\end{equation*}
Using Eq.~\eqref{eq:alloc-normalised}, $w_i = Z_i$. Thus, the stage-local unnormalised weight update in \Cref{sec:method:weight} is $\wt{W}_i^{(m)} = W_{i-1}^{(m)} \cdot Z_i^{(m)}$.
\endgroup

\newpage
\section{Stabilising update via temperature tempering}
\label{app:tempering}

The stabilising update used in \Cref{alg:complete-smc} is implemented as a within-step temperature tempering scheme for the incremental weight update in \Cref{sec:method:weight} \citep{jasra2011inference,ChopinPapaspiliopoulos2020,DaiHengJacobWhiteley2022}. After the particle-specific allocation normalising constants $Z_i^{(m)}$ have been computed for all particles in \Cref{alg:complete-smc}, this update is invoked when the diagnostic ESS indicates that the standard one-step weight update in Eq.~\eqref{eq:weight-update} would lead to severe weight degeneracy. It then replaces the one-step weight-update part of \Cref{alg:complete-smc}, namely the update of $\wt{W}_i^{(m)}$, weight normalisation, and the subsequent ESS-triggered resampling branch for observation $\D_i$. Within this stabilising route, the algorithm then performs the same remaining particle-state updates as in \Cref{alg:complete-smc}: updating $\alpha^{(m)}$, sampling $\omega_i^{(m)}$, and updating the sufficient statistics and current values of $(\beta^{(m)},\gamma^{(m)})$ before applying the standard rejuvenation move. The stabilising update changes only the computational path used to incorporate observation $\D_i$ and does not change the stage-$i$ posterior target.

The core of the stabilising update is to replace the one-step multiplication by the particle-specific incremental weight $Z_i^{(m)}$ with a sequence of tempered reweighting steps. A sharp drop in ESS indicates that the new observation $\D_i$ carries strong information about which particles are more consistent with the updated posterior target; however, incorporating this information in a single step may concentrate almost all weight on a small number of particles. We therefore introduce a finite temperature ladder and incorporate the corresponding $Z_i^{(m)}$ values gradually, using the conditional effective sample size (CESS) criterion to control the intermediate temperatures.

At each intermediate temperature, we also refresh the realised allocation label $\zeta_i$ by a lightweight Gibbs update. This allocation-label refresh \rev{allows resampled copies to acquire different current-observation allocations} and helps maintain allocation uncertainty for $\D_i$ along the tempered path. To describe both the tempered reweighting and the allocation-label refresh, we \rev{define the temporary allocation-stage particle state as $\Theta_i^\dagger=(\Theta_{i-1},\psi_i)$}, with $\psi_i=(\zeta_i,\wt{\beta}_{\mathrm{new}}^{(1:J)})$. This state contains the previous particle state, the realised allocation label for $\D_i$, and the auxiliary new-cluster candidates.

In \Cref{alg:complete-smc}, $Z_i$ is stored as a scalar after the allocation probabilities for observation $\D_i$ have been computed. To refer to the same allocation normalising constant while the temporary allocation-stage state is being updated, we \rev{write $Z_i(\Theta_i^\dagger)$ for the value computed from the previous particle state and the auxiliary new-cluster candidates contained in $\Theta_i^\dagger$}. Specifically, for a given allocation-stage state $\rev{\Theta_i^\dagger} = (\Theta_{i-1},\psi_i)$,
\begin{equation}
Z_i(\rev{\Theta_i^\dagger})
\coloneqq
\sum_{h \in \Hset_i(\rev{\Theta_i^\dagger})} p(\zeta_i = h \given \zeta_{1:i-1},\alpha) \, L_i(h),
\label{eq:tempering-normaliser}
\end{equation}
where $\Hset_i(\rev{\Theta_i^\dagger})$ is the candidate label set at stage $i$, and $L_i(h)$ is the likelihood contribution defined in \Cref{sec:method:alloc}, evaluated using the coefficients and auxiliary candidates contained in $\rev{\Theta_i^\dagger}$. Evaluating $Z_i(\rev{\Theta_i^\dagger})$ for the current temporary state gives the scalar $Z_i$ that would be stored in the main algorithm for that state.

Let $\wt{\pi}_i(\rev{\Theta_i^\dagger})$ denote the unnormalised target over this temporary allocation-stage state after observation $\D_i$ has been fully introduced. Since the incremental weight is exactly $Z_i(\rev{\Theta_i^\dagger})$ as shown in \appref{app:weight}, we define a continuum of tempered intermediate targets indexed by $\phi \in [0,1]$ as
\begin{equation}
\wt{\pi}_{i,\phi}(\rev{\Theta_i^\dagger})
\propto
\wt{\pi}_i(\rev{\Theta_i^\dagger})\, \bigl\{Z_i(\rev{\Theta_i^\dagger})\bigr\}^{\phi-1}.
\label{eq:tempered-target}
\end{equation}
Under this convention, $\phi=0$ represents the proposal-side state after the auxiliary candidate block and the initial allocation label have been sampled, but before the incremental-weight term $Z_i(\rev{\Theta_i^\dagger})$ has been applied to the particle weight. By contrast, $\phi=1$ represents the allocation-stage target after this term has been fully incorporated, so that $\wt{\pi}_{i,1}(\rev{\Theta_i^\dagger}) = \wt{\pi}_i(\rev{\Theta_i^\dagger})$. Moreover, for any $0 \le \phi_{t-1} < \phi_t \le 1$,
\begin{equation*}
\frac{\wt{\pi}_{i,\phi_t}(\rev{\Theta_i^\dagger})}{\wt{\pi}_{i,\phi_{t-1}}(\rev{\Theta_i^\dagger})} = \bigl\{Z_i(\rev{\Theta_i^\dagger})\bigr\}^{\phi_t - \phi_{t-1}}.
\end{equation*}
Choose a finite ladder $0 = \phi_0 < \phi_1 < \cdots < \phi_T = 1$. The ESS-based quantities used here have distinct roles. The diagnostic ESS in \Cref{alg:complete-smc} decides only whether this stabilising update should be invoked. Within the stabilising update, CESS is used to choose the next temperature, whereas the standard ESS defined in Eq.~\eqref{eq:ess} is computed after each actual reweighting step to decide whether resampling is required. At intermediate step $t$, suppose the current particles are $\{\rev{\Theta_{i,t}^{\dagger(m)}}\}_{m=1}^M$, with current normalised weights $\{W_{i,t}^{(m)}\}_{m=1}^M$. For any candidate $\phi \in (\phi_t,1]$, define the incremental factor
\begin{equation*}
g^{(m)}(\phi) = \bigl\{ Z_i\bigl(\rev{\Theta_{i,t}^{\dagger(m)}}\bigr) \bigr\}^{\phi - \phi_t},
\qquad m = 1,\dots,M.
\end{equation*}
We choose the next temperature $\phi_{t+1}$ adaptively using the CESS criterion \citep{ZhouJohansenAston2016}. Specifically, define
\begin{equation}
\CESS_t(\phi)
\coloneqq
M\,\frac{\left(\sum_{m=1}^M W_{i,t}^{(m)} g^{(m)}(\phi)\right)^2}{\sum_{m=1}^M W_{i,t}^{(m)} \bigl(g^{(m)}(\phi)\bigr)^2}.
\label{eq:tempering-cess}
\end{equation}
Given a pre-specified threshold $\mu M$ (e.g., $\mu = 0.9$), we choose $\phi_{t+1}$ as the largest value in $(\phi_t,1]$ such that
\begin{equation*}
\CESS_t(\phi_{t+1}) \ge \mu M.
\end{equation*}
In practice, $\phi_{t+1}$ can be found by the bisection method. Given $\phi_{t+1}$, the reweighting step is
\begin{equation*}
\wt{W}_{i,t+1}^{(m)} = W_{i,t}^{(m)} g^{(m)}(\phi_{t+1}),
\qquad m = 1,\dots,M,
\end{equation*}
followed by normalisation
\begin{equation*}
W_{i,t+1}^{(m)} = \frac{\wt{W}_{i,t+1}^{(m)}}{\sum_{\ell=1}^M \wt{W}_{i,t+1}^{(\ell)}}.
\end{equation*}
The standard ESS defined in Eq.~\eqref{eq:ess}, computed from the normalised weights after the tempered reweighting step, is then
\begin{equation*}
\ESS_{i,t+1} \coloneqq \frac{1}{\sum_{m=1}^M \bigl(W_{i,t+1}^{(m)}\bigr)^2}.
\end{equation*}
If $\ESS_{i,t+1}$ falls below a pre-specified resampling threshold $\rho M$ (e.g., $\rho = 0.5$), we resample the temporary particle states and their associated sufficient statistics and cluster counts according to $\{W_{i,t+1}^{(m)}\}_{m=1}^M$, and reset the weights to $W_{i,t+1}^{(m)} = 1/M$.

At every intermediate temperature $\phi_{t+1}$, the allocation-label refresh holds $\Theta_{i-1}$ and the auxiliary candidate block $\wt{\beta}_{\mathrm{new}}^{(1:J)}$ fixed and resamples only the realised label $\zeta_i$ from Eq.~\eqref{eq:alloc-normalised}. \rev{As established after Eq.~\eqref{eq:Zi}, this is an exact Gibbs update under the tempered target because the allocation normalising constant sums over all candidate labels and, conditional on the fixed candidate block, is unchanged by the realised draw of $\zeta_i$.}

Once $\phi_T = 1$, the temperature-tempering sequence is complete. The resulting allocation is treated as the final allocation for observation $\D_i$ within each particle. If the final allocation corresponds to an auxiliary new-cluster candidate, the selected candidate is instantiated as an occupied cluster using the same rule as in \Cref{sec:method:alloc}. We then complete the remaining within-particle updates for observation $\D_i$, namely, the updates of $\alpha$, $\omega_i$, sufficient statistics, and the current values of $(\beta,\gamma)$. Subsequently, we apply the standard rejuvenation move (\Cref{alg:rejuvenation}) and return to the complete algorithm. Because the tempered sub-updates have already replaced the one-step reweighting and any resampling required while incorporating observation $\D_i$, the one-step reweighting step in Eq.~\eqref{eq:weight-update} and the subsequent ESS check in \Cref{alg:complete-smc} are not repeated. After this update is completed, the algorithm proceeds to observation $\D_{i+1}$. The overall procedure is summarised in \Cref{alg:tempering}.

\begin{algorithm}[H]
\caption{Stabilising update via temperature tempering}
\label{alg:tempering}
\begin{algorithmic}[1]
\State Initialise: set $t = 0$, $\phi_0 = 0$, $\rev{\Theta_{i,0}^{\dagger(m)}} \coloneqq (\Theta_{i-1}^{(m)},\psi_i^{(m)})$, and $W_{i,0}^{(m)} \coloneqq W_{i-1}^{(m)}$ for $m = 1,\dots,M$.
\While{$\phi_t < 1$}
    \State For any candidate $\phi \in (\phi_t,1]$, compute $g^{(m)}(\phi)$ and $\CESS_t(\phi)$.
    \State Choose $\phi_{t+1}$ as the largest value such that $\CESS_t(\phi_{t+1}) \ge \mu M$.
    \State Reweight $\wt{W}_{i,t+1}^{(m)}$ and normalise $W_{i,t+1}^{(m)}$.
    \State Compute $\ESS_{i,t+1}$.
    \If{$\ESS_{i,t+1} < \rho M$}
        \State Resample the temporary particle states and reset the weights to $W_{i,t+1}^{(m)} = 1/M$.
    \EndIf
    \State With $\Theta_{i-1}^{(m)}$ and $\wt{\beta}_{\mathrm{new}}^{(1:J,m)}$ fixed, resample $\zeta_i^{(m)}$ from Eq.~\eqref{eq:alloc-normalised}.
    \State Set $t \leftarrow t+1$.
\EndWhile
\For{$m = 1$ to $M$}
    \State Finalise $\zeta_i^{(m)}$ and update the cluster counts, instantiating a selected new-cluster candidate as in \Cref{sec:method:alloc} when required.
    \State $W_i^{(m)} \coloneqq W_{i,T}^{(m)}$.
    \State Update the current value of $\alpha^{(m)}$.
    \State Sample $\omega_i^{(m)}$.
    \State Update the sufficient statistics and current values of $(\beta^{(m)},\gamma^{(m)})$.
\EndFor
\State Rejuvenation move (\Cref{alg:rejuvenation}).
\State Return to \Cref{alg:complete-smc} and continue with observation $\D_{i+1}$.
\end{algorithmic}
\end{algorithm}

\newpage
\section{Marginal posterior comparisons}
\label{app:marginal-comparisons}

\begin{figure}[htbp]
    \centering
    \includegraphics[width=0.9\textwidth]{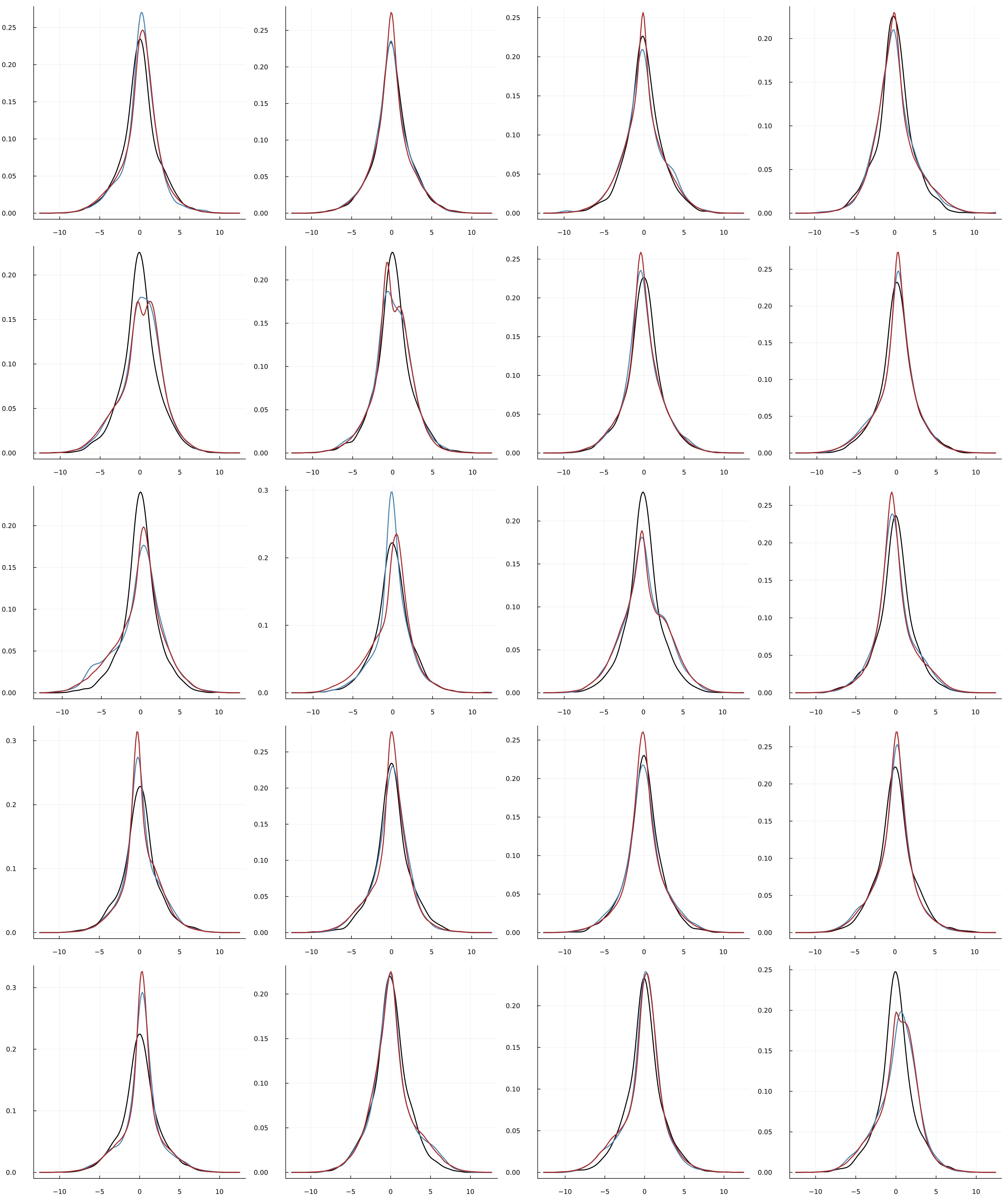}
    \caption{Replication-level comparison of marginal posterior densities for $\beta_1$ (intercept). Black: generated $\beta_1$; blue: baseline SMC; red: batch MCMC.}
    \label{fig:beta_const}
\end{figure}

\begin{figure}[htbp]
    \centering
    \includegraphics[width=0.9\textwidth]{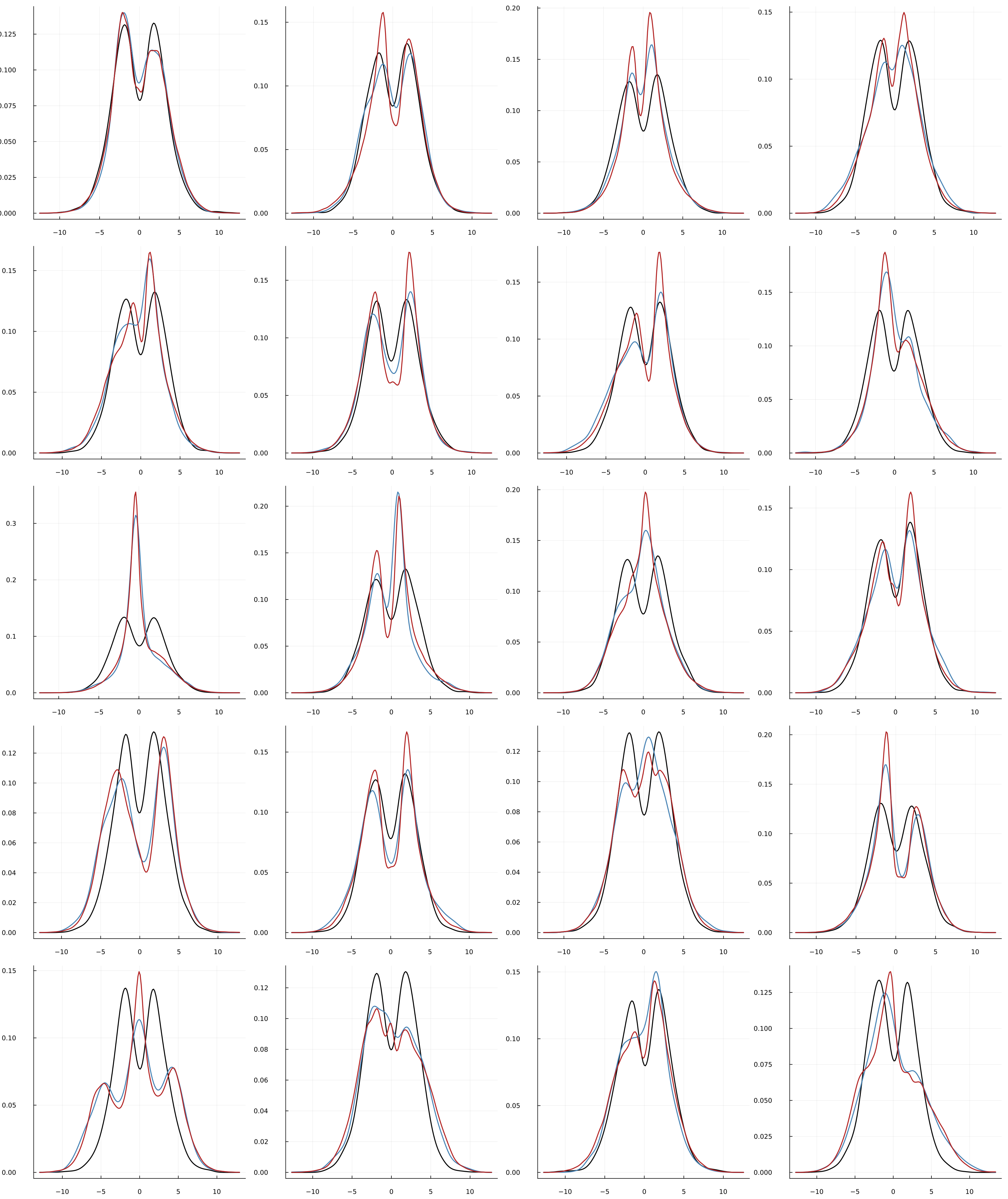}
    \caption{Replication-level comparison of marginal posterior densities for $\beta_2$. Black: generated $\beta_2$; blue: baseline SMC; red: batch MCMC.}
    \label{fig:beta_slope}
\end{figure}

\begin{figure}[htbp]
    \centering
    \includegraphics[width=1.0\textwidth]{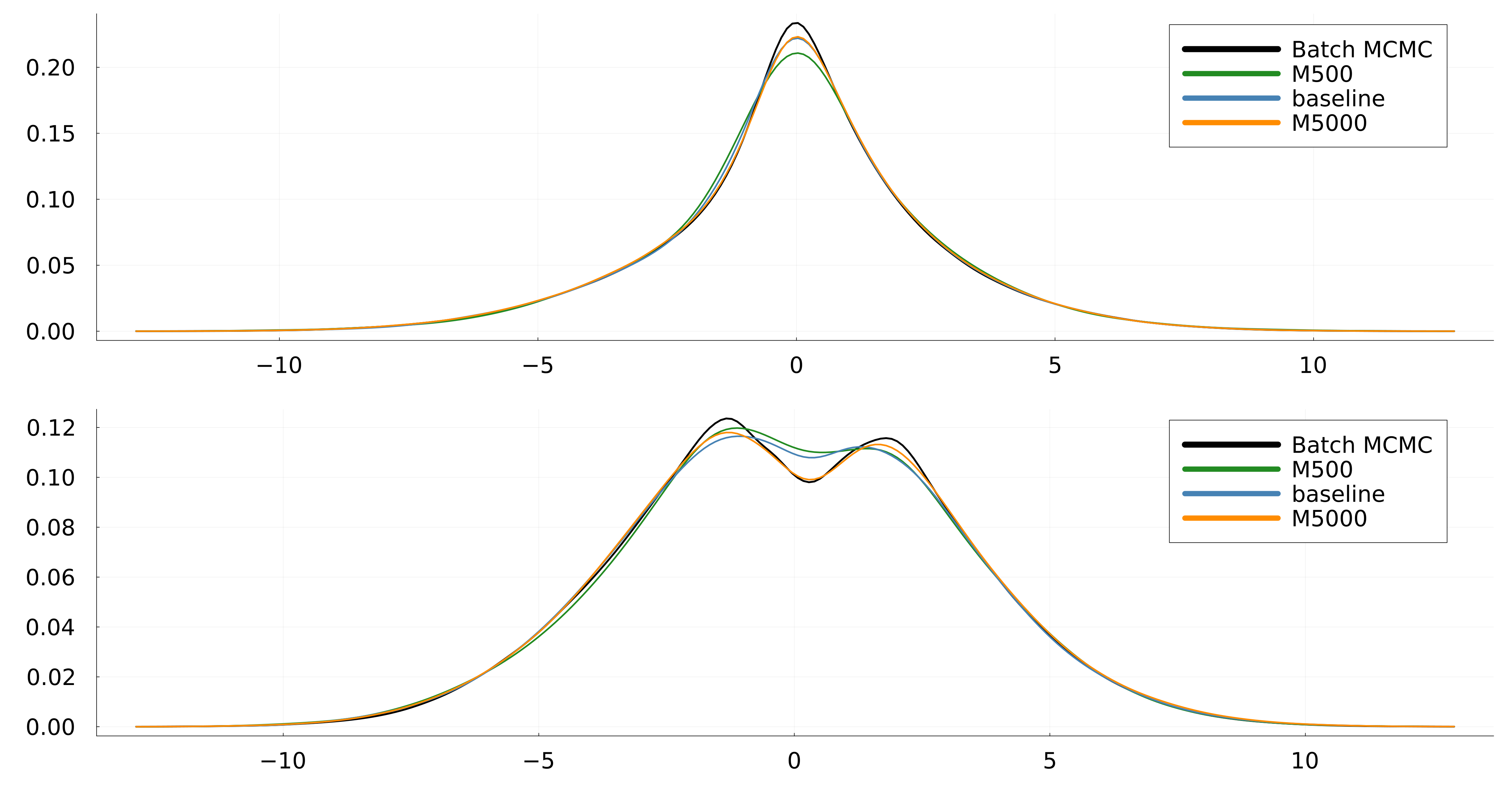}
    \caption{\rev{Effect of $M$ on posterior densities: the proposed SMC algorithm versus batch MCMC. Upper: $\beta_1$ (intercept). Lower: $\beta_2$.}}
    \label{fig:M_effect}
\end{figure}

\begin{figure}[htbp]
    \centering
    \includegraphics[width=1.0\textwidth]{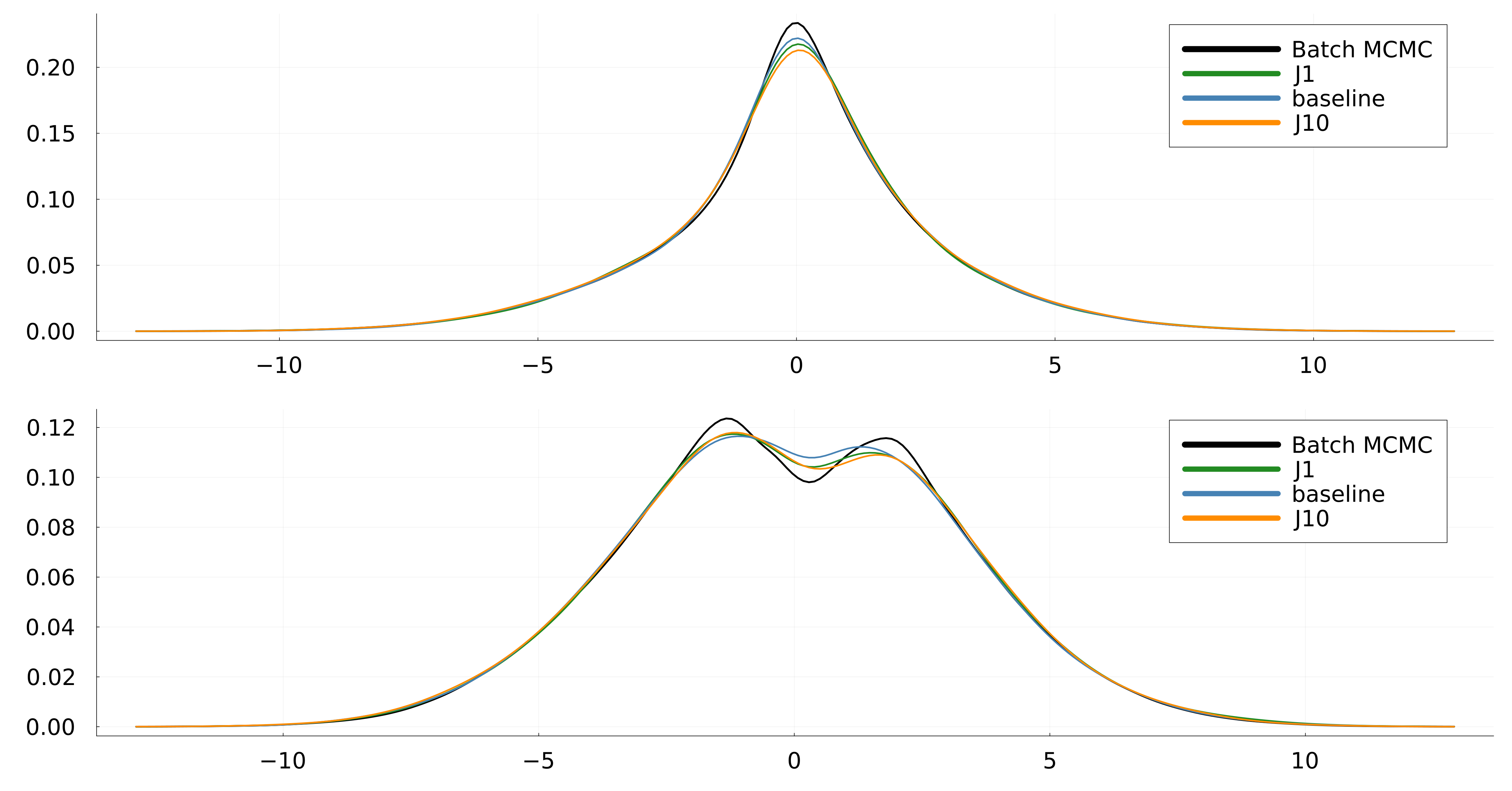}
    \caption{\rev{Effect of $J$ on posterior densities: the proposed SMC algorithm versus batch MCMC. Upper: $\beta_1$ (intercept). Lower: $\beta_2$.}}
    \label{fig:J_effect}
\end{figure}

\begin{figure}[htbp]
    \centering
    \includegraphics[width=1.0\textwidth]{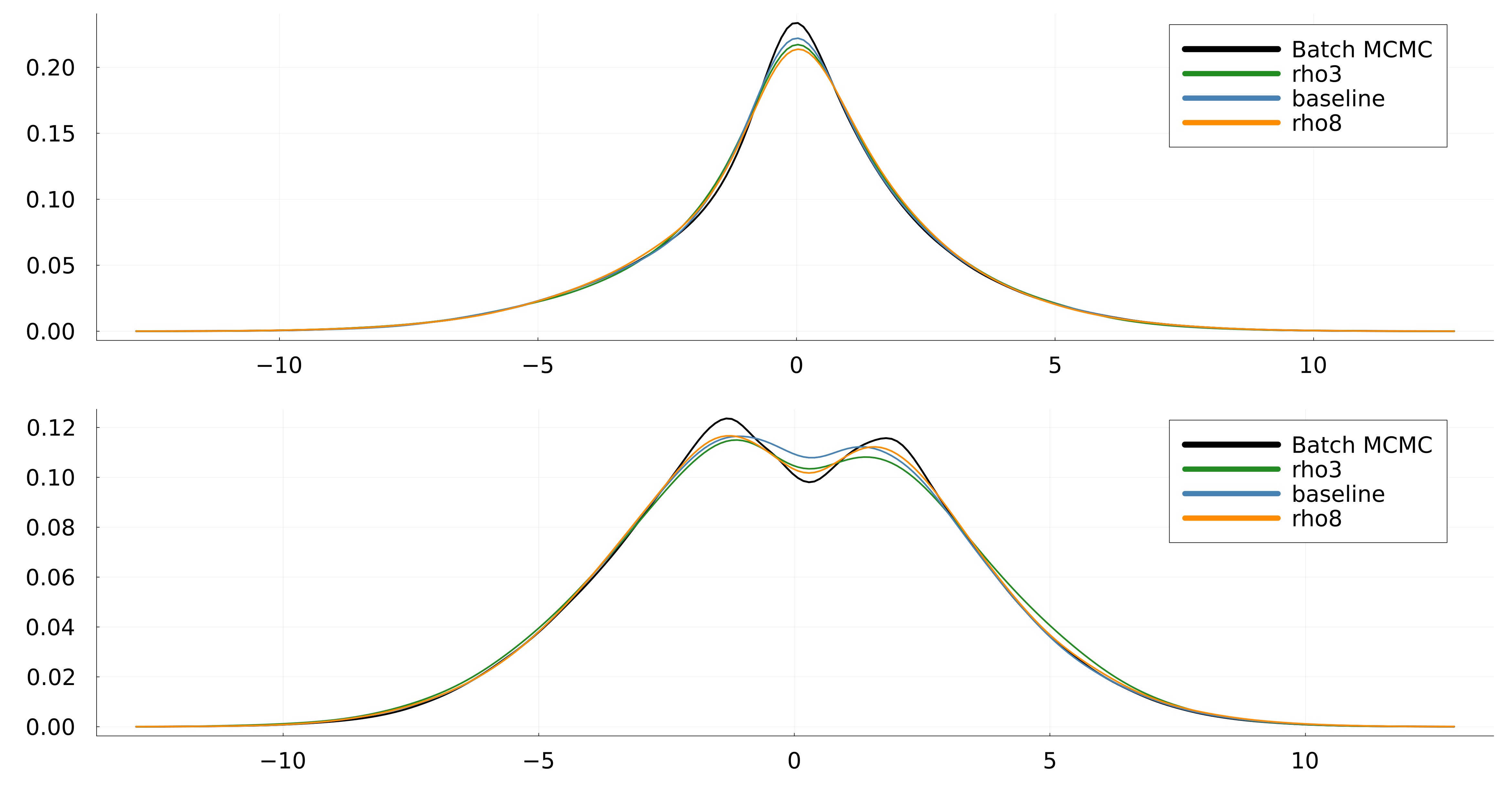}
    \caption{\rev{Effect of $\rho$ on posterior densities: the proposed SMC algorithm versus batch MCMC. Upper: $\beta_1$ (intercept). Lower: $\beta_2$.}}
    \label{fig:rho_effect}
\end{figure}

\begin{figure}[htbp]
    \centering
    \includegraphics[width=1.0\textwidth]{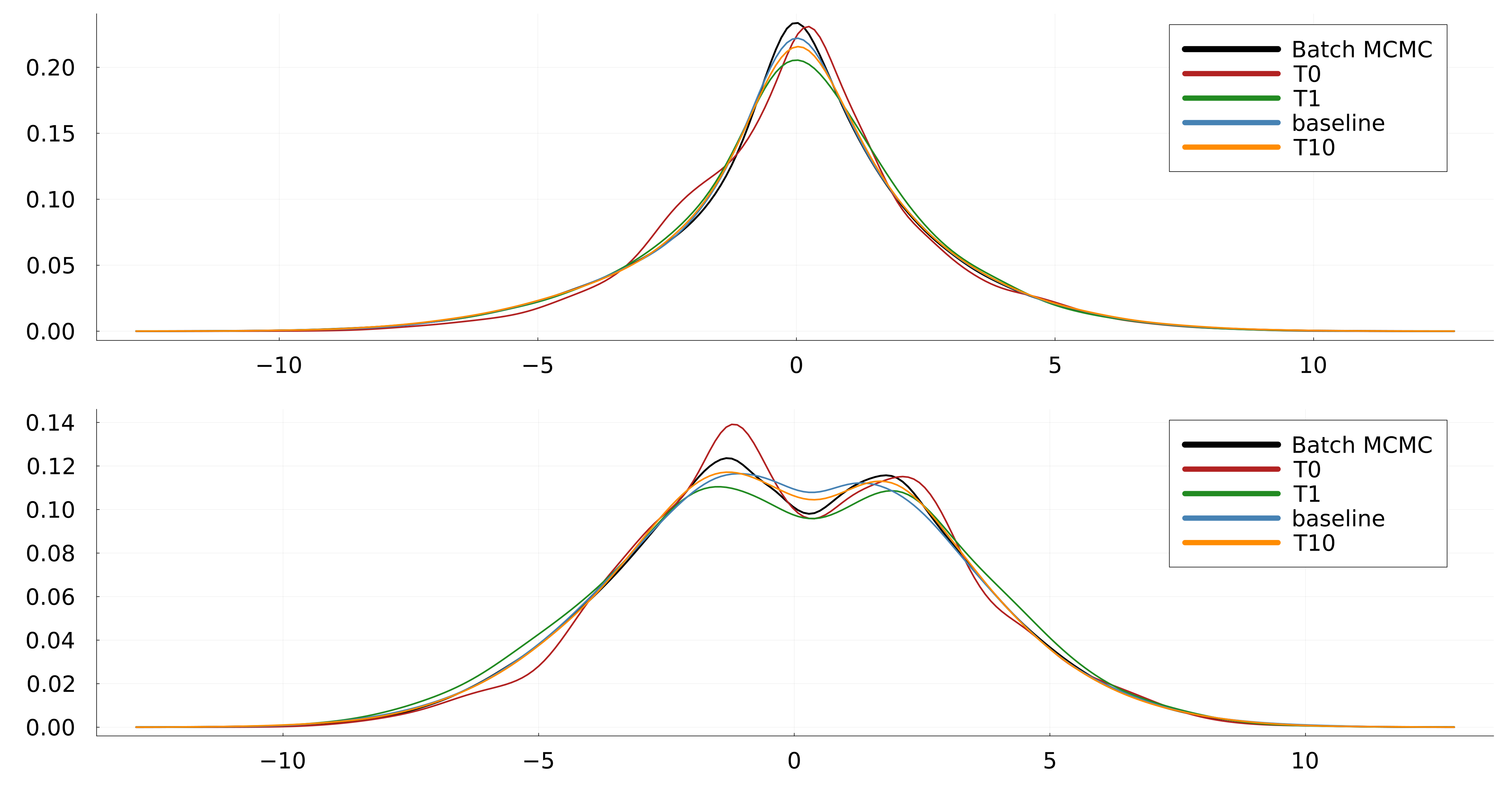}
    \caption{\rev{Effect of $T_{\mathrm{move}}$ on posterior densities: the proposed SMC algorithm versus batch MCMC, including the no-move setting ($T_{\mathrm{move}}=0$). Curves are averaged across 20 replications; baseline denotes $T_{\mathrm{move}}=5$. Upper: $\beta_1$ (intercept). Lower: $\beta_2$.}}
    \label{fig:T_effect}
\end{figure}

\begin{figure}[htbp]
    \centering
    \includegraphics[width=1.0\textwidth]{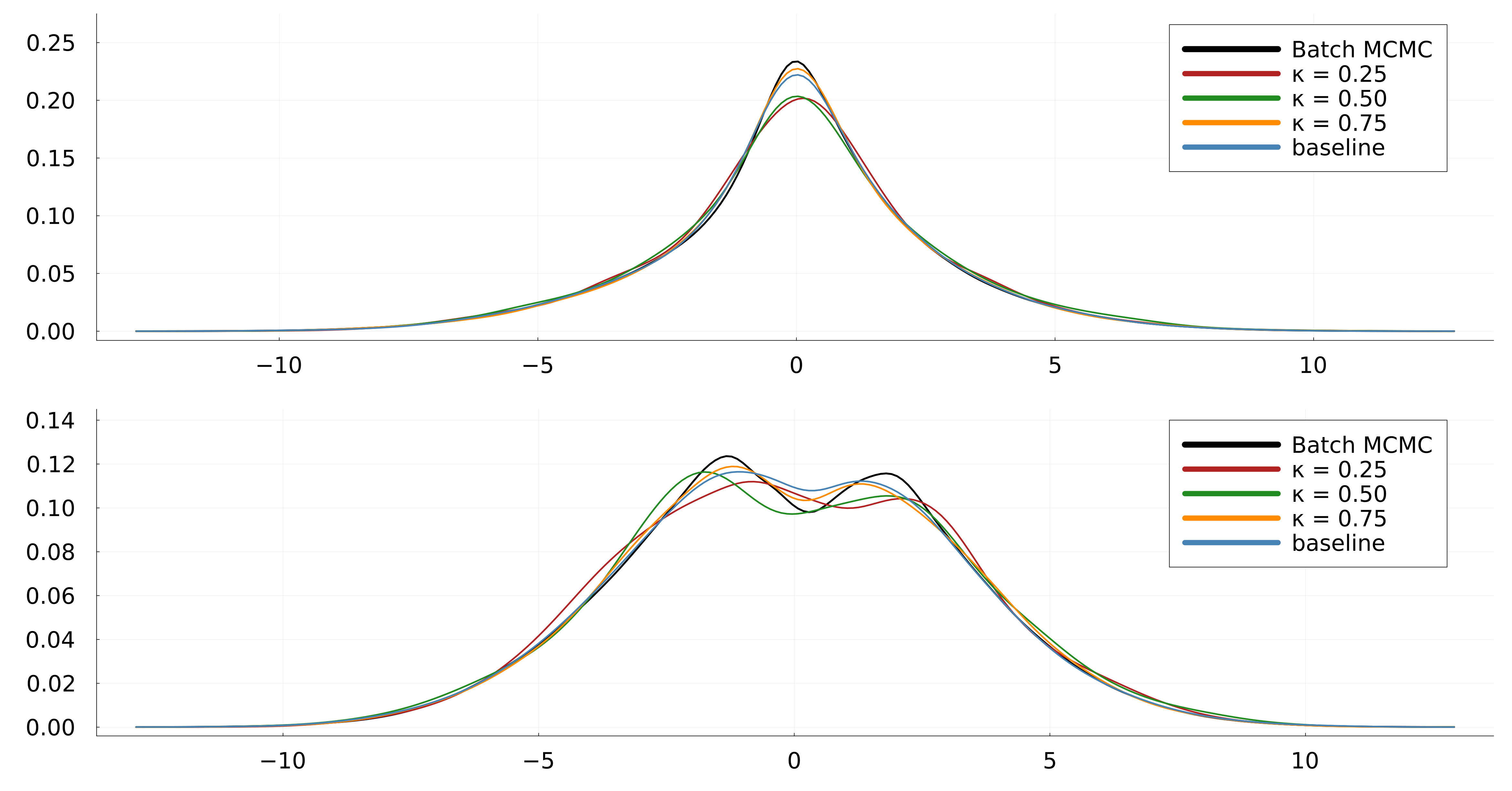}
    \caption{\rev{Effect of $\kappa$ on posterior densities: the proposed SMC algorithm versus batch MCMC. Curves are averaged across 20 replications; baseline denotes $\kappa=1$. Upper: $\beta_1$ (intercept). Lower: $\beta_2$.}}
    \label{fig:kappa_effect}
\end{figure}

\rev{\Cref{fig:beta_const,fig:beta_slope} show replication-level comparisons for the baseline setting. \Cref{fig:M_effect,fig:J_effect,fig:rho_effect,fig:T_effect,fig:kappa_effect} show replication-averaged density comparisons across tuning settings, complementing the runtime and posterior-fidelity comparisons in \Cref{sec:simres:base}. The Wasserstein-1 distances reported there are calculated separately for each replication before averaging, rather than from these averaged density curves.}

\clearpage
\section{\rev{Memory usage of the particle system}}
\label{app:additional-diagnostics}

\begin{revision}
This appendix reports memory usage under the baseline simulation setting in \Cref{sec:simulation}. \Cref{fig:simulation_memory} shows how the memory used by particle-specific quantities changes as observations are processed. Storage for the cluster labels $\zeta_{1:i}$ and PG variables $\omega_{1:i}$ increases linearly with the observation index, reaching 34.33~MiB at $i=3{,}000$. Memory for the parameters and sufficient statistics reaches 2.87~MiB and varies mainly with the total number of occupied clusters stored across particles. The figure excludes memory for the observed data $\D_{1:i}$, which are shared across all particles. These data must remain accessible for historical rejuvenation, either in memory or through external storage.

\end{revision}

\begin{revision}
\begin{figure}[H]
    \centering
    \includegraphics[width=0.78\textwidth]{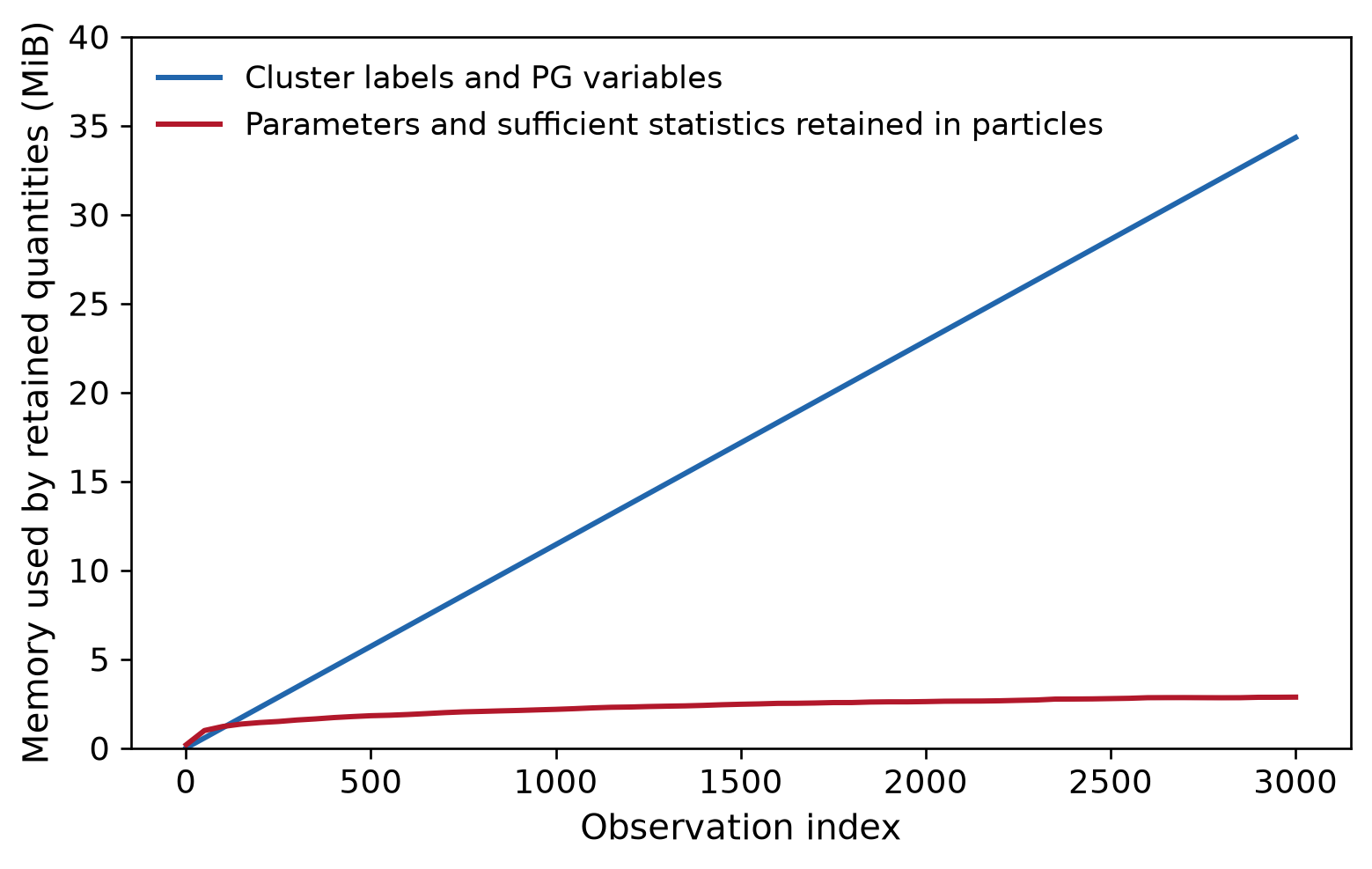}
    \caption{\rev{Memory used by particle-specific quantities retained over the sequential observation index in the baseline simulation. The lines show the cluster labels $\zeta_{1:i}$ and PG variables $\omega_{1:i}$, and the parameters and sufficient statistics retained in particles, averaged across 20 replications. Memory for the observed data, which are shared across all particles, is excluded.}}
    \label{fig:simulation_memory}
\end{figure}
\end{revision}

\clearpage
\end{appendices}

\section*{Declaration of Interest}
The authors declare no conflicts of interest.

\section*{Acknowledgements}

This work was supported by JSPS KAKENHI Grant Numbers JP23K19138 and JP24K17363. The first author also gratefully acknowledges support from the Kajima Foundation through its Researcher Exchange Support (Long-term Research Program), which enabled a research stay at University of Leeds in the United Kingdom during which part of this study was conducted.

The authors also thank Daisuke Kitahara of Nakatsugawa City and Hiroto Fukushima of Fukken Co., Ltd. for their assistance with survey implementation, data collection, and data processing for the illustrative empirical application.

\newpage
\bibliographystyle{plainnat}
\bibliography{references}

\end{document}